\documentclass[%
 reprint,prx,
superscriptaddress,
 amsmath,amssymb,
 aps,
]{revtex4-2}
\usepackage[dvipsnames]{xcolor}
\usepackage{tcolorbox}
\usepackage{wrapfig}
\usepackage{floatrow}
\usepackage{graphicx}%
\usepackage[english]{babel}
\usepackage{graphicx}
\usepackage{nccmath}
\usepackage{float}
\usepackage{mathrsfs}\newcommand{\rmd}{\mathrm{d}}
\newcommand{\lb}{\langle}
\newcommand{\rb}{\rangle}

\global\long\def\Cov{\mathrm{Cov}}

 \global\long\def\Var{\mathrm{Var}}
\pdfpageattr {/Group << /S /Transparency /I true /CS /DeviceRGB>>} 
\usepackage{mathtools}
\newlength{\arrow}
\newcommand*{\myrightarrow}[1]{\xrightarrow{\mathmakebox[\arrow]{#1}}}

\newlength{\extralongarrow}
\begin{document}

\preprint{APS/123-QED}

\title{Inferring gene regulation dynamics from static snapshots of gene expression variability}%

\author{Euan Joly-Smith}
\affiliation{Department of Physics, University of Toronto, 60 St.~George St., Ontario M5S 1A7, Canada}
\author{Zitong Jerry Wang}
\affiliation{Division of Biology and Biological Engineering, California Institute of Technology, Pasadena CA 91125, USA}
\author{Andreas Hilfinger}
\affiliation{Department of Physics, University of Toronto, 60 St.~George St., Ontario M5S 1A7, Canada}
\affiliation{Department of Mathematics, University of Toronto, 40 St.~George St., Toronto, Ontario M5S 2E4}
\affiliation{Department of Cell \& Systems Biology, University of Toronto, 25 Harbord St, Toronto, Ontario M5S 3G5}

\date{\today}%

\begin{abstract}
Inferring functional relationships within complex networks from static snapshots of a subset of variables is a ubiquitous problem in science. For example, a key challenge of systems biology is to translate cellular heterogeneity data obtained from single-cell sequencing or flow-cytometry experiments into regulatory dynamics. We show how static population snapshots of co-variability can be exploited to rigorously infer properties of gene expression dynamics when gene expression reporters probe their upstream dynamics on separate time-scales. This can be experimentally exploited in dual-reporter experiments with fluorescent proteins of unequal maturation times, thus turning an experimental bug into an analysis feature. We derive correlation conditions that detect the presence of closed-loop feedback regulation in gene regulatory networks. Furthermore, we show how genes with cell-cycle dependent transcription rates can be identified from the variability of co-regulated fluorescent proteins. %
Similar correlation constraints might prove useful in other areas of science in which static correlation snapshots are used to infer causal connections between dynamically interacting components.
\end{abstract}
\maketitle
\section{Introduction}
\vspace{-1.5em}
A large body of experimental work has quantified significant non-genetic variability in living cells \cite{Thattai2001, Ozbudak2002, Blake2003, Bar-Even2006, Newman2006, Eldar2010,elf2018,kaern2005}. Harnessing the information contained in this naturally occurring variability to infer molecular processes in cells without perturbation experiments is a long-standing goal of systems biology. However, 
measuring the spontaneous non-genetic variability of only one cellular component does not have sufficient discriminatory power to distinguish between models of complex cellular processes with many interacting components \cite{hilfinger2015b}.  Fortunately, progress in experimental techniques has made it possible to measure multiple components simultaneously. 
For example, covariances between mRNA and protein levels have been used to test hypotheses about translation rates in bacteria \cite{Taniguchi2010}. 

Despite improvements in experimental methods, it remains technically challenging to measure multiple different \emph{types} of molecules in the same cell. More feasible, and thus more common, are experiments that measure multiple levels of the same type of molecule, e.g., measuring different mRNA levels using sequencing techniques \cite{Cai2014}, or measuring abundances of proteins using fluorescence microscopy \cite{Elowitz2002}. Such experiments have motivated ``dual reporter'' approaches in which correlations between identical copies of reporters responding to a common upstream signal are used to characterize sources of variability within cellular processes \cite{Elowitz2002, Raser2004, maamar2007}.

Previous work focused on splitting the total observed variability into ``intrinsic'' and ``extrinsic'' contributions under the assumption that reporters are identical in all intrinsic properties. However, actual experimental reporters are never exactly identical. For example, commonly used fluorescent proteins differ enormously in their maturation half-lifes ranging from minutes to \mbox{hours \cite{Balleza2018}}. We show that despite such asymmetries, gene expression reporters can be used to rigorously detect closed-loop control networks from correlation measurements. Furthermore, we show that the inherent asymmetry of reporters can in fact be exploited to our advantage. Because reporters that differ in their intrinsic dynamics respond to their shared upstream input on different time-scales, their variability contains information about the unobserved upstream \emph{dynamics} even when we have only access to static population snapshots. For example, we show how asymmetric dual reporters can be used to distinguish periodically varying deterministic driving from stochastic upstream ``noise''. 

{The utility of these results lies in interpreting experimental data even when only a small part of a complex regulatory process can be observed directly. 
Instead of trying to model all of the many direct and indirect steps of gene expression regulation, we analyze entire classes of systems in which we specify only some steps but leave all other details unspecified. 
This approach allows us to derive inequalities that constrain the space of behaviour that could possibly be observed across a population of genetically identical cells within these classes, regardless of the details of the unspecified parts. 
These inequalities are in terms of coefficients of variation (CVs) and correlation coefficients of reporter levels, $x$ and $y$, engineered to readout components-of-interest, 
\begin{equation*}
CV_x: = \frac{\sqrt{\Var(x)}}{\lb x\rb} \quad 
\rho_{xy}:=\frac{\Cov(x,y)}{\sqrt{\Var(x)\Var(y)}}
\end{equation*}
where angular brackets denote population averages. 
Such population statistics are experimentally accessible from static snapshots of cellular populations that have reached a time-independent distribution of cell-to-cell variability. If a measurement violates our inequalities, one of our assumptions must be false, regardless of how the unspecified part of the system behaves. This way mathematical constraints can be used to deduce features of gene expression.}

While our results are motivated by methods in experimental cell biology to understand gene expression dynamics, they apply to any ``reporters'' embedded in a dynamic interaction network, subject to the specified production and elimination fluxes considered here and may thus be more broadly applicable.

\section{Detecting gene regulation feedback from static snapshots of population heterogeneity}
\vspace{-1em}
Cells employ both open-loop regulation or closed-loop feedback to control cellular processes \cite{shen2002network}. Here we show how to infer the presence of closed-loop feedback for any molecule within a network, by introducing an additional reporter molecule into the system. After describing the key theoretical result we detail the experimental set-up to detect feedback control in gene regulation from mRNA levels or fluorescent protein measurements. {In brief, our results apply to ``dual reporter'' genes that are engineered to share an unspecified but identical transcription rate. This assumption defines our class of models and needs to be experimentally ensured through appropriate genetic engineering in combination with self-consistency checks, such as indistinguishable reporter distributions, as for example reported in \cite{Raser2004, Bar-Even2006, Elowitz2002, baudrimont2019contribution}. Further experimental considerations are discussed in Sec.~\ref{SEC: Practical Applicability}}.

\vspace{-.5em}
\subsection{Mathematical correlation constraints for open-loop ``dual reporters''}
\begin{figure*}[!hbt]
\centering
  \includegraphics[width=0.95\columnwidth]{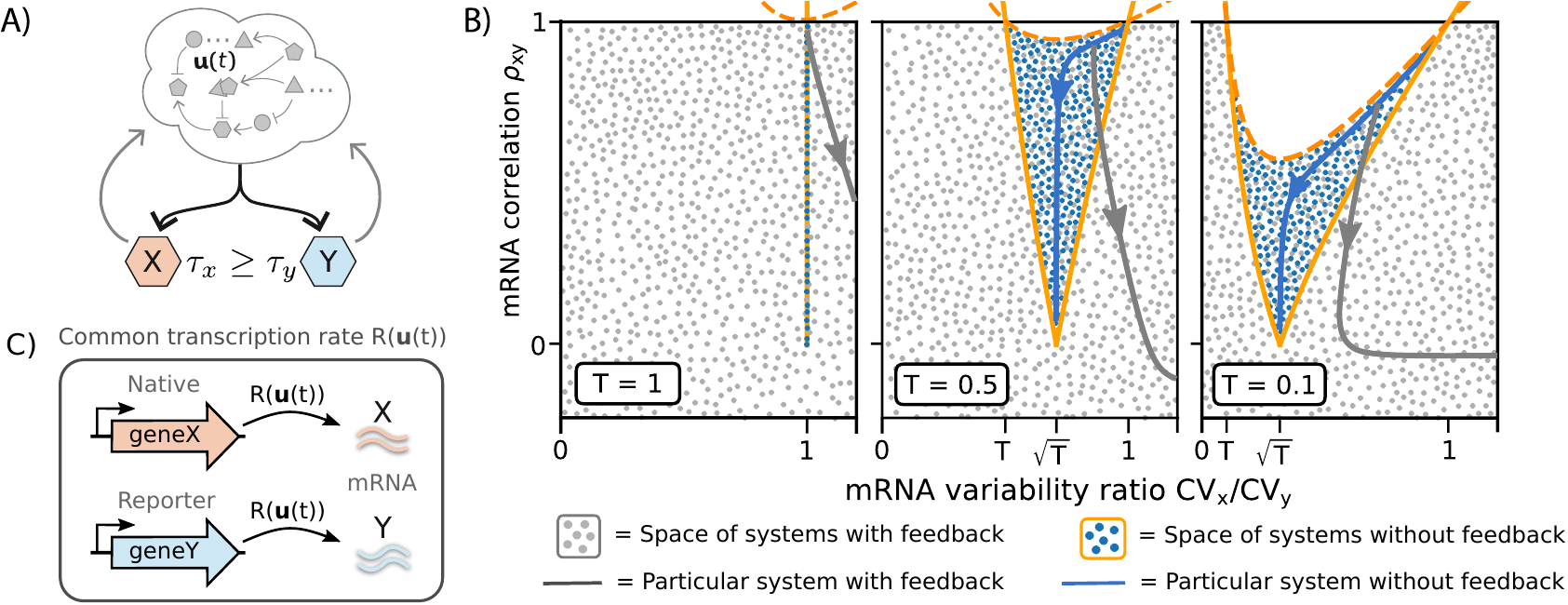}
   \caption{
    \textbf{Feedback in gene regulation affects the space of possible mRNA covariability.}
A) We consider all stochastic processes in which two components, X and Y, are made with an identical, but unspecified, rate. This rate can depend in any way on a cloud of unknown components $\mathbf{u}(t)$, which in turn can depend in arbitrary ways on the number of X and Y molecules. The shared production rate of X and Y, together with first-order degradation of X and Y with respective lifetimes $\tau_x$ and $\tau_y$,  are the only specified parts within this arbitrarily large network, as defined in Eq.~\eqref{EQ: Definition of 1-step Dual-Reporter System}.
B) Space of possible covariability for different values of the life-time ratio $T:=\tau_y / \tau_x$. All systems must satisfy \mbox{$\rho_{xy} (1+T) \leqslant CV_{x}/CV_{y}  + TCV_{y}/CV_{x}$} corresponding to the area below the dashed orange line. Allowing for feedback (grey dots), the entire space below the dashed orange line is accessible. In the absence of feedback (blue dots), $\rho_{xy}$ is additionally constrained by Eq.~\eqref{EQ: No feedback constraint (manuscript)}, corresponding to  the region between the solid orange lines. Dots are stochastic simulation data for specific models of co-regulated genes within the class defined in Eq.~\eqref{EQ: Definition of 1-step Dual-Reporter System}. {Plotted  are a subset of simulations with arbitrary density to demonstrate the accessibility of the  constrained regions. Blue and grey curves are exemplary toy models (Appendix \ref{SEC: Appendix particular systems}) that illustrate the effect of decreasing transcription rate variability (blue) or increasing feedback strength (grey)}. 
C) Experimental set-up to detect feedback in the regulation of a native gene of interest \emph{geneX}. The native and reporter genes are regulated by identical promoters in the same cell. If the covariability of the transcripts X and Y of such genes falls outside the open-loop constraint of Eq.~\eqref{EQ: No feedback constraint (manuscript)}, we can conclude that the gene of interest \emph{geneX} regulates its own transcription. {To maximize the discriminatory power of the approach the reporter \emph{geneY} should be engineered to be a passive read-out of transcription without significantly affecting gene expression.}}
    \label{FIG: Class 1 space of solutions -- feedback (manuscript)}
    \vspace{-.8ex}
\end{figure*}

\vspace{-.75em}
Motivated by the transcriptional dynamics of co-regulated genes, we consider a generic class of systems in which two cellular components, X and Y, are produced with a common (but unspecified) time-varying rate, and are degraded in a first order reaction, with average life-times $\tau_x$ and $\tau_y$, respectively 
\begin{align}
\vspace{-.8ex}
x & \myrightarrow{R(\mathbf{u}(t))}x+1\qquad & y\myrightarrow{R(\mathbf{u}(t))}y+1 \nonumber\\
x & \myrightarrow{x/\tau_{x}}x-1\qquad & y\myrightarrow{y/\tau_{y}}y-1
\label{EQ: Definition of 1-step Dual-Reporter System}
\vspace{-.8ex}
\end{align}
where the transcription rate can depend in any way on a cloud of unknown components $\mathbf{u}(t)$, which in turn can depend in arbitrary ways on the number of X and Y molecules, denoted by $x$ and $y$. While we characterize the stochastic reactions of X and Y, the dynamics of all other cellular components remain unspecified, see Fig.~\ref{FIG: Class 1 space of solutions -- feedback (manuscript)}A, {and the resulting dynamics need not be Markovian or ergodic in X and Y. A related class of stochastic processes has been previously considered to analyze mRNA-protein correlations in gene expression \cite{hilfinger2015b}, whereas here we analyze correlations between co-regulated transcripts.}

{Previous work established universal probability balance relations that constrain the stationary state distributions of stochastic processes \cite{hilfinger2015a}. For the above class of systems, these relations translate the  specified reactions of Eq.~\eqref{EQ: Definition of 1-step Dual-Reporter System} into an underdetermined system of equations for  (co)variances (see Appendix \ref{SEC: Appendix fluctuation balance relations})}. Though this system of equations cannot be solved due to the unspecified parts of the dynamics, not all dual reporter correlations $\rho_{xy}$ are accessible for all variability ratios $CV_x/CV_y$. For example, the correlation of any system is bounded by the correlation of two reporters that respond deterministically to their upstream input (Appendix \ref{SEC: Appendix fluctuation balance relations}). When the two reporters respond to their upstream input on unequal timescales, their correlations are constrained as illustrated by the dashed orange lines in Fig.~\ref{FIG: Class 1 space of solutions -- feedback (manuscript)}B that bound all systems for a given value of $T:= \tau_y/\tau_x$. {In Fig.~\ref{FIG: Class 1 space of solutions -- feedback (manuscript)} and throughout the paper, numerical simulations of specific stochastic models and parameters establish that the inequalities are tight, i.e., that the entire bounded regions are accessible. For illustration, we plot a subset of simulations with arbitrarily chosen sampling density, along with the analytically proven constraints.}

The space of possible correlations is restricted much further for open-loop systems in which upstream variables regulate the reporter production rates but are not affected by them, corresponding to all possible systems in  Fig.~\ref{FIG: Class 1 space of solutions -- feedback (manuscript)}A in which X and Y do not affect the unspecified cloud. 
{For all such systems, we can derive additional constraints by considering the hypothetical average of an ensemble of stochastic dual reporters conditioned on the history of their upstream influences.
While these conditional averages are typically experimentally inaccessible, they mathematically constrain the measurable (co)variances}. We find  {(see Appendix \ref{Appendix section on open-loop constraint})} that the correlations of cellular components X and Y that are regulated through an open-loop process must satisfy
\begin{equation}
\frac{\Big|\dfrac{CV_{x}}{CV_{y}}  - T\dfrac{CV_{y}}{CV_{x}}\Big|}{1-T}  \leqslant  \rho_{xy}  \quad \text{with} \quad T:= \tau_y/\tau_x,
\label{EQ: No feedback constraint (manuscript)}
\end{equation}
where without loss of generality we assume that $T\leqslant1$, i.e., that Y is the faster reporter. Fig.~\ref{FIG: Class 1 space of solutions -- feedback (manuscript)}B shows the above ``open-loop constraint'' of Eq.~\eqref{EQ: No feedback constraint (manuscript)} as solid orange lines, for specific values of $T$. Note that, in the symmetric limit  $T\to1$, Eq.~\eqref{EQ: No feedback constraint (manuscript)} reduces to $\rho_{xy} \geqslant 0$ and $CV_{x} = CV_{y}$ as intuitively expected.

Any system whose measured (co)variability falls outside the region defined by Eq.~\eqref{EQ: No feedback constraint (manuscript)} must be regulated through closed-loop feedback. Violations of this constraint can thus be used to experimentally detect the presence of feedback based on static population variability measurements without the need for perturbations. 
{Fig.~\ref{FIG: Class 1 space of solutions -- feedback (manuscript)}B shows the (co)variability of simulated systems with feedback (grey dots) and without feedback (blue dots), illustrating that only systems with feedback can fall outside the region bounded by solid orange lines, violating the ``open-loop constraint'' of Eq.~\eqref{EQ: No feedback constraint (manuscript)}.} 
The position of a system outside the ``open-loop constraint'' 
can be used to quantify a heuristic ``feedback strength'' and provides additional information to distinguish positive from negative feedback  (see Appendix \ref{SEC: Appendix types of feedback}).

\subsection{Experimentally exploiting mRNA correlations to detect feedback}
\vspace{-.75em}
{Eq.~\eqref{EQ: No feedback constraint (manuscript)} can be exploited through an experimental set-up analogous to previously engineered circuits in which co-regulated genes reportedly satisfied the assumptions of Eq.~\eqref{EQ: Definition of 1-step Dual-Reporter System} and transcripts were counted with single molecule fluorescence in-situ hybridization (smFISH) \cite{baudrimont2019contribution, raj2006stochastic}. To detect feedback regulation of a gene of interest \emph{geneX}, a reporter \emph{geneY} should be introduced whose expression is under the control of an identical copy of the promoter of \emph{geneX}, see Fig.~\ref{FIG: Class 1 space of solutions -- feedback (manuscript)}C.} The precise sequence of the reporter gene is unimportant as long as its transcript Y is sufficiently different from the transcript X of the gene of interest to avoid cross-hybridization by RNA probes. This can be achieved, e.g., by making the reporter gene sequence a scrambled version of the gene of interest. 

Using smFISH, transcripts X and Y can be measured simultaneously at the single cell level \cite{golding2013}. Simple population snapshots of cell-to-cell variability then determine whether experimentally observed $CV_{x}$, $CV_{y}$, $\rho_{xy}$ violate Eq.~\eqref{EQ: No feedback constraint (manuscript)}. If this open-loop constraint is violated we can conclude that \emph{geneX} must directly or indirectly affect its own production rate.

{If a system falls inside the region defined by Eq.~\eqref{EQ: No feedback constraint (manuscript)} we cannot say whether it is regulated through feedback or not, see Fig.~\ref{FIG: Class 1 space of solutions -- feedback (manuscript)}B. That is because systems with infinitesimally weak feedback are fundamentally indistinguishable from open-loop processes. To detect significant feedback in \emph{geneX}, it is advantageous to ensure the reporter \emph{geneY} is not involved in the same feedback regulation. This could, e.g., be achieved by removing the start codon from \emph{geneY} so its mRNA is not translated, thus preventing the protein of \emph{geneY} from exerting any feedback control. Furthermore, by ensuring that the life-time of the second reporter transcript is comparable to that of the gene of interest we can minimize the accessible area defined by Eq.~\eqref{EQ: No feedback constraint (manuscript)}  and thus maximize the discriminatory power of the approach.}

Note that only relative abundances are necessary to determine $CV_{x}$, $CV_{y}$, and $\rho_{xy}$. The above steps  can thus be applied to data from single cell sequencing techniques. Additionally, severe violations of Eq.~\eqref{EQ: No feedback constraint (manuscript)} can potentially be detected already from sequential rather than simultaneous measurements of X and Y: if their ratio of CVs falls outside the interval $[T,1]$ then Eq.~\eqref{EQ: No feedback constraint (manuscript)} must be violated regardless of the value of $\rho_{xy}$. When analyzing genes with potentially unknown mRNA life-time, the ratio of mRNA life-times $T$ can be inferred from pairwise correlation measurements between three reporter genes, as detailed in Sec.~\ref{SEC: Practical Applicability}.

\begin{figure*}[htb!]
\centering
\includegraphics[width=0.95\columnwidth]{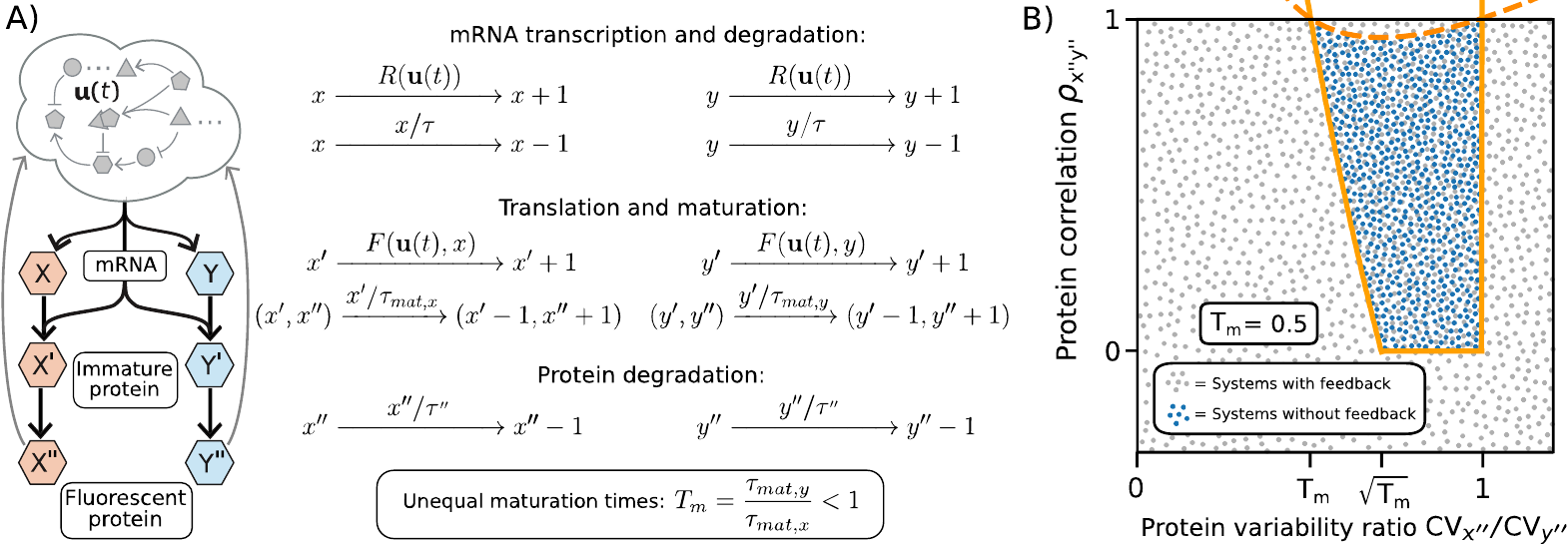}
\caption{ 
\textbf{Feedback in gene regulation affects the possible covariability of fluorescent protein measurements.}
A) 
As before X and Y correspond to co-regulated mRNA, but we now explicitly include the dynamics of immature fluorescent proteins denoted as X' and Y', as well as mature fluorescent proteins X'' and Y''. 
Because fluorescent proteins are typically stable and are thus effectively diluted with a common ``degradation'' time set by the cell-cycle, we focus on gene expression dynamics that is symmetric apart from the maturation step. The asymmetry between the co-regulated genes is then entirely characterized by the ratio of average fluorescent maturation times $T_\mathrm{m} := \tau_{\mathrm{mat},y} / \tau_{\mathrm{mat},x}$. B) Without feedback control (blue dots), fluorescence correlations are constrained to the region between the dashed and solid orange lines, the latter corresponding to the bound of Eq.~\eqref{EQ: two-step no-feedback constraint (manuscript)}.
Allowing for feedback (grey dots), the entire region becomes available. Correlations in fluorescence levels can thus be used to detect causal feedback in gene regulation from static population snapshots. {Dots are selected numerical simulations of specific fluorescent reporter systems within this class that illustrate the full accessibility of the region constrained by the analytically proven bounds.}}
\label{FIG: 3 step class of systems with maturation (manuscript)}
\end{figure*}

\vspace{-.5em}
\subsection{Experimentally exploiting fluorescent protein correlations to detect feedback}
\vspace{-.75em}
Similar constraints can be derived to interpret gene expression data in which fluorescent fusion proteins are used to quantify gene expression {\cite{Bar-Even2006}}. In this case the experimental read-out involves a fluorescent maturation step in addition to transcription and translation. While this maturation step is often well approximated as exponential \cite{Balleza2018}, its life-time differs significantly between commonly used fluorescent proteins. %
{For example, the maturation half-lifes of mCerulean, mEYFP, mEGFP, and mRFP1 are 6.6 min, 9 min, 14.5 min and 21.9 min respectively \cite{Balleza2018}}. 
In order to detect gene regulatory feedback from (co)variability data in such an experimental set-up, we extend our previous class of systems to explicitly account for protein translation and maturation events, see Fig.~\ref{FIG: 3 step class of systems with maturation (manuscript)}A. In this parallel cascade system X$''$ and Y$''$ represent the level of mature fluorescent proteins.

Crucially, open-loop systems in which none of the gene products directly, or indirectly, affect their transcription rate must additionally satisfy
\begin{equation}
\text{max}\left(\frac{T_{m}\mfrac{CV_{y''}}{CV_{x''}} - \mfrac{CV_{x''}}{CV_{y''}}}{1-T_m},0\right)  \leqslant  \rho_{x''y''}
 \hspace{.8em}\text{\&}\hspace{.86em}\frac{CV_{x''}}{CV_{y''}} \leqslant 1 ,
\label{EQ: two-step no-feedback constraint (manuscript)}
\end{equation}
where $T_\mathrm{m} := \tau_{\mathrm{mat},y} / \tau_{\mathrm{mat},x}$ is the ratio of maturation times between the two fluorescent proteins (see Appendix \ref{Appendix section on open-loop constraint for fluorescent reporters}).
The ``left boundary'' of this region is mathematically identical to the previous bound of Eq.~\eqref{EQ: No feedback constraint (manuscript)}, but now applied to the statistics of fluorescence levels X$''$ and Y$''$. The new right-hand bound $CV_{x''}\leqslant CV_{y''}$ broadens the region accessible to open-loop processes due the additional intrinsic degrees of freedom of this class of systems. {Fig.~\ref{FIG: 3 step class of systems with maturation (manuscript)}B illustrates the tightness of these bounds (solid orange lines) for $T_m = 0.5$, where dots correspond to simulated systems with (grey) and without feedback (blue).} 

Moreover,  analogously to the mRNA-correlations, all possible fluorescence correlations {for open-loop systems} are constrained by $\rho_{x''y''} (1+T_{m}) \leqslant T_{m} CV_{x''}/CV_{y''}  + T_{m}CV_{y''}/CV_{x''}$. {Correlations of systems with feedback can break this bound, as shown by the grey dots in Fig.~\ref{FIG: 3 step class of systems with maturation (manuscript)}B. This is because in the limit of infinitesimally small maturation times and mRNA fluctuations, the system becomes identical to Fig.~\ref{FIG: Class 1 space of solutions -- feedback (manuscript)}A with $T = 1$, which has unbounded correlations as shown in Fig.~\ref{FIG: Class 1 space of solutions -- feedback (manuscript)}B.}  

Fluorescent proteins can thus be {used} to detect whether a given gene regulates its own production as follows. Considering \emph{geneZ} as the native gene of interest, two recombinant genes, \emph{geneZ-GFP}  and \emph{geneZ-RFP} (or other spectrally distinguishable pairs) would be engineered into an isogenic cell population under the control of the same (but distinct) promoter as \emph{geneZ}. The transcripts of \emph{geneZ-GFP} and \emph{geneZ-RFP} then correspond to X and Y in Fig.~\ref{FIG: 3 step class of systems with maturation (manuscript)}A, as they are transcribed with identical rates. The level of mature fusion protein, X$''$ and Y$''$, can be read out at the single-cell level with fluorescence microscopy, and from the observed $CV_{x''}$, $CV_{y''}$, and $\rho_{x''y''}$ we can detect violations of the open-loop constraint Eq.~\eqref{EQ: two-step no-feedback constraint (manuscript)}.

If necessary, the discriminatory power of this approach can be increased by introducing a third fusion protein with a different fluorescent maturation time to eliminate the unknown internal degrees of freedom, and determine how much variability is generated through transcriptional, translational, or maturation events (see Appendix \ref{SEC: Stronger constraints}).

In modeling this parallel cascade system, we assumed the gene expression dynamics of the two fusion proteins is identical apart from the fluorescent maturation step. This is motivated by the experimental set-up in which we use the same native protein fused to two different fluorescent proteins. {The absence of further asymmetries, e.g., caused by codon usage, translation initiation, or reporter cross talk, would need to be established experimentally. Traditionally, this has been done by comparing the distributions of reporter variability for reporters that are claimed to be identical in their transcriptional and translational dynamics \cite{Raser2004, Bar-Even2006, Elowitz2002, baudrimont2019contribution}.  Note that the class of systems defined in Fig.~\ref{FIG: 3 step class of systems with maturation (manuscript)}A is in fact a subset of a much larger class of systems in which the key specified part is that there is a final maturation step in a symmetric but otherwise unspecified intrinsic cascade of arbitrary steps (see Appendix \ref{SEC: General class of fluorescent proteins}), thus allowing for a cascade of sequential post translational modifications that occur before the maturation step.}

{So far we considered arbitrary values of $T$ to allow for the experimental reality that reporters are never completely identical. Our next results show that $T\neq1$ is not just a nuisance but can be exploited to infer the dynamics of the transcription rate of open-loop systems.}

\section{Distinguishing stochastic from deterministic transcription rate variability} 
\vspace{-1em}
\begin{figure*}
\floatbox[{\capbeside\thisfloatsetup{capbesideposition={right,bottom},capbesidewidth=0.42\columnwidth}}]{figure}
{\caption{
    \textbf{
    Periodic transcriptional variability can be distinguished from stochastic variability {without experimental access to transcription rates and without following individual cells over time.}}
A) In a noisy cellular milieu, the auto-correlation of a periodic signal is not perfectly periodic but decays. We thus operationally define a signal as periodic when its periodicity is strong enough such that its auto-correlation function becomes negative at some point. Conversely, 
we classify signals with non-negative auto-correlations as stochastic. 
B) The left-side of the ``open-loop'' region defined by Eq.~\eqref{EQ: No feedback constraint (manuscript)} is only accessible by mRNA reporters that are driven by oscillatory production rates rather than purely stochastic upstream variability. The boundary (dashed black) line is defined by Eq.~\eqref{EQ: No oscillation constraint (manuscript)}. 
C) 
Sequential measurements of mRNA reporters X and Y, or fluorescent proteins X'' and Y'', can be used to discriminate between stochastic and oscillatory transcription rates if a system violates the respective bounds of Eq.~\eqref{EQ: No oscillation constraint (manuscript)} or Eq.~\eqref{EQ: two-step no oscillation constraint (manuscript)} (indicated by dashed black line). 
{Selected numerical simulation (dots) illustrate the achievability of the constrained regions, and the arrowed curves indicate how specific models (Appendix \ref{SEC: Appendix particular systems}) behave as the downstream response becomes slower.
For these systems, we find that oscillatory systems cross the black dashed line when $2\pi \sqrt{\tau_{x}\tau_{y}}$ or $2\pi \sqrt{\tau_{\mathrm{mat},x}\tau_{\mathrm{mat},y}}$ become slower than the period of the driving oscillation. To detect oscillations it is thus advantageous to choose slow reporters. }
 D) 
Periodic transcription rates of a gene of interest \textit{geneZ} can be experimentally detected either utilizing mRNA reporters, X and Y (left panel), or fluorescent protein reporters, X'' and Y'' (right panel), driven by the promoter of \textit{geneZ}. If experimental reporters violate Eq.~\eqref{EQ: No oscillation constraint (manuscript)} or \eqref{EQ: two-step no oscillation constraint (manuscript)}, they land to the left of the dashed black line (panel C), and \emph{geneZ} must be driven by a periodically varying transcription rate.}
    \label{FIG: Class 1 space of solutions -- oscillations (manuscript)}}
{\includegraphics[width=1\columnwidth]{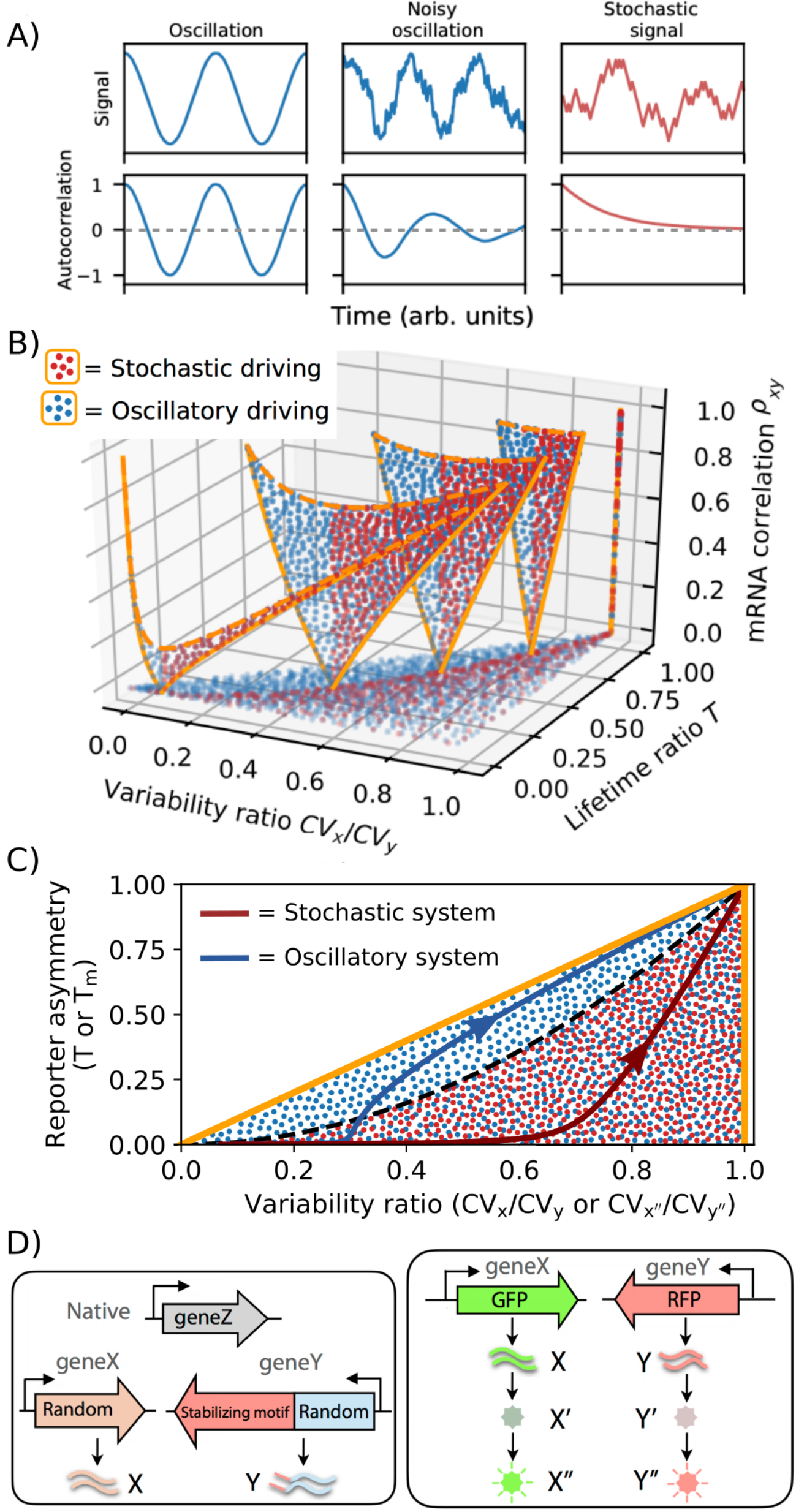} \hspace{2cm}}
\end{figure*}

A fundamental issue when interpreting cell-to-cell variability is that we generally do not know whether a component's variability is due to stochastic upstream noise or whether a component is driven by deterministic variability {\cite{eling2019challenges}}. Next, we show how we can distinguish the two types of dynamics from static population snapshots of asymmetric gene expression reporters. 
\vspace{-.5em}
\subsection{Mathematical correlation constraints for stochastic transcriptional noise}
\vspace{-.75em}
Focusing on the class of systems defined in Eq.~\eqref{EQ: Definition of 1-step Dual-Reporter System} in the absence of feedback, we next show how snapshots of dual reporters can be used to infer temporal properties of the unobserved production rates. To discuss periodic driving in cells we consider the stationary state auto-correlation of the transcription rate
\begin{equation*}
A(s) : = \frac{\lb R(t+s) R(t) \rb - \lb R(t+s)\rb \lb R(t) \rb}{\Var[R(t)]}.
\end{equation*}
We define a production rate as periodic if this auto-correlation $A(s)$ becomes negative for some $s$. In other words, the periodicity of the driving has to be strong enough such that the rate of production is negatively correlated with itself some time later. 
Conversely, we define upstream variability as stochastic if the auto-correlation of the unobserved transcription rate is non-negative everywhere, see Fig.~\ref{FIG: Class 1 space of solutions -- oscillations (manuscript)}A.

Fourier analysis of the dynamics of X and Y conditioned on the histories of their production rates shows (Appendix \ref{Appendix section on time-scales}) that not all systems can exhibit fluctuations everywhere within the region defined by Eq.~\eqref{EQ: No feedback constraint (manuscript)}. Components that are stochastically driven are additionally constrained by
\begin{equation}
\sqrt{T} \leqslant \frac{CV_{x}}{CV_{y}}
\label{EQ: No oscillation constraint (manuscript)} ,
\end{equation}
Sequential measurements of $CV_x$ and $CV_y$ from static snapshots of X and Y can thus discriminate between deterministic and stochastic transcription rates when $CV_x < CV_y \sqrt{T}$ without access to time-series data or directly measuring the unobserved upstream dynamics.
{Fig.~\ref{FIG: Class 1 space of solutions -- oscillations (manuscript)} illustrate this discriminatory power with simulated systems in which genes that are periodically driven (blue dots) can fill the entire region defined by Eq.~\eqref{EQ: No feedback constraint (manuscript)} (solid yellow lines), whereas genes that are driven stochastically (red dots) are further constrained by Eq.~\eqref{EQ: No oscillation constraint (manuscript)} (dashed black line).}

A pair of asymmetric ($T < 1$), co-regulated reporters in an open-loop system can exhibit perfect correlations ($\rho_{xy} = 1$) in two distinct regimes, see Fig.~\ref{FIG: Class 1 space of solutions -- oscillations (manuscript)}B. First, when the upstream variability is much slower than both $\tau_{x}$ and $\tau_{y}$, the reporters adjust rapidly to their quasi-stationary states such that $y(t) = T x(t)$ at all times, and thus $CV_x = CV_y$.
The second regime occurs when production rates oscillates rapidly, such that the upstream signal enslaves the reporters into a transient regime where their different response times %
simply shift their average dynamics. 
This second regime is not accessible by reporters that are driven stochastically. Instead, in the limit of infinitely fast stochastic variability, dual reporter systems approach $CV_{x}/CV_{y} \to \sqrt{T}$ corresponding to the bound of Eq.~\eqref{EQ: No oscillation constraint (manuscript)}. %
Where on the right-hand side a stochastically driven system falls can be used to infer the time-scale of the upstream fluctuations (Appendix \ref{Appendix section on time-scales}).

Inferring upstream dynamics from static snapshots is possible because the reporters probe their upstream dynamics on different time-scales. In fact, in the hypothetical limit of an infinite number of reporters responding to an upstream signal on all time-scales, the auto-correlation of the upstream signal is entirely determined by static measurements of its downstream variability. That is because knowing a signal's downstream variability on all time-scales effectively determines the Laplace transform of its auto-correlation, see Appendix \ref{Appendix section on time-scales}.

\subsection{Experimentally exploiting mRNA correlations to detect periodic transcription rates}
\vspace{-.75em}
The constraint of Eq.~\eqref{EQ: No oscillation constraint (manuscript)} can be exploited to detect {periodic transcription rates}, in experiments in which the {promoter} of a gene of interest is used to independently drive the expression of two non-native passive gene expression reporters {\cite{baudrimont2019contribution, raj2006stochastic}}. The details of the two reporter genes are not important as long as their transcripts have unequal life-times, and their products are not involved in any cellular control. Both criteria can be satisfied, e.g., by using a random sequence to encode the first reporter, and combining another random sequence with a mRNA stabilizing motif to encode the second reporter. This motif could be a 5' stem-loop structure in the case of prokaryotic cells or a carbohydrate recognition domain (CRD) sequence in eukaryotes \cite{emory19925,cheneval2010review}. Transcript levels of the two reporter genes then correspond to our components X and Y. Fig.~\ref{FIG: Class 1 space of solutions -- oscillations (manuscript)}D (left panel) illustrates this mRNA reporter set-up. 

Because sequential measurements of $CV_x$ and $CV_y$ suffice to detect violations of Eq.~\eqref{EQ: No oscillation constraint (manuscript)}, the two reporters X and Y do not need to be expressed simultaneously and can be measured independently at the single cell level, e.g., using smFISH \cite{Raj2013}.
Systems satisfying Eq.~\eqref{EQ: No oscillation constraint (manuscript)} can be either driven by stochastic or periodic transcription rates, but any violation strictly implies the existence of periodic upstream variability. To detect oscillations it is advantageous to use sufficiently long-lived mRNA reporters such that $2\pi\sqrt{\tau_{x}\tau_{y}}$ 
is comparable {or larger} to the period of the upstream signal. { This is demonstrated by the arrowed curves in Fig.~\ref{FIG: Class 1 space of solutions -- oscillations (manuscript)}C corresponding to exemplary oscillating and stochastic systems (defined in Appendix \ref{SEC: Appendix particular systems}) for varying $\tau_{y}$ and fixed $\tau_{x}$. This oscillating system crosses into the discriminatory region when $2\pi \sqrt{\tau_{x}\tau_{y}}$ is greater than the period of the upstream oscillation. Furthermore, Fig.~\ref{FIG: Class 1 space of solutions -- oscillations (manuscript)}C suggests an advantageous range for $T$ to detect oscillations:  $0.25 \leq T \leq 0.5$.} 

\subsection{Experimentally exploiting fluorescent protein correlations to detect periodic transcription rates}
\vspace{-.75em}
To detect transcriptional oscillations using co-regulated fluorescent reporter proteins we consider systems as defined in Fig.~\ref{FIG: 3 step class of systems with maturation (manuscript)}A in the absence of feedback. Just like transcript levels, we can prove (see Appendix \ref{SEC: Appendix dynamics fluorescent proteins}) 
that correlations between non-identical fluorescent proteins in the absence of periodic driving are constrained by
\begin{equation}
\sqrt{T_\mathrm{m}} \leqslant \frac{CV_{x''}}{CV_{y''}} ,
\label{EQ: two-step no oscillation constraint (manuscript)}
\end{equation}
where X$''$ and Y$''$ denote the fluorescence levels of the two reporter protein levels with a ratio of maturation times $T_\mathrm{m} := \tau_{\mathrm{mat},y} / \tau_{\mathrm{mat},x}$. In contrast, systems that are periodically driven can fill the entire region defined by Eq.~\eqref{EQ: two-step no-feedback constraint (manuscript)}. The constraint of Eq.~\eqref{EQ: two-step no oscillation constraint (manuscript)} can be used to detect oscillations in transcription rates from measures fluorescent protein variability analogous to the mRNA method discussed above, see Fig.~\ref{FIG: Class 1 space of solutions -- oscillations (manuscript)}C. The corresponding experimental set-up simply requires two different fluorescent proteins under the same transcriptional control as our gene of interest as illustrated in Fig.~\ref{FIG: Class 1 space of solutions -- oscillations (manuscript)}D (right panel).

\section{Applying theoretical bounds to experimental data}
\vspace{-1em}
\label{SEC: Practical Applicability}
{
The variables in our constraints are experimentally accessible: mRNA dual reporter life-time ratios, CVs, and correlations have been reported \cite{baudrimont2019contribution, raj2006stochastic} with measurements in the range $0.14 \leq T \leq 1$, $0.3 \leq CV_{x,y} \leq 3$, and $0.056 \leq \rho_{xy} \leq 0.89$. CVs from fluorescent protein reporters tend to be smaller, with CVs typically ranging from 0.1 to 1 \cite{Raser2004, Bar-Even2006, Elowitz2002}. Next we discuss real-world challenges when our constraints meet experimental data and analyze recent gene expression data quantifying population variability of constitutively expressed fluorescent proteins in \emph{E.~coli} \cite{Balleza2018}.
}
\vspace{-.5em}
\subsection{Unknown life-time ratios}\vspace{-.75em}
Our constraints depend on the ratio of reporter life-times or maturation times. For many fluorescent proteins, maturation times can be obtained from the literature \cite{Balleza2018}, but mRNA life-times may not be precisely known or might be highly context dependent. This problem can be overcome through the addition of a third reporter: by measuring pairwise correlations between three dual reporter transcripts within the class of Eq.~\eqref{EQ: Definition of 1-step Dual-Reporter System}, we can determine the ratio of life-times between any pair of three reporters (Appendix \ref{SEC: Appendix inferring unknown lifetime ratios}). In particular, the ratio $T:=\tau_y/\tau_x$ is given by 
 \begin{equation}
     T=\frac{\eta_{xx}-\eta_{xw}}{\eta_{yy}-\eta_{yw}},
     \label{EQ: T (manuscript)}
 \end{equation}
 where $w$ denotes the abundance of a third co-regulated mRNA reporter (with unknown/arbitrary life-time), {and \mbox{$\eta_{fg}:= \text{Cov}(f,g)/(\lb f \rb \lb g \rb)$} denote normalized covariances obtained from population snapshots.}

Similarly, for fluorescent proteins within the class of Fig.~\ref{FIG: 3 step class of systems with maturation (manuscript)}A, we can determine the ratio of maturation times $T_m$ by measuring the correlations of X$''$ and Y$''$ with two other fluorescent proteins of known maturation time (see Appendix \ref{SEC: Appendix inferring unknown lifetime ratios}). Utilizing additional reporters to determine unknown asymmetries between X and Y is a successful general strategy because the number of constraints on a system increases faster than the number of parameters when adding additional reporters and observing their (co)variance. For example, next we specify how unknown constants of proportionality in transcription or experimental detection can be inferred from pairwise correlation measurements with additional reporters.

\begin{figure*}[hbt!]
\begin{center}
\includegraphics[width=1\columnwidth]{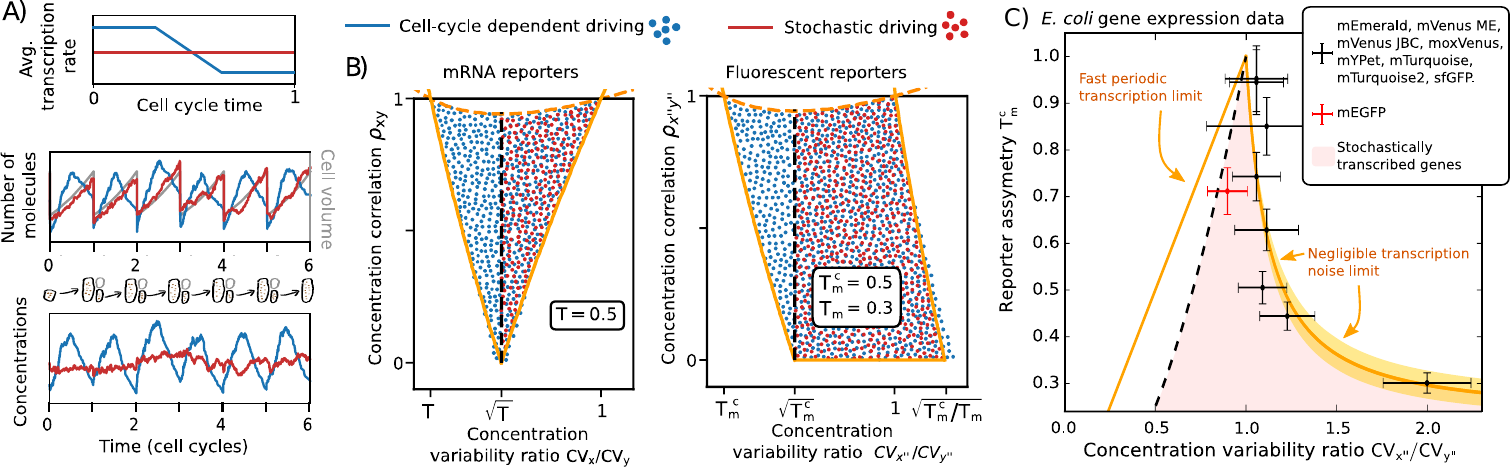}
\end{center}
\vspace{-0.5em}
\caption{ \textbf{Utilizing bounds to analyze concentrations in growing and dividing cells.}
A) 
We want to distinguish stochastically varying transcription rates whose average remains constant throughout the cell-cycle (red line) from transcription rates that change periodically during the cell-cycle (exemplified by blue line). 
Numbers of molecules oscillate in both types of systems but only systems with cell-cycle dependent driving oscillate in \emph{concentrations}. {Our constraints can identify such periodic gene expression from static snapshots of population variability without access to time-series data.} 
B) Solid orange lines denote open-loop constraints proven for systems in which the reporter concentrations are independent of the cell volume, under the assumption that the cell volume grows exponentially between division events. For mRNA reporters (left panel) concentration correlations are then bounded by the same constraints derived for absolute numbers in the absence of feedback. When transcription rates are periodic, the open-loop constraint of Eq.~\ref{EQ: No feedback constraint (manuscript)} for absolute numbers does not strictly apply to concentrations. Numerical simulations of example systems (blue dots) suggest that such systems only fall fall marginally outside the orange bounds. For fluorescent reporters (right panel), the feedback bounds (orange lines) for concentrations depend on the additional parameter \mbox{$T_{m}^{c} := (1/\tau_{\mathrm{mat},x} + \ln(2)/\tau_c)/(1/\tau_{\mathrm{mat},y} + \ln(2)/\tau_c)$} where $\tau_c$ is the average cell-cycle time, see Eq.~\eqref{EQ: two-step no-feedback constraint concentrations (manuscript)}. 
However, for both mRNA and fluorescence levels, Eqs.~\eqref{EQ: No oscillation constraint (manuscript)} and \eqref{EQ: two-step no oscillation constraint (manuscript)} (dashed black lines) can be used to detect cell cycle dependent genes from asynchronous static snapshots of dual reporter concentrations.
{
C) Previously reported gene expression variability data for constitutively expressed fluorescent proteins that exhibit first order maturation dynamics in \emph{E.~coli} \cite{Balleza2018}. 
Plotted are variability ratios and reporter asymmetries with respect to the observed dynamics of mEYFP. 
As expected for constitutively expressed fluorescent proteins, the experimental data is consistent with the class of gene expression models that do not exhibit feedback and are not periodically driven (pink region) bounded by Eq.\eqref{EQ: two-step no oscillation constraint (manuscript)} and Eq.\eqref{EQ: two-step sequential no-feedback constraint (manuscript)}. 
With the exception of the indicated mEGFP outlier, all data fall along the right hand boundary that is only accessible for systems whose fluorescence variability is dominated by a translation, maturation, or machine read-out step. The yellow corridor indicates the estimated uncertainty in reported cell-cycle time $\tau_c =(28.5\pm2)$~min and the reference mEYFP maturation half-life $\tau_{mat,y}\ln(2)  = (9.0\pm0.7)$~min. Variability ratios with respect to all other reference fluorescence proteins confirm the above picture with the exception of the observed mEGFP variability which violates the expected behaviour, see Appendix \ref{SEC: Appendix Experimental data analysis}.
}
}
\vspace{-1em}
\label{FIG: Concentrations (manuscript)}
\end{figure*} 

\vspace{-.5em}
\subsection{Proportional transcription rates}\vspace{-.75em}
So far we considered co-regulated genes that have identical (though probabilistic) transcription rates. This is motivated by experimental synthetic reporter systems in which two fluorescent proteins are expressed under two identical copies of a known transcriptional promoter. Additionally, the above results can be extended to co-regulated genes in which the transcription rates are not identical but proportional to each other, i.e., $R_{y} = \alpha R_{x}$ for some $\alpha$. Similar bounds hold for this new class of systems (see Appendix \ref{SEC: Proportional rates}) but they depend on the proportionality factor $\alpha$ which is potentially unknown. For the class of systems defined in Eq.~\eqref{EQ: Definition of 1-step Dual-Reporter System} this issue can be resolved by introducing an additional reporter to determine the constant of proportionality between the transcription rates. If the life-time ratio $T$ is not known \emph{a priori}, then four reporters would be needed in order to determine all unknowns. Additionally, for synthetic fluorescent protein reporters, the constant of proportionality in transcription can be inferred from absolute numbers of molecules. If measurements only report relative numbers, $\alpha$ can be inferred through additional reporter correlations or sequential single-color experiments in which the same fluorescent protein is expressed under the two different promoters (see Appendix \ref{SEC: Measuring alpha and detection probabilities}).

\vspace{-.5em}
\subsection{Systematic undercounting}\vspace{-.75em}
Experimental data in cell biology might not reflect absolute abundances. For example, mRNA molecules are often counted using fluorescence in situ hybridization (FISH), a method that can lead to a systematic undercounting of the copy number due to probabilistic binding between mRNA molecules and the fluorescent probes. Similarly, sequencing methods rely on probabilistic amplification steps to detect transcripts. %
However, as long as each type molecule is experimentally detected with the same probability, the above bounds and accessible regions are strictly unaffected by the detection probabilities (Appendix \ref{SEC: Undercounting}). This can be intuitively understood because undercounting corresponds to a binomial read-out step which de-correlates measurements just like the already accounted for  stochastic reaction steps. While reported averages, variances, and correlations strongly depend on detection probabilities, their accessible region is bounded by the same inequalities described above.

If different reporters have a different probability of being detected we can generalize the above constraints to account for that and infer the missing parameter from pairwise correlations between three reporters (see Appendix \ref{SEC: Measuring alpha and detection probabilities}).

\vspace{-.5em}
\subsection{Concentration measurements instead of absolute numbers}\vspace{-.75em}
Growing and dividing cells exhibit a natural cycle in which, on average, cell-division ``eliminates'' half the molecules of a cell while also reducing the cell volume by a factor of two. Instead of considering absolute numbers, many experiments thus report concentrations of molecules which (on average) are unaffected by cell-division. In the preceding discussion, the specified reactions describe the birth and death of molecules, but exactly the same constraints of Eqs.~\eqref{EQ: No oscillation constraint (manuscript)} and \eqref{EQ: two-step no oscillation constraint (manuscript)} can be used to analyze concentration measures of growing and dividing cells. When using concentrations to determine CVs, violations of Eqs.~\eqref{EQ: No oscillation constraint (manuscript)} and \eqref{EQ: two-step no oscillation constraint (manuscript)} can then detect genes whose transcription rate varies periodically during the cell-cycle, see Fig.~\ref{FIG: Concentrations (manuscript)}. In this analysis of concentrations we allow for binomial splitting noise, division time fluctuations, and asymmetric divisions, and we assume that cellular volume grows exponentially between divisions (Appendix \ref{SEC: Appendix concentrations}).

{When reporter concentrations are independent of cell volume, open-loop constraints for concentrations similar to the ones derived for absolute numbers can be analytically proven (Appendix \ref{SEC: Appendix concentrations}).
For co-regulated mRNA this constraint is exactly identical to Eq.~\eqref{EQ: No feedback constraint (manuscript)}, but for fluorescent proteins the equivalent of Eq.~\eqref{EQ: two-step no-feedback constraint (manuscript)} changes because concentrations are subject to an additional degradation rate from dilution governed by the average cell-cycle time $\tau_{c}$. In the absence of feedback, CVs in fluorescence \emph{concentrations} that are volume independent are constrained by
\begin{equation}
\text{max}\left(\frac{T_{m}^{c}\frac{CV_{y''}}{CV_{x''}}-\frac{CV_{x''}}{CV_{y''}}}{1-T_m^{c}},0\right)  \leqslant \rho_{x''y''}\leqslant 
\frac{\frac{T_{m}^{c}}{T_{m}}\frac{CV_{y''}}{CV_{x''}} - \frac{CV_{x''}}{CV_{y''}}}{\frac{T_{m}^{c}}{T_{m}} - 1}
\label{EQ: two-step no-feedback constraint concentrations (manuscript)}
\end{equation}
where $T_{m}^{c} = (1/\tau_{mat,x} + \ln(2)/\tau_{c})/ (1/\tau_{mat,y} + \ln(2)/\tau_{c})$, and these bounds are indicated by the solid orange lines in Fig.~\ref{FIG: Concentrations (manuscript)}B (right panel).
}

{
Systems in which the reporter concentrations are not independent of the volume are not bound by the above constraints.
However, numerical simulations suggest that CVs in concentrations violate inequalities Eqs.~\eqref{EQ: No feedback constraint (manuscript)},
\eqref{EQ: two-step no-feedback constraint concentrations (manuscript)} only marginally, see blue dots in Fig.~\ref{FIG: Concentrations (manuscript)}B. An exact bound can be derived to strictly constrain this class of systems if we have experimental access to a third reporter (see Appendix \ref{SEC: Appendix exact constraints on volume dependent genes}).
}

{
\subsection{Data from constitutively expressed fluorescent proteins fall within the expected bounds}\vspace{-.75em}
Ultimate proof that our constraints can be used to identify periodically expressed genes will consist of experimentally observing CVs of cell-cycle dependent promoters that violate Eq.~\eqref{EQ: No oscillation constraint (manuscript)} or Eq.~\eqref{EQ: two-step no oscillation constraint (manuscript)} in physiologically relevant regimes. However, even in the absence of such direct verification, the applicability of our method can be supported through a self-consistency check, i.e., whether data for fluorescent proteins expressed through a constitutive promoter fall into the expected region of gene expression dynamics that is stochastically driven and not subject to feedback control. Such data exists in \emph{E.~coli} for fluorescent proteins with precisely determined maturation dynamics \cite{Balleza2018}. Here, we analyze this cell-to-cell variability data for fluorescent reporters that exhibited clear first-order maturation dynamics by determining the CV in protein concentration after subtracting volume variability of growing and dividing cells (Appendix~\ref{SEC: Appendix Experimental data analysis}). 
}

{
In the absence of simultaneous fluorescence measurements, sequential CV measurements in concentrations are constrained by 
\begin{align}
    T_{m}^{c} \leqslant \frac{CV_{x''}}{CV_{y''}} \leqslant \sqrt{\frac{T_{m}^{c}}{T_{m}}},
    \label{EQ: two-step sequential no-feedback constraint (manuscript)}
\end{align}
which corresponds to the bounds of Eq.~\eqref{EQ: two-step no-feedback constraint concentrations (manuscript)} when allowing for any value of the correlation coefficient (unobserved for sequential variability measurements). As discussed in the previous section, the no-oscillation constraint of Eq.~\eqref{EQ: two-step no oscillation constraint (manuscript)} still applies directly even when we analyze concentrations rather than absolute numbers.
}

{
In Fig.~\ref{FIG: Concentrations (manuscript)}C we present experimentally determined variability data \cite{Balleza2018} for fluorescent proteins under control of the constitutive promoter \emph{proC} in \emph{E.~coli} with mEYFP variability and maturation dynamics as a reference point. As expected for constitutively expressed fluorescent proteins, the data for these biologically ``boring'' systems is consistent with the class of gene expression models that do not exhibit feedback and are not periodically driven (pink region), bounded by Eq.\eqref{EQ: two-step no oscillation constraint (manuscript)} and Eq.\eqref{EQ: two-step sequential no-feedback constraint (manuscript)}.
Error bars for $T_{m}^{c}$ in Fig.~\ref{FIG: Concentrations (manuscript)}C are taken from \cite{Balleza2018} with error propagation, and error bars in CVs and their ratios are the standard error of the mean from three independent cell cultures with error propagation respectively. 
}
{
Furthermore, all data (with the exception of mEGFP) follow the right boundary of the allowed region. 
The observed variability  of mEGFP is confirmed as an outlier when using the mEGFP measurements as a reference point for which the observed CV-ratios violate our predicted bounds (see Appendix \ref{SEC: Appendix Experimental data analysis}). One possible explanation of this outlier would be that mEGFP cells grew more slowly than the experimentally reported \mbox{28.5 min} cell-division times for all strains.
}

{
Data that falls on the right hand boundary is consistent with negligible biological noise upstream of translation or with variability that is dominated by technical measurement noise whose normalized variance decreases with the inverse of the mean population signal (see Appendix \ref{SEC: Appendix Experimental data analysis}).}

\vspace{-.5em}
{
\subsection{Measurement noise \& technological limitations}\vspace{-.75em}
Theoretical constraints can be experimentally exploited as long as measurement uncertainties are sufficiently small. When considering the sampling error of cell-to-cell variability, 95\% confidence intervals for CVs have been reported to be around 5-10\% of the respective CV value \cite{Elowitz2002, raj2006stochastic}. %
Similarly, experiments have shown high biological reproducibility, e.g., three repeats of identical flow-cytometry experiments exhibited a standard error of the CVs of around 10\% \cite{Balleza2018}. 
However, a significant limitation remains for current experimental high-throughput methods like flow-cytometry or single-cell sequencing: the measurements techniques themselves introduce significant noise, especially when used with bacteria \cite{galbusera2020using,grun2014validation}, which means that technical variability can introduce a systematic error in estimates of biological variability. 
In such cases, one can attempt to deconvolve true biological variability from measurement noise using experimental and analytical techniques, e.g., by using calibration beads \cite{galbusera2020using} or by using noise models \cite{grun2014validation}. Alternatively, one can opt for methods of lower throughput but greater precision, for example, mRNA measurements by single-molecule FISH (smFISH) are well-suited for validation of our method given its accuracy and high sensitivity \cite{raj2008imaging}. 
Rapid improvements in high throughput quantification tools such as sequential FISH (seqFISH) will provide exciting opportunities for future applications of our work \cite{eng2019transcriptome}.
}

\section{Discussion}
\vspace{-1em}

While our results are motivated by the analysis of gene expression dynamics, utilizing correlation constraints to characterize the dynamics within complex processes may prove useful in other areas of science in which systems involve many interacting components whose dynamics is difficult or impossible to track completely. 
Our results show that it is possible to rigorously analyze correlations within classes of incompletely specified physical models without resorting to statistical inference methods. Crucially, our approach does not require time-resolved data, which is often unavailable for complex systems. For example, following individual cells over time is much more challenging, and thus less common, than taking static population snapshots of cellular abundances. 

Our correlation constraints provide a framework to detect the presence of feedback strictly from observations of a small subset of components within a much larger cellular process. This could, e.g., be utilized to pinpoint molecular components that are involved in feedback regulation of a gene by observing how our proposed signature of feedback is affected in knock-out experiments. Additionally, our results highlight that using unequal reporters can reveal dynamic properties of regulation even in the absence of temporal data. These constraints are fundamentally due to the dynamics of interactions and are not apparent in previous work that considered asymmetric dual reporters as static random variables \cite{Rhee2014}.
For example, we show that theoretically cell-cycle dependent transcription rates can be detected from static population measurements of asymmetric downstream products of gene expression. Additionally, we explicitly show that our mathematical framework can be utilized even when experimental techniques detect individual molecules only probabilistically or when key parameters are unknown.  
{Finally, we report an experimental ``negative control'' in which we confirm that the measured variability of constitutively expressed fluorescence proteins falls into the expected region of gene expression variability for genes that are not subject to feedback regulation or periodic driving.}

\begin{acknowledgments}
\vspace{-1em}
We thank Raymond Fan, Brayden Kell, Seshu Iyengar, Timon Wittenstein, Sid Goyal, Ran Kafri, and Josh Milstein for many helpful discussions. We thank Laurent Potvin-Trottier and Nathan Lord for valuable feedback on the manuscript. This work was supported by the Natural Sciences and Engineering Research Council of Canada and a New Researcher Award from the University of Toronto Connaught Fund. %
AH gratefully acknowledges funding through grant NSF-1517372 while in Johan Paulsson's group at Harvard Medical School.
\end{acknowledgments}

\appendix

\section*{Appendix}
\vspace{-1em}
Here we detail the mathematical derivations and illustrate some of the results in greater depth. {Throughout the appendix we denote the normalized (co)variance as \mbox{$\eta_{fg}:= \text{Cov}(f,g)/(\lb f \rb \lb g \rb)$}, where the statistical measures are stationary population averages.}

First we go over the mathematical framework in which we model reaction networks in cells.
The state of an intracellular biochemical network at a given moment in time is given by the integer numbers $\{x_{i}\}$ of the chemical species $\{X_{i}\}$, which we can write as 
$\mathbf{x} = (x_{1}, \dots, x_{n})$.
This system is dynamic, and so this state $\mathbf{x}$ will undergo discreet changes over time as reactions occur. If there are $m$ possible reactions in the system, we can write them as
\begin{align*}
    \mathbf{x} \xrightarrow[]{\quad r_{k}(\mathbf{x}) \quad} \mathbf{x} + \mathbf{d}_{k}  \qquad k = 1,\dots,m 
\end{align*}
where the rate $r_{k}(\mathbf{x})$ corresponds to the probability per unit time of the $k$-th reaction occurring, and the step size $\mathbf{d}_{k} = (d_{1k}, \dots, d_{nk})$ 
corresponds to the change in the chemical species numbers $\mathbf{x}$ from the $k$-th reaction (the $d_{ik}$ are positive or negative integers).
It then follows that the system probability distribution $P(\mathbf{x},t)$ evolves according to the chemical master equation \cite{vank1992, Lestas2008}:
\begin{align*}
    \frac{d}{dt}P(\mathbf{x},t) = \sum_{k}\left[r_{k}(\mathbf{x} - \mathbf{d}_{k})P(\mathbf{x} - \mathbf{d}_{k},t) - r_{k}(\mathbf{x})P(\mathbf{x},t)\right] .
\end{align*}
Time evolution equations for the moments $\langle x_{i}^{k} \rangle$ of each component $X_{i}$ follow directly from the chemical master equation  \cite{Lestas2008}.

\vspace{-.5em}
\section{Fluctuation balance relations for systems as defined in Eq.~\eqref{EQ: Definition of 1-step Dual-Reporter System}}\vspace{-.75em}
\label{SEC: Appendix fluctuation balance relations}
Previous work established general relations that constrain fluctuations of components within incompletely specified reaction networks \cite{hilfinger2015a}.
In particular, any two components Z$_{1}$ and Z$_{2}$ in an arbitrarily complex network that reach wide-sense stationarity, must satisfy the following flux balance relations 
\begin{align}
    \langle R_{1}^{+} \rangle = \langle R_{2}^{-}\rangle,
    \label{EQ: Appendix general flux balance equation}
\end{align}
as well as the following fluctuation balance relations
\begin{align}
    \text{Cov}(z_{1},R_{2}^{-}\hspace{-.2em}-\hspace{-.2em} R_{2}^{+}) + \text{Cov}(z_{2},R_{1}^{-} \hspace{-.2em}-\hspace{-.2em} R_{1}^{+})  \hspace{-.2em}=\hspace{-.2em} \sum_{k}d_{1k}d_{2k}\langle r_{k}\rangle,
    \label{EQ: Appendix general covariance balance equation}
\end{align}
where $R_{i}^{+}$ and $R_{i}^{-}$ are the net birth and death fluxes of component Z$_{i}$, and  the summation  is over all reactions in the network, with $r_{k}$ the rate of the $k$-th reaction, and $d_{ik}$ the step-size of Z$_{i}$ of the $k$-th reaction. For the class of systems in Eq.~\eqref{EQ: Definition of 1-step Dual-Reporter System}, 
Eqs.~\eqref{EQ: Appendix general flux balance equation} and \eqref{EQ: Appendix general covariance balance equation} imply
\begin{align}
 \langle R \rangle = \langle x \rangle /\tau_{x} \quad \text{\&} \quad \langle R \rangle = \langle y \rangle /\tau_{y} ,
 \label{EQ: 1-step flux-balance relations}
\end{align}
and
\begin{align}
\begin{split}
&\eta_{xx} = \frac{1}{\lb x \rb} + \eta_{xR} ,\quad \eta_{yy} = \frac{1}{\lb y \rb} + \eta_{yR} , \\	 
&\eta_{xy} = \frac{1}{1+T}\eta_{xR}  + \frac{T}{1+T}\eta_{yR}  . 
\end{split}
\label{EQ: 1-step fluctuation-balance relations}
\end{align}
These relations must be satisfied by all systems in the class, regardless of the unspecified details. This allows us to set general constraints 
that hold for the entire class of systems. For example, the above relations lead to
\begin{align*}
\eta_{xy} 
	  &= \frac{1}{1+T}\left(\eta_{xx} - \frac{1}{\lb x \rb}\right) + \frac{T}{1+T}\left(\eta_{yy} - \frac{1}{\lb y \rb}\right) \\
	  &\leq \frac{1}{1+T}\eta_{xx} + \frac{T}{1+T}\eta_{yy} ,
\end{align*}
where in the 2nd step we used  fact that the averages are positive. Dividing this bound by $\sqrt{\eta_{xx}\eta_{yy}}$ leads to $
\rho_{xy}(1+T) \leqslant \frac{CV_{x}}{CV_{y}}  + T\frac{CV_{y}}{CV_{x}}$, 
which is the bound indicated by the dashed orange line in Fig.~\ref{FIG: Class 1 space of solutions -- feedback (manuscript)}B. The inequality becomes an equality when $\frac{1}{\lb x \rb}, \frac{1}{\lb y \rb} \to 0$, which 
corresponds to systems where $X$ and $Y$ exhibit no intrinsic stochasticity and follow the upstream signal deterministically.

\section{Constraints on open-loop systems for systems as define in Eq.~\eqref{EQ: Definition of 1-step Dual-Reporter System}} \label{Appendix section on open-loop constraint}

\subsection{Derivation of  the open-loop constraint  Eq.~\eqref{EQ: No feedback constraint (manuscript)}}
We consider a hypothetical ensemble of systems from the class of Eq.~\eqref{EQ: Definition of 1-step Dual-Reporter System} that all share the same upstream history $\mathbf{u}[-\infty,t]$.  We can then consider the average stochastic dual reporters conditioned on
the history of their upstream influences \cite{Hilfinger2011, Hilfinger2012}, which corresponds to the averages at a moment in time in this hypothetical ensemble
\begin{align*}
\bar{x}(t) = \mathrm{E}\big[ X_{t}|\mathbf{u}[-\infty,t] \big] \quad  \text{\&} \quad  \bar{y}(t) = \mathrm{E}\big[ Y_{t}|\mathbf{u}[-\infty,t] \big].
\end{align*}
These are time-dependent variables that depend on the upstream history $\mathbf{u}[-\infty,t]$. We can take averages $\langle \bar{x} \rangle$ and (co)variances $\eta_{\bar{x}\bar{x}}$ over the distribution of all possible histories. 

This conditional system is not stationary because of the synchronised rate $R(t) = R(\mathbf{u}(t))$, and so the time-evolution of the averages will follow the following differential 
equations~\cite{Hilfinger2011}
\begin{align}
\begin{split}
    \frac{d\Delta \bar{x}}{dt} = \Delta R(t) - \Delta \bar{x}(t)/\tau_{x},  \\ \frac{d \Delta \bar{y}}{dt} = \Delta R(t) - \Delta \bar{y}(t)/\tau_{y},
\end{split}
\label{EQ: no-feedback 1-step differential equations}
\end{align}
where $\Delta \bar{x} = \bar{x} - \langle x \rangle$, $\Delta \bar{y} = \bar{y} - \langle y \rangle$, 
and $\Delta R(t) = R(t) - \langle R \rangle$. These differential equations correspond to a linear response problem in which $X$ and $Y$ both respond to $R(t)$ on different timescales.
To relate the results to the actual ensemble we  multiply the left differential equation 
by $\bar{x}$ and take the expectation over all histories $\mathbf{u}[-\infty,t]$ to get 
\begin{align*}
    \mathrm{E}\left[\frac{1}{2}\frac{d \Delta \bar{x}^{2}}{dt} \right] = \mathrm{E}\big[\Delta \bar{x}R(t) \big] - \mathrm{E}\big[\Delta \bar{x}^{2}\big]/\tau_{x} .
\end{align*}
Note that $\mathrm{E}\left[\frac{d \Delta \bar{x}^{2}}{dt}\right] = \frac{d}{dt} \mathrm{E} \big[ \Delta \bar{x}^{2} \big] = \frac{d}{dt}\text{Var}(\bar{x})$, which equals zero 
at wide-sense stationarity. We thus have 
\begin{align*}
    \text{Cov}( \bar{x} , \Delta R(t) ) = \text{Var}( \bar{x} ) /\tau_{x} .
\end{align*}
Moreover, we have
\begin{align*}
&\text{Cov}(\bar{x},R(t)) = \mathrm{E}\big[ \bar{x}R(t) \big] - \mathrm{E}\big[\bar{x}\big] \cdot \mathrm{E}\big[R(t)\big] \\
    &= \mathrm{E}\big[ \mathrm{E}[ X_{t}|\mathbf{u}[-\infty,t]] \cdot R(t) \big] - \langle x \rangle \langle R \rangle  \\
    &= \mathrm{E}\big[ \mathrm{E}[ X_{t} R(\mathbf{u}(t))|\mathbf{u}[-\infty,t]] \big] - \langle x \rangle \langle R \rangle   \\
    &=  \langle xR \rangle \nonumber -  \langle x \rangle \langle R \rangle = \text{Cov}(X,R)  .
\end{align*}
Putting these results together and normalizing we find 
\begin{align}
 \eta_{\bar{x}\bar{x}} = \eta_{xR} \quad \text{\&} \quad \eta_{\bar{y}\bar{y}} = \eta_{yR} ,
 \label{EQ: no-feedback 1-step barbar equals xR}
\end{align}
where the flux-balance relations Eq.~\eqref{EQ: 1-step flux-balance relations} were used and the expression on the right follows by symmetry.
Similarly, it has been shown \cite{Hilfinger2011} that $\eta_{\bar{x}\bar{y}} = \eta_{xy}$.
Comparing with the fluctuation-balance relations Eq.~\eqref{EQ: 1-step fluctuation-balance relations}, we find 
\begin{align}
\begin{split}
&\eta_{xx} = \frac{1}{\lb x \rb} + \eta_{\bar{x}\bar{x}} ,\quad \eta_{yy} = \frac{1}{\lb y \rb} + \eta_{\bar{y}\bar{y}} , \\
&\eta_{\bar{x}\bar{y}} = \eta_{xy} = \frac{1}{1+T}\eta_{\bar{x}\bar{x}}  + \frac{T}{1+T}\eta_{\bar{y}\bar{y}}  . 
\end{split}
\label{EQ: 1-step no-feedback fluctuation-balance}
\end{align}
These relations allow us to translate results derived from the deterministic dynamics to the stochastic dynamics. We now use Cauchy-Schwarz as follows
\begin{align*}
 \left(\frac{1}{1+T}\eta_{\bar{x}\bar{x}}  + \frac{T}{1+T}\eta_{\bar{y}\bar{y}}\right)^{2} = \eta_{\bar{x}\bar{y}}^{2} \leq \eta_{\bar{x}\bar{x}}\eta_{\bar{y}\bar{y}} .
\end{align*}
This inequality leads to a quadratic that can be solved to obtained the following inequality
\begin{equation}
T^2 \eta_{\bar{y}\bar{y}} \leq \eta_{\bar{x}\bar{x}} \leq \eta_{\bar{y}\bar{y}}  .
\label{EQ: 1-step no-feedback conditioned averages}
\end{equation}
We need to write this inequality in terms of the measurable (co)variances $\eta_{xx}$, $\eta_{yy}$, and $\eta_{xy}$. To do this, note that the flux-balance equations 
Eq.~\eqref{EQ: 1-step flux-balance relations} and the fluctuation-balance equations Eq.~\eqref{EQ: 1-step fluctuation-balance relations} comprise a linear system of 5 equations 
and 5 unknowns. We can thus solve for $\eta_{xR}$ and $\eta_{yR}$ in terms of the measurable (co)variances
\begin{align}
\begin{split}
 \eta_{xR} = \frac{(1+T)\eta_{xy} - T\eta_{yy}+\eta_{xx}}{2}, \\
 \eta_{yR} = \frac{(1+T)\eta_{xy} + T\eta_{yy}-\eta_{xx}}{2T}  .
\end{split}
\label{EQ: 1-step eta bar in terms of covariances}
\end{align}
From Eq.~\eqref{EQ: no-feedback 1-step barbar equals xR} we find that $\eta_{\bar{x}\bar{x}} = \eta_{xR}$ and $\eta_{\bar{y}\bar{y}} = \eta_{yR}$, 
so we substitute Eq.~\eqref{EQ: 1-step eta bar in terms of covariances} into Eq.~\eqref{EQ: 1-step no-feedback conditioned averages}, which leads to the open-loop constraint of Eq.~\eqref{EQ: No feedback constraint (manuscript)}. 

\subsection{Discriminating types of feedback}
\label{SEC: Appendix types of feedback}
The open-loop constraint of Eq.~\eqref{EQ: No feedback constraint (manuscript)} constrains all systems from the class of Eq.~\eqref{EQ: Definition of 1-step Dual-Reporter System} in which $X$ and $Y$ are not connected in some kind of feedback loop. Here we derive similar constraints on systems where only one of the components undergoes open-loop regulation, while the other can still be connected in a feedback loop. We will also show how co-regulated reporters can be used to infer whether or not the feedback is negative, 
and if so, how to measure the noise suppression from this negative feedback using only (co)variance measurements.

\begin{figure}[hbt!]
\centering
  \includegraphics[width=0.99\columnwidth]{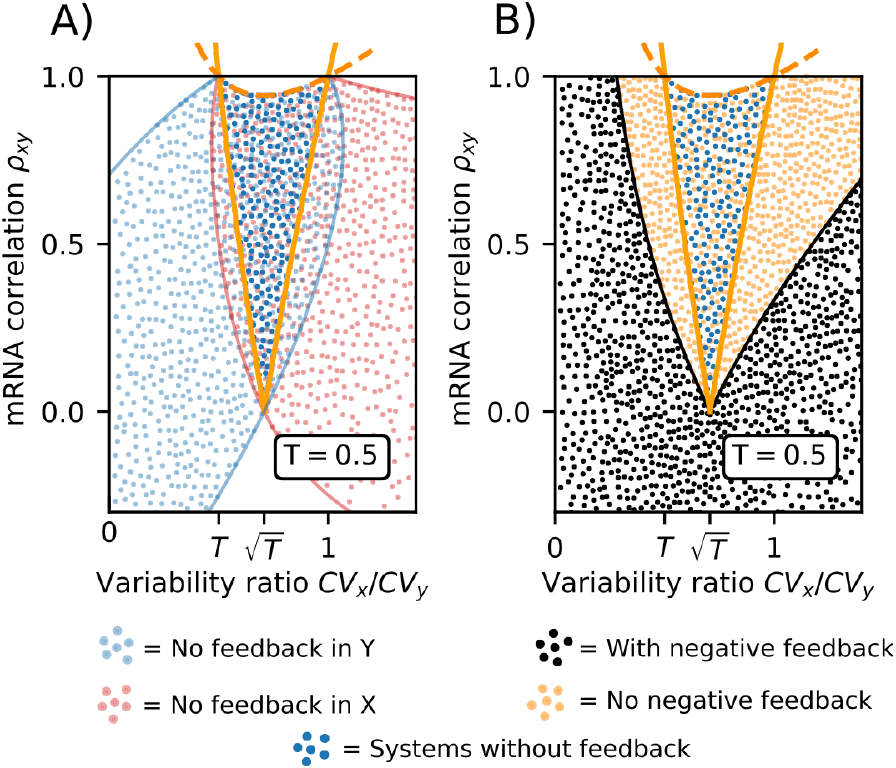}
   \vspace{-.8em}
   \caption{
    \textbf{The space of possible covariability between mRNA-levels of co-regulated genes depends on the type of feedback} 
    A) Systems that have no feedback in $X$ must lie in the region bounded by the light-red solid lines which corresponds to the region 
    bounded by Eq.~\eqref{EQ: no-feedback 1-step no feedback in X}. Systems that lie outside of this region must have feedback in X. 
    Similarly, systems that have no feedback in $Y$ must lie in the region
    bounded by the light-blue solid lines which corresponds to the region 
    bounded by Eq.~\eqref{EQ: no-feedback 1-step no feedback in Y}. 
    B) Here we define feedback to be negative when the mRNA transcription rate is negatively correlated with the mRNA levels: $\eta_{xR} < 0$. 
    This would correspond to highly regulated systems that exhibit noise suppression. Using co-regulated reporters, $\eta_{xR}$ and 
    $\eta_{yR}$ can be measured from static (co)variance measurements of $X$ and $Y$ using Eq.~\eqref{EQ: 1-step eta bar in terms of covariances}.}
    \label{FIG: Types of feedback}
\end{figure}

First, we consider systems in which there is no feedback in one of the components, say $X$. We can then condition on the history of the upstream variables $\mathbf{u}(t)$ 
and the history of $Y$ --- together they make a larger cloud of components that can affect $X$ but are not affected by $X$. Just like the previous section, we 
then consider the conditional average $\bar{x} = \mathrm{E}[X_{t}|\mathbf{u}[-\infty,t], y[-\infty,t]]$, from which we have
\begin{align*}
 &\mathrm{E}[\bar{x}(t)y(t)] - \mathrm{E}[\bar{x}]\mathrm{E}[y(t)] =\dots\\
 &= \mathrm{E}\Big[ \mathrm{E}\big[X_{t}y(t)|\mathbf{u}[-\infty,t], y[-\infty,t]\big] \Big] - \langle x \rangle \langle y \rangle \\
  &= \mathrm{E}\Big[ \mathrm{E}\big[X_{t}Y_{t}|\mathbf{u}[-\infty,t], y[-\infty,t]\big] \Big] - \langle x \rangle \langle y \rangle \\
  &= \langle xy \rangle - \langle x \rangle \langle y \rangle \Rightarrow \eta_{\bar{x}y} = \eta_{xy} \quad . 
\end{align*}
We can then use the Cauchy-Schwarz inequality in the following way: $\eta_{xy}^{2} = \eta_{\bar{x}y}^{2} \leq \eta_{\bar{x}\bar{x}}\eta_{yy}$. This inequality bounds systems in which there is no feedback in $X$. We would now like to write it in terms of measurable (co)variances. 
To do this, we note that the relation $\eta_{\bar{x}\bar{x}} = \eta_{xR}$ from Eq.~\eqref{EQ: no-feedback 1-step barbar equals xR} still holds here as we made no assumptions about 
$Y$ in that derivation. We can thus use Eq.~\eqref{EQ: 1-step eta bar in terms of covariances} to write $\eta_{\bar{x}\bar{x}}$ in terms of the measurable (co)variances 
and subsitute the results in the above inequality. This leads to the following constraint
\begin{align}
 \rho_{xy}^{2} \leq \frac{1}{2}\left( (1+T)\rho_{xy}\frac{CV_{y}}{CV_{x}} - T \left(\frac{CV_{y}}{CV_{x}}\right)^{2} + 1 \right)  .
 \label{EQ: no-feedback 1-step no feedback in X}
\end{align}
Systems that break this constraint must have some kind of feedback in $X$. Similarly, we can derive the analogous constraint on systems with no feedback in $Y$:
\begin{align}
 \rho_{xy}^{2} \leq \frac{1}{2T}\left( (1+T)\rho_{xy}\frac{CV_{x}}{CV_{y}} + T - \left(\frac{CV_{x}}{CV_{y}}\right)^{2} \right)  .
 \label{EQ: no-feedback 1-step no feedback in Y}
\end{align}
Systems that break this constraint must have some kind of feedback in $Y$. These bounds are plotted in Fig.~\ref{FIG: Types of feedback}A.

Next, we show how to detect negative feedback. Here we define feedback to be negative when $\eta_{xR} < 0$. That is, the birthrate $R$ acts to suppress noise in $X$
below poisson noise. From Eq.~\eqref{EQ: 1-step eta bar in terms of covariances}, $\eta_{xR}$ and $\eta_{yR}$ can be solved for in terms of the measurable (co)variances.
(Co)variance measurements between co-regulated mRNA can thus be used to measure $\eta_{xR}$ and $\eta_{yR}$ using Eq.~\eqref{EQ: 1-step eta bar in terms of covariances}, see Fig.~\ref{FIG: Types of feedback}B. Moreover, we can quantify how strong this negative feedback 
is through the noise suppression :
$\eta_{xx}/(1/\langle x \rangle) = \eta_{xx}/(\eta_{xx} - \eta_{xR})$.
As the negative feedback gets stronger, the noise suppression will get smaller and quantifies the strength of the negative feedback. 

\section{Dynamics from static transcript variability}\label{Appendix section on time-scales}

The stationary solution for the variance of $\bar{x}(t)$ follows from the linear response problem Eq.~(\ref{EQ: no-feedback 1-step differential equations}) with
\begin{align}
 \eta_{\bar{x}\bar{x}} = \frac{1}{\tau_{x}}\eta_{RR}\int_{0}^{\infty} A(s) e^{-s/\tau_{x}}  \rmd s,
 \label{EQ: no-oscillation 1-step eta bar autocorrelation integral}
\end{align}
where $A(s)$ is the auto-correlation of the time-varying transcription rate $R(t)$ (see below for a detailed derivation). For a given input $R(t)$, $\eta_{\bar{x}\bar{x}}$ depends on $\tau_{x}$, and so measuring $\tau_{x}\eta_{\bar{x}\bar{x}}$ as a function of $\tau_{x}$ effectively determines the Laplace Transform of the auto-correlation of $R(t)$ from static variance measurements of downstream reporters. Our results exploit the fact that knowing static downstream variability for just two values of $\tau_x$ constrains the possible dynamics of $R(t)$.

In particular, with $\tau_{y} < \tau_{x}$ we have 
\begin{align}
    \eta_{\bar{x}\bar{x}} - T\eta_{\bar{y}\bar{y}} = \frac{\eta_{RR}}{\tau_{x}}\int_{0}^{\infty}\hspace{-0.5em}A(s)\left(e^{-s/\tau_{x}} - e^{-s/\tau_{y}}\right) \rmd s . 
    \label{EQ: Appendix stochastic Laplace}
\end{align}
For stochastic upstream signals with $A(s) \geq 0$ the integrand of Eq.~\eqref{EQ: Appendix stochastic Laplace} is non-negative such that 
\begin{equation}
T \eta_{\bar{y}\bar{y}} \leq \eta_{\bar{x}\bar{x}} .
\label{EQ: Appendix eta time-scale result}
\end{equation}
Eq.~\eqref{EQ: Appendix eta time-scale result} gives the bound of Eq.~\eqref{EQ: No oscillation constraint (manuscript)} after substituting $\eta_{\bar{x}\bar{x}}= \eta_{xx}-1/\lb x \rb$ and $\eta_{\bar{y}\bar{y}}= \eta_{yy}-1/\lb y \rb$ together with the flux balance given by Eq.~\eqref{EQ: 1-step flux-balance relations}.

\begin{figure}[htb!]
\centering
  \includegraphics[width=0.95\columnwidth]{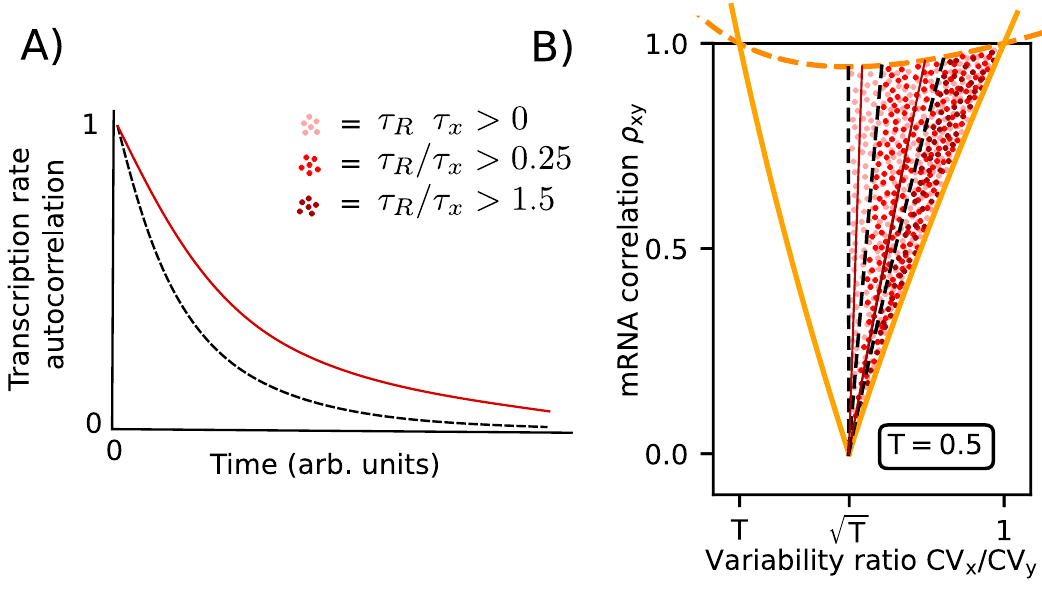}
   \caption{
    \textbf{The space of possible correlations between co-regulated reporters depends on the timescale of the upstream fluctuations.} A) The timescale of stochastic fluctuations is characterized by how quickly the auto-correlation of the signal (red) decays to zero. By bounding the auto-correlation from below by $e^{-t/\tau_{R}}$ (black dashed curve) we set a lower bound on how quickly the signal auto-correlation can decay to zero. This thus sets a lower bound on the "speed" of the signal fluctuations. B) Open-loop systems in which the upstream auto-correlation is bounded below by $e^{-t/\tau_{R}}$ are bounded further towards the right. The solid dark red lines correspond to the loose bound given by Eq.~\eqref{EQ: Fluctuation timescale bound (manuscript)}, whereas the black dashed lines are given by the stricter nuerical bound presented in the supplement \cite{SI}.  These constraints can be used to gain information on the timescale of upstream fluctuations with only access to static snapshots of dual reporter transcript concentrations obtained, e.g., from single-cell sequencing methods that report snapshots of transcript abundances rather than temporal information.}
    \label{FIG: Timescales}
\end{figure}

{We can further bound the class of stochastic systems based on the timescale of the upstream fluctuations. If  upstream fluctuations are slower than $\tau_R$ such that 
$A(s) \geq e^{-s/\tau_R}$, 
we have 
\begin{align*}
 \eta_{\bar{x}\bar{x}} - T \eta_{\bar{y}\bar{y}} 
 &\geq 
 \frac{\eta_{RR}}{\tau_{x}}\int_{0}^{\infty} e^{-t/\tau_{R}} \left( e^{-t/\tau_{x}} - e^{-t/\tau_{y}}\right)  \rmd t \\
 &= \eta_{RR}\tau_{R}\left(\frac{1}{\tau_{R} + \tau_{x}} - \frac{T}{\tau_{R} + \tau_{y}}\right) \\
 &\geq \eta_{\bar{y}\bar{y}}\tau_{R}\left(\frac{1}{\tau_{R} + \tau_{x}} - \frac{T}{\tau_{R} + \tau_{y}}\right),
\end{align*}
and analogously to the above derivation of Eq.~\eqref{EQ: No oscillation constraint (manuscript)}, it follows that
\begin{align}
\beta(1-T) \rho_{xy} \leqslant  \dfrac{CV_{x}}{CV_{y}} - T\dfrac{CV_{y}}{CV_{x}} ,
 \label{EQ: Fluctuation timescale bound (manuscript)} 
\end{align}
where $\beta= (1+T)/\left(2T(1+\frac{\tau_{x}}{\tau_{R}})(1+T\frac{\tau_{x}}{\tau_{R}})+(1-T)\right)$. In the limit where $\tau_{R} \to 0$ we have $\beta \to 0$, and as a result Eq.~(\ref{EQ: Fluctuation timescale bound (manuscript)}) converges to Eq.~(\ref{EQ: No oscillation constraint (manuscript)}). Conversely, for slow stochastic upstream fluctuations as $\tau_{R} \to \infty$ we have $\beta \to 1$, and Eq.~(\ref{EQ: Fluctuation timescale bound (manuscript)}) converges to the right boundary of the open-loop constraint (see Fig.~\ref{FIG: Timescales}). 
The analytical bound of Eq.~(\ref{EQ: Fluctuation timescale bound (manuscript)}) is marginally loose. A tight bound can be derived numerically as presented in the Supplemental Material \cite{SI}.}\\

\noindent \textbf{Derivation of Eq.~\eqref{EQ: no-oscillation 1-step eta bar autocorrelation integral}:} Taking the Fourier transforms of Eq.~\eqref{EQ: no-feedback 1-step differential equations} and using the Wiener-Khinchin theorem \cite{davenport1958introduction} which states that the spectral density of a random signal is equal to the Fourier transform of its autocorrelation, we find
 \begin{align*}
 \mathcal{F}[\mathcal{R}_{\bar{x}}] = \frac{\mathcal{F}[\mathcal{R}_{R}]}{(1 + \omega^{2})} \quad \text{\&} \quad \mathcal{F}[\mathcal{R}_{\bar{y}}] = \frac{\mathcal{F}[\mathcal{R}_{R}]/T^{2}}{(1/T^{2} + \omega^{2})}
,
\end{align*}
where $\mathcal{R}_{R}$ is the autocovariance of $R$. 
Since $\mathcal{R}_{R}(t = 0) = \text{Var}(R)$, the variances can be found by taking the inverse Fourier transforms at $t = 0$
\begin{align}
\begin{split}
 &\text{Var}(\bar{x}) = \frac{1}{2\pi}\int_{-\infty}^{\infty}\frac{\mathcal{F}[\mathcal{R}_{R}]}{(1 + \omega^{2})}d\omega ,
 \\ &\text{Var}(\bar{y}) = \frac{1}{2\pi}\int_{-\infty}^{\infty}\frac{\mathcal{F}[\mathcal{R}_{R}]}{(1/T^{2} + \omega^{2})}d\omega
 .
\end{split}
\label{EQ: no-feedback 1-step single bar variances from Fourier}
\end{align}
We thus have
\begin{equation*}
\begin{split}
    &\text{Var}(\bar{y}) = \dots\\
    &= \frac{1}{2\pi}\int_{-\infty}^{\infty}\mathcal{F}[\mathcal{R}_{R}](\omega)\left( \frac{T^{2}}{1 + (\omega T)^{2}} \right)d\omega \\
    &= \frac{1}{2\pi}\int_{-\infty}^{\infty}\left[\int_{-\infty}^{\infty}\mathcal{R}_{R}(t)e^{-i\omega t}dt\right] \left( \frac{T^{2}}{1 + (\omega T)^{2}} \right)d\omega\\
    &= \frac{1}{2\pi}\int_{-\infty}^{\infty}\left[\int_{-\infty}^{\infty}\mathcal{R}_{R}(t)\text{cos}(\omega t)dt\right] \left( \frac{T^{2}}{1 + (\omega T)^{2}} \right)d\omega\\
    &= \frac{1}{2\pi}\int_{-\infty}^{\infty}\int_{-\infty}^{\infty}\mathcal{R}_{R}(t)\text{cos}(\omega t) \left( \frac{T^{2}}{1 + (\omega T)^{2}} \right)dt d\omega \\
    &= \frac{1}{2\pi}\int_{-\infty}^{\infty}\mathcal{R}_{R}(t)\left[\int_{-\infty}^{\infty}\text{cos}(\omega t) \left( \frac{T^{2}}{1 + (\omega T)^{2}} \right)d\omega\right] dt \\
    &= \frac{1}{2}\int_{-\infty}^{\infty}\mathcal{R}_{R}(t)T e^{-|t|/T}  dt \\
    &= \int_{0}^{\infty}\mathcal{R}_{R}(t)T e^{-t/T}  dt ,
    \end{split}
\end{equation*}
where in the third step we use the fact that $\mathcal{R}_{R}(t)$ is real and symmetric, and in the last step we use the fact that the integrand is symmetric in $t$. Normalizing by the averages, we have  Eq.~\eqref{EQ: no-oscillation 1-step eta bar autocorrelation integral}. Note that this expression can also be derived by writing down the general solution of $\Delta \bar{y}(t)$ from the differential equation Eq.~\eqref{EQ: no-feedback 1-step differential equations}, squaring the solution to get $\Delta \bar{y}(t)^{2}$, and taking the ensemble average over all histories, where ergodicity is not assumed.

\subsection*{Setting bounds on the spectral density of upstream influences using co-regulated reporters}
Numerical simulations of sinusoidal driving indicate that slow oscillations lie along the right hand side of the open-loop region in Fig.~\ref{FIG: Class 1 space of solutions -- feedback (manuscript)}B, and move towards the left as the frequency of oscillation increases. Here we derive approximate conditions for how large an oscillation frequency needs to be to break the constraint given by Eq.~\eqref{EQ: No oscillation constraint (manuscript)} of the main text. In particular, systems in which the spectral density of the upstream signal $|\mathcal{F}[R]|^{2}(\omega)$ is zero for all $\omega \geq \omega_{u}$ 
are bounded by the following inequality 
\begin{align}
\begin{split}
 \left(T\frac{CV_{y}}{CV_{x}} - \frac{CV_{x}}{CV_{y}}\right)\frac{(1+T)}{(\gamma+T)} \geq (T - \gamma)\rho_{xy} \\ \text{where} \quad \gamma = \left(\frac{1 + \omega_{u}^{2}\tau_{y}^{2}}{1 + \omega_{u}^{2}\tau_{x}^{2}}\right). 
 \end{split}
 \label{EQ: spec 1}
\end{align}
Systems that break the above inequality cannot satisfy the requirement that $|\mathcal{F}[R]|^{2}(\omega) = 0$ for all $\omega \geq \omega_{u}$. 
Similarly, systems in which the spectral density of the upstream signal $|\mathcal{F}[R]|^{2}(\omega)$ is zero for all $\omega \leq \omega_{u}$  
are bounded by the following inequality
\begin{align}
\begin{split}
 (T - \gamma)\rho_{xy} \leq \left(T\frac{CV_{y}}{CV_{x}} - \frac{CV_{x}}{CV_{y}}\right)\frac{(1+T)}{(\gamma+T)}  \\ \text{where} \quad \gamma = \left(\frac{1 + \omega_{u}^{2}\tau_{y}^{2}}{1 + \omega_{u}^{2}\tau_{x}^{2}}\right) .
\end{split}
 \label{EQ: spec 2}
\end{align}
Systems that break the above inequality cannot satisfy the requirement that $|\mathcal{F}[R]|^{2}(\omega) = 0$ for all $\omega \leq \omega_{u}$. \\

Eqs.~\eqref{EQ: spec 1} and \eqref{EQ: spec 2} bound fast oscillations towards the left and slow oscillations towards the right of the open-loop region in Fig.~\ref{FIG: Class 1 space of solutions -- feedback (manuscript)}B. As an oscillatory signal will have a peak in its spectral density centered at the angular frequency of the oscillation, Eqs.~\eqref{EQ: spec 1} and \eqref{EQ: spec 2} provide us with an estimation for how fast an upstream oscillation needs to be relative to the reporter lifetimes in order for the system to cross the no-oscillation line and be fully discriminated from stochastic signals. In particular, setting $\gamma = T$ turns \eqref{EQ: spec 2} into the  "no-oscillation constraint" given by Eq.~\eqref{EQ: No oscillation constraint (manuscript)} of the main text, and turns Eq.~\eqref{EQ: spec 1} into the opposite constraint which constrains systems to be in the region only accessible by oscillations. This  $\gamma = T$ is achieved when $\omega_{u} = 1/\sqrt{\tau_{x}\tau_{y}}$. We thus have the following approximate requirement for how fast the upstream signal needs to oscillate relative to the reporter lifetimes in order to break the no-oscillation bound 
\begin{align}
 \frac{1}{\sqrt{\tau_{x}\tau_{y}}} \lesssim \omega_{R} = 2 \pi f_{R} ,
\end{align}
where $f_{R}$ is the frequency of the upstream oscillation. 

We will now derive the bounds that were just presented.
We normalize Eq.~\eqref{EQ: no-feedback 1-step single bar variances from Fourier} by the averages 
\begin{align}
\begin{split}
 &\eta_{\bar{x}\bar{x}} = \frac{1}{2\pi\langle x \rangle^{2}}\int_{-\infty}^{\infty}\frac{|\mathcal{F}[R]|^{2}}{1+\omega^{2}}d\omega \\
 &\eta_{\bar{y}\bar{y}} = \frac{1}{2\pi \langle x \rangle^{2}}\int_{-\infty}^{\infty}\frac{|\mathcal{F}[R]|^{2}}{1+\omega^{2}T^{2}}d\omega ,
\end{split}
\end{align}
where without loss of generality we work in units where $\tau_{x} = 1$ and $\tau_{y} = T$, and where we use the fact that $\langle y \rangle = T \langle x \rangle$. We thus have 
\begin{widetext}
\begin{align*}
 \eta_{\bar{x}\bar{x}} - \left(\frac{1 + \omega_{u}^{2}T^{2}}{1 + \omega_{u}^{2}}\right) \eta_{\bar{y}\bar{y}} =  
 \frac{1}{2\pi\langle x \rangle^{2}}\int_{-\infty}^{\infty}|\mathcal{F}[R]|^{2}\left(\frac{1}{1+\omega^{2}} -  \left(\frac{1 + \omega_{u}^{2}T^{2}}{1 + \omega_{u}^{2}}\right) \frac{1}{1 + \omega^{2}T^{2}} \right) d\omega .
\end{align*}
\end{widetext}
The expression in square brackets is negative for $\omega > \omega_{u}$ and positive otherwise. Thus, if the spectral density of $R$ is zero for $\omega \geq \omega_{u}$, we have 
\begin{align*}
  \eta_{\bar{x}\bar{x}} - \left(\frac{1 + \omega_{u}^{2}T^{2}}{1 + \omega_{u}^{2}}\right) \eta_{\bar{y}\bar{y}} \geq 0 .
\end{align*}
Using Eq.~\eqref{EQ: no-feedback 1-step barbar equals xR} and Eq.~\eqref{EQ: 1-step eta bar in terms of covariances} to write $\eta_{\bar{x}\bar{x}}$ and $\eta_{\bar{y}\bar{y}}$ in terms of measurable (co)variances, this inequality becomes Eq.~\eqref{EQ: spec 1}. Similarly, the same arguments show that systems in which the spectral density of $R$ is zero for all $\omega \leq \omega_{u}$ must satisfy Eq.~\eqref{EQ: spec 2}.

\section{The general class of co-regulated fluorescent proteins}
\label{SEC: General class of fluorescent proteins}
\begin{figure*}[hbt!]
\vspace{-0.5cm}
\centering
  \includegraphics[width=0.9\columnwidth]{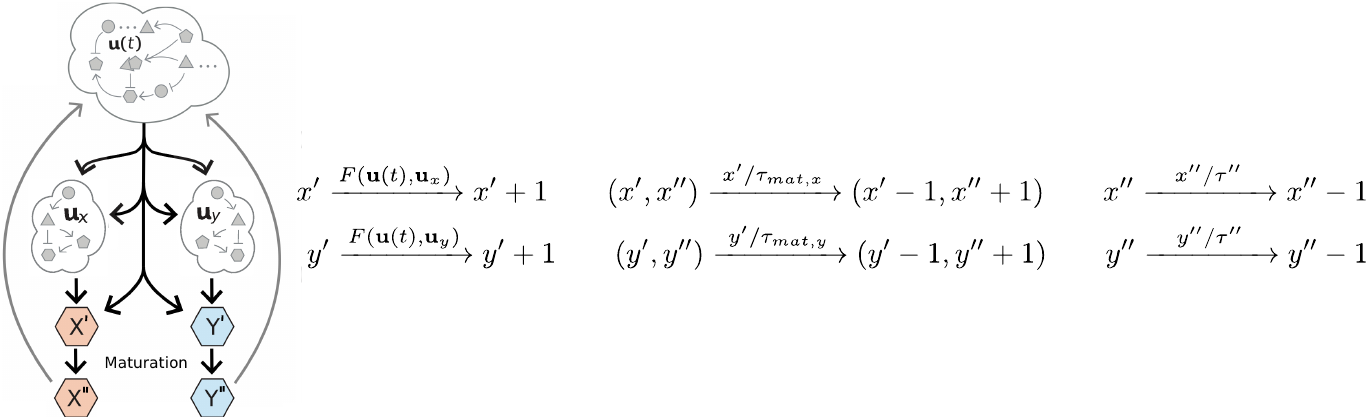}
   \caption{
    \textbf{General class of dual reporter systems with co-regulated fluorescent proteins} 
    We consider a class of systems identical to Fig.~\ref{FIG: 3 step class of systems with maturation (manuscript)}A with the exception that we now include multiple unspecified steps in the intrinsic dynamics of gene expression. 
    Here $\mathbf{u}_{x}$ and $\mathbf{u}_{y}$ are identical (though independent) systems of components that are affected by the upstream cloud of components $\mathbf{u}(t)$ in the same way.
    These two smaller clouds of components model the intrinsic steps in gene expression and allow for a wide range of possible mRNA dynamics {and post-translational modifications that occur before the maturation step. They are left unspecified except for the fact that we assume they do not form circuits of components that can create oscillations, so that any oscillatory variability is caused by the shared environment $\mathbf{u}(t)$.}
    We model the dynamics of immature fluorescent proteins denoted as X' and Y', as well as mature fluorescent proteins X'' and Y''. The birthrate of X' and Y' can now depend on the components in $\mathbf{u}_{x}$ and $\mathbf{u}_{y}$ respectively, in an arbitrary way. The asymmetry between the co-regulated genes is characterized by the ratio of average fluorescent maturation times $T_\mathrm{m} := \tau_{\mathrm{mat},y} / \tau_{\mathrm{mat},x}$. }
    \label{FIG: Class of systems for FPs}
\end{figure*}
For the class of fluorescent proteins defined in Fig.~\ref{FIG: 3 step class of systems with maturation (manuscript)}A we can derive similar results. First, however, we will define a more general class of systems from which 
Fig.~\ref{FIG: 3 step class of systems with maturation (manuscript)}A is a subset. This class consists of the class of systems in Fig.~\ref{FIG: 3 step class of systems with maturation (manuscript)}A when we don't make any assumptions about the mRNA intrinsic system, see Fig.~\ref{FIG: Class of systems for FPs}.
Here, $\mathbf{u}_x$ and $\mathbf{u}_y$ are systems of variables that model the mRNA dynamics after transcription {or any post-translational modifications that occur before the maturation step}.
Together with $\mathbf{u}(t)$ these form a larger ``cloud'' of 
components that affect the proteins $X'$ and $Y'$. We require that $\mathbf{u}_x$ and  $\mathbf{u}_y$ are identical, though independent {(in the sense that they are not equal, as the shared influence of $\mathbf{u}(t)$ will make them statistically dependent)}.  Aside from this they are left almost completely 
unspecified. We will make one additional requirement on these non-specified intrinsic systems when we prove the open-loop constraint given by Eq.~\eqref{EQ: two-step no-feedback constraint (manuscript)} of the main text.
In particular, we require that the fluctuations that originate from these intrinsic systems be non-oscillatory, {so that any oscillatory variability is caused by variability in the shared upstream environment $\mathbf{u}(t)$.}
Specifically, the birthrates of X' and Y' will not be entirely equal due to random fluctuations that originate from the intrinsic systems. This intrinsic noise can be quantified as the difference between the two translation rates: $F(\mathbf{u}(t), \mathbf{u}_x) - F(\mathbf{u}(t), \mathbf{u}_y)$. This difference is a stochastic signal, and here we require that the autocorrelation of this signal be non-negative. This assumption holds for the class in Fig.~\ref{FIG: 3 step class of systems with maturation (manuscript)}A
and in systems where the intrinsic systems consist of an otherwise unspecified cascade of arbitrary steps. The assumption excludes systems in which $\mathbf{u}_{x}$ and $\mathbf{u}_{y}$ form circuits of components that create oscillations.

Similarly to Appendix \ref{SEC: Appendix fluctuation balance relations}, at wide-sense stationarity the following flux-balance relations must hold~\cite{hilfinger2015a}
\begin{align}
\begin{split}
 &\langle x'' \rangle / \tau'' = \langle x' \rangle /\tau_{max,x} = \langle F_x \rangle, \quad \\ &\langle y'' \rangle / \tau'' = \langle y' \rangle /\tau_{max,y} = \langle F_y \rangle ,
\end{split}
\label{EQ: 3-step flux-balance}
\end{align}
where $F_{x} := F(\mathbf{u}(t), \mathbf{u}_x)$ and $F_{y} := F(\mathbf{u}(t), \mathbf{u}_y)$. In addition, the following fluctuation-balance relations must hold, 
\begin{widetext}

\begin{align}
\begin{split}
 \eta_{x'x'}   = \frac{1}{\langle x' \rangle} + \eta_{x'F_{x}} , \quad \eta_{y'y'} = \frac{1}{\langle y' \rangle} + \eta_{y'F_{y}} , \quad \eta_{x'y'} = \frac{1}{1 + T_{m}}\eta_{x'F_{y}} + \frac{T_{m}}{1 + T_{m}}\eta_{y'F_{x}} ,\\
 \eta_{x''x''} = \frac{1}{\langle x'' \rangle} + \eta_{x'x''} , \quad \eta_{y''y''} = \frac{1}{\langle y'' \rangle} + \eta_{y'y''} , \quad \eta_{x''y''} = \frac{1}{2}\eta_{y'x''} + \frac{1}{2}\eta_{x'y''} , \\
 \eta_{x'x''}  = \frac{1}{1 + \frac{\tau''}{\tau_{mat,x}}}\eta_{x'F_{x}} + \frac{\frac{\tau''}{\tau_{mat,x}}}{1 + \frac{\tau''}{\tau_{mat,x}}}\eta_{x''F_{x}} ,\quad \eta_{y'y''} = \frac{1}{1 + \frac{\tau''}{\tau_{mat,y}}}\eta_{y'F_{y}} + \frac{\frac{\tau''}{\tau_{mat,y}}}{1 + \frac{\tau''}{\tau_{mat,y}}}\eta_{y''F_{y}} , \\
 \eta_{y'x''}  = \frac{1}{1 + \frac{\tau''}{\tau_{mat,y}}}\eta_{x'y'} + \frac{\frac{\tau''}{\tau_{mat,y}}}{1 + \frac{\tau''}{\tau_{mat,y}}}\eta_{x''F_{y}}  ,\quad \eta_{x'y''} =  \frac{1}{1 + \frac{\tau''}{\tau_{mat,x}}}\eta_{x'y'} + \frac{\frac{\tau''}{\tau_{mat,x}}}{1 + \frac{\tau''}{\tau_{mat,x}}}\eta_{y''F_{x}}  . 
\end{split}
 \label{EQ: 3-step fluctuation-balance}
\end{align}
\end{widetext}

{
\section{Constraints on open-loop systems for systems as define in Fig.~\ref{FIG: 3 step class of systems with maturation (manuscript)}A } \label{Appendix section on open-loop constraint for fluorescent reporters}

Here we prove the inequality Eq.~\eqref{EQ: two-step no-feedback constraint (manuscript)}. The inequality Eq.~\eqref{EQ: two-step no-feedback constraint (manuscript)} has three parts, which are represented by the three solid orange lines in Fig.~\ref{FIG: 3 step class of systems with maturation (manuscript)}B in the paper. 
We will first prove these three constraints one at a time, followed by the upper bound on correlations illustrated by the dashed orange line in Fig.~\ref{FIG: 3 step class of systems with maturation (manuscript)}B. Refer back to Fig.~\ref{FIG: Class of systems for FPs} for illustration of the system being studied. \\

\noindent \textbf{Proof of the bottom bound $\rho_{x''y''} \geq 0$:} Just like in the previous sections, we consider the average stochastic dual reporter dynamics conditioned on
the history of their upstream influences 
\begin{align*}
\bar{x}''(t) = \mathrm{E}\big[ X''_{t}|\mathbf{u}[-\infty,t] \big] \quad \text{\&} \quad  \bar{y}''(t) = \mathrm{E} \big[ Y''_{t}|\mathbf{u}[-\infty,t] \big] .
\end{align*}
When the cloud of components $\mathbf{u}(t)$ is not affected by the downstream components we can write the time evolution of the conditional averages using the chemical 
master equation of the conditional probability space
\begin{align*}
 \frac{d\Delta \bar{x}'}{dt} = \Delta \bar{F}(t) - \Delta \bar{x}' \qquad \frac{d \Delta \bar{y}'}{dt} = \Delta \bar{F}(t) - \Delta \bar{y}'/T_{m} \quad\\
 \frac{d\Delta \bar{x}''}{dt} = \Delta \bar{x}' - \Delta \bar{x}''/\tau''  \qquad \frac{d\Delta \bar{y}''}{dt} = \Delta \bar{y}'/T_{m} - \Delta \bar{y}''/\tau'' 
\end{align*}
where without loss of generality we work in units where $\tau_{mat,x} = 1$ and $\tau_{mat,y} = T_{m}$, and where 
\begin{align*}
\bar{F}(t) &= \mathrm{E}\big[F(\mathbf{u}(t),\mathbf{u}_{x})|\mathbf{u}[-\infty,t]\big] \\
&=  \mathrm{E}\big[F(\mathbf{u}(t),\mathbf{u}_{y})|\mathbf{u}[-\infty,t]\big] ,
\end{align*}
is the translation rate after averaging out the mRNA fluctuations. Note that the equality on the right hand side is from the fact that the unspecified intrinsic systems $\mathbf{u}_{x}$ and $\mathbf{u}_{y}$ are identical and thus the conditional averages are the same. 
Taking the Fourier transform of the above differential equations, and then using the Wiener-Khinchin theorem \cite{davenport1958introduction} which states that the spectral density of a random signal is equal to the Fourier transform of its autocorrelation, we find
\begin{align*}
 &\mathcal{F}[\mathcal{R}_{\bar{x}}] = \frac{\mathcal{F}[\mathcal{R}_{\bar{F}}]}{(1/\tau''^{2} + \omega^{2})(1 + \omega^{2})}, \\
 &\mathcal{F}[\mathcal{R}_{\bar{y}}] = \frac{\mathcal{F}[\mathcal{R}_{\bar{F}}]}{(1/\tau''^{2} + \omega^{2})(1 + \omega^{2}T_{m}^{2})},
\end{align*}
where $\mathcal{R}_{\bar{F}}$ is the autocovariance of $\bar{F}(t)$. 
Since $\mathcal{R}_{z}(t = 0) = \text{Var}(z)$, the variances can be found by taking the inverse Fourier transforms at $t = 0$
\begin{align}
\begin{split}
 &\text{Var}(\bar{x}'') = \frac{1}{2\pi}\int_{-\infty}^{\infty}\frac{\mathcal{F}[\mathcal{R}_{\bar{F}}]}{(1/\tau''^{2} + \omega^{2})(1 + \omega^{2})}d\omega \\
 &\text{Var}(\bar{y}'') = \frac{1}{2\pi}\int_{-\infty}^{\infty}\frac{\mathcal{F}[\mathcal{R}_{\bar{F}}]}{(1/\tau''^{2} + \omega^{2})(1 + \omega^{2}T_{m}^{2})}d\omega.
\end{split}
\label{EQ: no-feedback 3-step single bar variances from Fourier}
\end{align}
Moreover, we then use the Wiener-Khinchin theorem \cite{davenport1958introduction} which states that the Fourier transform of the cross power spectral density of two signals is equal to the Fourier transform of their cross-correlation to write 
\begin{align*}
 \mathcal{F}[\mathcal{R}_{\bar{x}'',\bar{y}''}] = \frac{\mathcal{F}[\mathcal{R}_{\bar{F}}](1/T_{m} +\omega^{2} +i\omega(1 - 1/T_{m}))/T_{m}}{(1/\tau''^{2} + \omega^{2})(1 + \omega^{2})(1/T_{m}^{2} + \omega^{2})},
 \end{align*}
where $\mathcal{R}_{\bar{x}'',\bar{y}''}$ is the cross-covariance of $\bar{x}''(t)$ and $\bar{y}''(t)$. Taking the inverse Fourier transform at $t=0$ gives us the covariance
\begin{align*}
 &\text{Cov}(\bar{x}'', \bar{y}'') =  \\ &\frac{1}{2\pi}\int_{-\infty}^{\infty}\mathcal{F}[\mathcal{R}_{\bar{F}}]\frac{(1/T_{m} + \omega^{2})/T_{m}}{(1/\tau'^{2} + \omega^{2})(1+\omega^{2})(1/T_{m}^{2} + \omega^{2})}d\omega,
\end{align*}
where the term proportional to $i\omega$ integrated to zero because it was odd in $\omega$ whereas the rest of the integrand is even in $\omega$ through $\mathcal{F}[\mathcal{R}_{\bar{F}}] = |\mathcal{F}[\Delta \bar{F}]|^{2}$.
We thus have
\begin{align*}
    &(1+T_{m})\text{Cov}(\bar{x}'', \bar{y}'') = 
    \\
    =& \frac{1}{2\pi}\int_{-\infty}^{\infty}\mathcal{F}[\mathcal{R}_{\bar{F}}]\frac{(1+T_{m})(1/T_{m} + \omega^{2})/T_{m}}{(1/\tau'^{2} + \omega^{2})(1+\omega^{2})(1/T_{m}^{2} + \omega^{2})}d\omega\\
    =& \frac{1}{2\pi}\int_{-\infty}^{\infty}\mathcal{F}[\mathcal{R}_{\bar{F}}]\frac{1}{(1+\omega^{2})(1/\tau'^{2} + \omega^{2})}d\omega \\
    & \quad +  \frac{T_{m}}{2\pi}\int_{-\infty}^{\infty}\mathcal{F}[\mathcal{R}_{\bar{F}}]\frac{1}{(1+\omega^{2}T_{m}^{2})(1/\tau'^{2} + \omega^{2})}d\omega\\
    =& \text{Var}(\bar{x}'') + T_{m}\text{Var}(\bar{y}'')  .
\end{align*}
Upon normalizing with the averages we have 
\begin{align*}
 \eta_{\bar{x}''\bar{y}''} = \frac{1}{1+T_{m}}\eta_{\bar{x}''\bar{x}''} + \frac{T_{m}}{1+T_{m}}\eta_{\bar{y}''\bar{y}''}  .
\end{align*}
Moreover, 
\begin{align*}
\text{Cov} &(\bar{x}'', \bar{y}'') = \dots\\
 &= \mathrm{E}\big[ \mathrm{E}[X''_{t}|\mathbf{u}[-\infty,t]]\cdot \mathrm{E}[Y''_{t}|\mathbf{u}[-\infty,t]] \big ] \\
 & \qquad - \mathrm{E}\big[ \mathrm{E}[X''_{t}|\mathbf{u}[-\infty,t]] \big ]\cdot\mathrm{E}\big[ \mathrm{E}[Y''_{t}|\mathbf{u}[-\infty,t]] \big ] \\
 &= \mathrm{E}\big[ \mathrm{E}[X''_{t}Y''_{t}|\mathbf{u}[-\infty,t]] \big ]
 - \langle x'' \rangle \langle y'' \rangle \\
 &= \langle x''y'' \rangle - \langle x'' \rangle \langle y'' \rangle = \text{Cov}(x'',y'') ,
\end{align*}
where the 3rd step comes from the fact that $X''$ and $Y''$ are independent when we condition on the upstream history $\mathbf{u}[-\infty,t]$ \cite{Hilfinger2012}. 
Upon normalizing by the averages we have $\eta_{\bar{x}''\bar{y}''} = \eta_{x''y''}$, and so 
\begin{align}
 \eta_{x''y''} = \frac{1}{1+T_{m}}\eta_{\bar{x}''\bar{x}''} + \frac{T_{m}}{1+T_{m}}\eta_{\bar{y}''\bar{y}''}  \geq 0  .
\label{EQ: no-feedback 3-step etaxy covariance equation}
\end{align}
We thus have the lower bound $\rho_{x''y''} \geq 0$ . \\

\noindent \textbf{Proof of the right bound $CV_{x''}/CV_{y''} \leq 1$ :} Here we will need to consider the 
average stochastic dual reporter dynamics conditioned on
the history of all their upstream influences 
\begin{align*}
&\bar{\bar{x}}''(t) := \mathrm{E} \big[ X''_{t}|\mathbf{u}[-\infty,t], \mathbf{u}_{x}[-\infty,t] \big] \quad \\
&\bar{\bar{y}}''(t) := \mathrm{E} \big[ Y''_{t}|\mathbf{u}[-\infty,t],   \mathbf{u}_{y}[-\infty,t] \big] ,
\end{align*}
where now we also condition on the trajectory of the mRNA dynamics. 
When the cloud of components $\mathbf{u}(t)$ does not depend on the downstream components we can write down the time evolution of the conditional averages using the chemical 
master equation of the conditional probability space. The time evolution of the conditional averages $\bar{\bar{y}}'$ and $\bar{\bar{y}}''$ is given by 
\begin{align*}
 \frac{d \bar{\bar{y}}'}{dt} = F_{y}(t) - \bar{\bar{y}}'/T_{m} \quad \text{\&} \quad \frac{d \bar{\bar{y}}''}{dt} = \bar{\bar{y}}'/T_{m} - \bar{\bar{y}}''/\tau'' , 
\end{align*}
where now $F_{y}(t)$ corresponds to a particular translation rate trajectory as we no longer average out the mRNA dynamics, and without loss of generality we let $\tau_{mat,x} = 1$
and $\tau_{mat,y} = T_{m}$. In terms of the deviations from the means these differential equations become
\begin{align}
\begin{split}
 \frac{d \Delta \bar{\bar{y}}'}{dt} &= \Delta F_{y}(t) - \Delta \bar{\bar{y}}'/T_{m}, \\
 \frac{d \Delta \bar{\bar{y}}''}{dt} &= \Delta \bar{\bar{y}}'/T_{m} - \Delta \bar{\bar{y}}''/\tau''  ,
 \end{split}
\label{EQ: no-feedback 3-step double bar differential equations}
\end{align}
Multiplying the left and right equations with $\Delta \bar{\bar{y}}''$ and $\Delta \bar{\bar{y}}'$ respectively, summing the results, and taking ensemble averages of the different upstream histories 
gives us
\begin{align*}
 &\mathrm{E}\left[ \frac{d}{dt}(\Delta \bar{\bar{y}}'\bar{\bar{y}}'') \right] = \\
 &\mathrm{E}\big[\Delta \bar{\bar{y}}'' \Delta F_{y} \big] + \mathrm{E}\big[\Delta \bar{\bar{y}}'^{2} \big]/T_{m} 
	      - \left(\frac{1}{T_{m}} + \frac{1}{\tau''}\right) \mathrm{E}\big[\Delta \bar{\bar{y}}' \Delta \bar{\bar{y}}''\big]  .
\end{align*}
At stationarity the left hand side is zero, so we have 
\begin{align*}
\left(\frac{1}{T_{m}} + \frac{1}{\tau''}\right) \mathrm{E} \big[ \Delta \bar{\bar{y}}' \Delta \bar{\bar{y}}'' \big] = \mathrm{E}\big[ \Delta \bar{\bar{y}}'' \Delta F_{y} \big] + \mathbf{E}\big[(\bar{\bar{y}}')\big]/T_{m} .	      
\end{align*} 
Similarly, multiplying the first equation in Eq.~\eqref{EQ: no-feedback 3-step double bar differential equations} by $\Delta \bar{\bar{y}}'$, taking the ensemble average, and using the fact that the left hand side will be zero at stationarity, we have  
\begin{align*}
 \mathrm{E}\big[\Delta \bar{\bar{y}}' \Delta \bar{\bar{y}}'\big] = T_{m}\mathrm{E}\big[ \Delta\bar{\bar{y}}' \Delta F_{y}\big]  .
\end{align*}
Now mutiplying the second equation in Eq.~\eqref{EQ: no-feedback 3-step double bar differential equations} by $\Delta \bar{\bar{y}}''$ and following the same steps we have 
\begin{align*}
 \mathrm{E}\big[\bar{\bar{y}}''\bar{\bar{y}}''\big]/\tau'' = T_{m}\mathrm{E}\big[\bar{\bar{y}}'\bar{\bar{y}}''\big] .
\end{align*}
Combining these expressions gives us 
\begin{align*}
 \left(\frac{1}{T_{m}} + \frac{1}{\tau''}\right) \text{Var}( \bar{\bar{y}}'') = \text{Cov}( \bar{\bar{y}}'', F_{y} ) + \text{Cov}( \bar{\bar{y}}', F_{y}) . 
\end{align*}
Now note that 
\begin{widetext}
\begin{align*}
 &\text{Cov}( \bar{\bar{y}}'', F_{y}) =  \mathrm{E}\big[ \bar{\bar{y}}''\cdot F_{y}(t) \big] - \mathrm{E}\big[\bar{\bar{y}}''\big] \cdot \mathrm{E}\big[F_{y}\big] \\
 &= \mathrm{E}\big[ \mathrm{E}[Y''_{t}|\mathrm{u}[-\infty,t], \mathbf{u}_{y}[-\infty,t]]\cdot F_{y}(t) \big] - \mathrm{E}\big[\mathrm{E}[Y_{t}|\mathrm{u}[-\infty,t], \mathbf{u}_{y}[-\infty,t]]\big] \cdot \mathrm{E}\big[\mathrm{E}[F(\mathbf{u}(t), \mathbf{u}_{y}(t)|\mathrm{u}[-\infty,t], \mathbf{u}_{y}[-\infty,t]]\big] \\
 &= \mathrm{E}\big[ \mathrm{E}[Y''_{t}\cdot F_{y}(\mathbf{u}(t), \mathbf{u}_{y}(t))|\mathrm{u}[-\infty,t], \mathbf{u}_{y}[-\infty,t]] \big] - \langle y'' \rangle \langle F_{y} \rangle = \text{Cov}(y'',F_{y}) ,
\end{align*}
\end{widetext}
and similarly $ \text{Cov}( \bar{\bar{y}}', F_{y} ) = \text{Cov}( y', F_{y} )$. Thus, after normalizing with the averages, we have
\begin{align*}
\eta_{\bar{\bar{y}}'\bar{\bar{y}}'} =   \frac{1}{1 + \frac{\tau''}{T_{m}}}\eta_{y'F_{y}} + \frac{\frac{\tau''}{T_{m}}}{1 + \frac{\tau''}{T_{m}}}\eta_{y''F_{y}} .
\end{align*}
This is the expression for $\eta_{y'y''}$ in the fluctuation-balance equations Eq.~\eqref{EQ: 3-step fluctuation-balance}, and so
\begin{align}
    \eta_{\bar{\bar{x}}''\bar{\bar{x}}''}  = \eta_{x''x''} - 1/\langle x''\rangle \quad \text{\&} \quad \eta_{\bar{\bar{y}}''\bar{\bar{y}}''}  = \eta_{y''y''} - 1/\langle y'' \rangle ,
\label{EQ: 3-step no-feedback double bar extrinsic}
\end{align}
where the $x''$ expression follows by symmetry.
Moreover, we apply the same analysis that was done in the previous proof to write 
\begin{align}
\begin{split}
 &\text{Var}(\bar{\bar{x}}'') = \frac{1}{2\pi}\int_{-\infty}^{\infty}\frac{\mathcal{F}[\mathcal{R}_{F_{x}}]}{(1/\tau''^{2} + \omega^{2})(1 + \omega^{2})}d\omega, \\ &\text{Var}(\bar{\bar{y}}'') = \frac{1}{2\pi}\int_{-\infty}^{\infty}\frac{\mathcal{F}[\mathcal{R}_{F_{y}}]}{(1/\tau''^{2} + \omega^{2})(1 + \omega^{2}T_{m}^{2})}d\omega,
\end{split}
\label{EQ: no-feedback 3-step double bar variances from Fourier}
\end{align}
where $\mathcal{R}_{F_{x}}$ and $\mathcal{R}_{F_{y}}$ are the autocovariances of the translation rates $F(\mathbf{u}(t),\mathbf{u}_{x})$ and $F(\mathbf{u}(t),\mathbf{u}_{y})$. Note that since the two intrinsic systems $\mathbf{u}_{x}$ and $\mathbf{u}_{y}$ are statistically identical, we have $\mathcal{R}_{F_{x}} = \mathcal{R}_{F_{y}}$, and so the $\text{Var}( \bar{\bar{y}}'')$ expression 
has a larger integrande for all $\omega$ (recall that $\mathcal{F}[\mathcal{R}_{F_{y}}] = \mathcal{F}[\Delta F_{y}]^{2}$ and so is positive). We thus have 
$\text{Var}( \bar{\bar{x}}'') < \text{Var}( \bar{\bar{y}}'')$, which after normalizing gives us $\eta_{\bar{\bar{x}}''\bar{\bar{x}}''}  \leq \eta_{\bar{\bar{y}}''\bar{\bar{y}}''}$.
From Eq.~\eqref{EQ: 3-step flux-balance} we have $\langle x'' \rangle = \langle y'' \rangle$, and so by using 
Eq.~\eqref{EQ: 3-step no-feedback double bar extrinsic} we find the right bound $CV_{x''x''} \leq CV_{y''y''}$.\\

\noindent \textbf{Proof of the left bound:} Recall from the previous two proofs we derived the following equations
\begin{align}
\begin{split}
 \eta_{x''x''} = \frac{1}{\langle x'' \rangle} +\eta_{\bar{\bar{x}}''\bar{\bar{x}}''}  \quad\quad 
 \eta_{y''y''} = \frac{1}{\langle y'' \rangle} +\eta_{\bar{\bar{y}}''\bar{\bar{y}}''}  \\
 \eta_{x''y''} = \eta_{\bar{x}''\bar{y}''} = \frac{1}{1+T_{m}}\eta_{\bar{x}''\bar{x}''} + \frac{T_{m}}{1+T_{m}}\eta_{\bar{y}''\bar{y}''}.
\end{split}
 \label{EQ: no-feedback 3-step fluctuation-balance in terms of bared variables}
\end{align}
We now use the Cauchy-Schwarz inequality with the last expression
\begin{align*}
 \left(\frac{1}{1+T_{m}}\eta_{\bar{x}''\bar{x}''} + \frac{T_{m}}{1+T_{m}}\eta_{\bar{y}''\bar{y}''}\right)^{2} \leq \eta_{\bar{x}''\bar{x}''}\eta_{\bar{y}''\bar{y}''} ,
\end{align*}
which leads to 
\begin{align}
 T_{m}^{2}\eta_{\bar{y}''\bar{y}''} \leq \eta_{\bar{x}''\bar{x}''} \leq \eta_{\bar{y}''\bar{y}''}  .
 \label{EQ: no-feedback 3-step intuitive no-feedback bound}
\end{align}
Unlike the mRNA system of equations, we cannot solve Eq.~\eqref{EQ: no-feedback 3-step fluctuation-balance in terms of bared variables} for $\eta_{\bar{x}''\bar{x}''} $ and $\eta_{\bar{y}''\bar{y}''}$ in terms of the measurable (co)variances 
because the system is underdetermined. This is due to the fact that we have not specified the mRNA intrinsic system. 
Nevertheless, we can derive an additional bound that will allow us to close the system of equations to write Eq.~\eqref{EQ: no-feedback 3-step intuitive no-feedback bound}
in terms of the measurable (co)variances. \\

From Eqs.~\eqref{EQ: no-feedback 3-step single bar variances from Fourier} and ~\eqref{EQ: no-feedback 3-step double bar variances from Fourier} we have
\begin{align*}
 \text{Var}(\bar{\bar{x}}'') - \text{Var}(\bar{x}'') &= \frac{1}{2\pi}\int_{-\infty}^{\infty}\frac{\mathcal{F}[\mathcal{R}_{F_{x}} - \mathcal{R}_{\bar{F}}]}{(1/\tau''^{2} + \omega^{2})(1 + \omega^{2})}d\omega.
\end{align*}
Now, note that 
\begin{widetext}
\begin{align*}
  \mathcal{R}_{\bar{F}}(t) &= \mathrm{E}\big[ \mathrm{E}[\Delta F(\mathbf{u}(t'), \mathbf{u}_{x})|\mathbf{u}[-\infty,t']] \cdot \mathrm{E}[\Delta F(\mathbf{u}(t'+t), \mathbf{u}_{x})|\mathbf{u}[-\infty,t'+t]] \big] \\
    &= \mathrm{E}\big[ \mathrm{E}[\Delta F(\mathbf{u}(t'), \mathbf{u}_{x})|\mathbf{u}[-\infty,t']] \cdot \mathrm{E}[\Delta F(\mathbf{u}(t'+t), \mathbf{u}_{y})|\mathbf{u}[-\infty,t'+t]] \big] \\
    &= \mathrm{E}\big[ \mathrm{E}[\Delta F(\mathbf{u}(t'), \mathbf{u}_{x}) \cdot \Delta F(\mathbf{u}(t'+t), \mathbf{u}_{y}) |\mathbf{u}[-\infty,t'+t]] \big] = \text{Cov}(F_{x}(t'), F_{y}(t' + t)) = \mathcal{R}_{F_{x},F_{y}}(t) ,
\end{align*}
\end{widetext}
where the second step comes from the fact that $F_{x}$ and $F_{y}$ are statistically equivalent and the third step comes from the fact that they are independent when we condition 
on the upstream history $\mathbf{u}[-\infty,t]$. Taking the autocovariance of $F_{x} - F_{y}$ gives us
\begin{align*}
 \mathcal{R}_{F_{x} - F_{y}} = 2\mathcal{R}_{F_{x}} - 2\mathcal{R}_{F_{x}, F_{y}} = 2(\mathcal{R}_{F_{x}} - \mathcal{R}_{\bar{F}}) .
\end{align*}
Thus the requirement that we make for this class of systems --- that the autocorrelation of $F_{x} - F_{y}$ be non-negative --- is equivalent to saying that 
$\mathcal{R}_{F_{x}} - \mathcal{R}_{\bar{F}}$ is non-negative. Thus, we have 
\begin{align*}
 &\text{Var}(\bar{\bar{x}}'') - \text{Var}(\bar{x}'') = \frac{1}{2\pi}\int_{-\infty}^{\infty}\frac{\mathcal{F}[\mathcal{R}_{F_{x}} - \mathcal{R}_{\bar{F}}]}{(\frac{1}{\tau''^{2}} + \omega^{2})(1 + \omega^{2})}d\omega \\
  &\text{Var}(\bar{\bar{y}}'') - \text{Var}(\bar{y}'') = \frac{1}{2\pi}\int_{-\infty}^{\infty}\frac{\mathcal{F}[\mathcal{R}_{F_{y}} - \mathcal{R}_{\bar{F}}]}{(\frac{1}{\tau''^{2}} + \omega^{2})(1 + T_{m}^{2}\omega^{2})}d\omega .
 \end{align*}
Since the two intrinsic systems $\mathbf{u}_{x}$ and $\mathbf{u}_{y}$ are identical, we have $\mathcal{R}_{F_{x}} = \mathcal{R}_{F_{y}}$.
Defining $f := \mathcal{R}_{F_{x}} - \mathcal{R}_{\bar{F}}$, we have 
\begin{widetext}
\begin{align*}
    &(\text{Var}(\bar{\bar{x}}) - \text{Var}(\bar{x})) - T_{m}(\text{Var}(\bar{\bar{y}}) - \text{Var}(\bar{y})) = \dots \\
    &= \frac{1}{2\pi}\int_{-\infty}^{\infty}\mathcal{F}[f]\left\{\frac{1}{(1 + \omega^{2})(1/\tau''^{2} + \omega^{2})}- \frac{T_{m}}{(1+(\omega T_{m})^{2})(1/\tau''^{2} + \omega^{2})}\right\} d\omega   \\
    &= \frac{1}{2\pi}\int_{-\infty}^{\infty}\int_{-\infty}^{\infty}f(t)\text{cos}(\omega t)\left\{\frac{1}{(1/\tau''^{2} + \omega^{2})(1 + \omega^{2})}- \frac{T_{m}}{(1/\tau''^{2} + \omega^{2})(1+(\omega T_{m})^{2})}\right\} dt d\omega   \\
    &= \frac{1}{2\pi}\int_{-\infty}^{\infty}f(t) \left(\int_{-\infty}^{\infty}\frac{\text{cos}(\omega t)}{(1/\tau''^{2} + \omega^{2})}\left\{\frac{1}{(1 + \omega^{2})}- \frac{T_{m}}{1+(\omega T_{m})^{2}}\right\} d\omega \right)dt   \\
    &= \frac{1}{\pi}\int_{0}^{\infty}f(t) \left(\int_{-\infty}^{\infty}\frac{\text{cos}(\omega t)}{(1/\tau''^{2} + \omega^{2})}\left\{\frac{1}{(1 + \omega^{2})}- \frac{T_{m}}{1+(\omega T_{m})^{2}}\right\} d\omega \right)dt   \\
    &= \int_{0}^{\infty}f(t) \tau''^{2} \left(\frac{ (e^{-t} -  \tau''e^{-t/\tau''})}{(1 - \tau''^2)} - \frac{ (T_{m}e^{-t/T_{m}} - \tau''e^{-t/\tau''} )}{(T_{m}^{2} - \tau''^2)} \right)dt  ,
\end{align*}
\end{widetext}
where in the second step we used the fact that $f(t)$ is symmetric which lets us omit the $\text{sin}(\omega t)$ part of the Fourier transform, and
in the fourth step we use the fact that the integrand is symmetric in $t$. The expression in parentheses is always positive, and since $f(t)$ is non-negative, this means that $
 T_{m}\left(\text{Var}(\bar{\bar{y}}) - \text{Var}(\bar{y}) \right) \leq \left(\text{Var}(\bar{\bar{x}}) - \text{Var}(\bar{x})\right)$,
which in terms of the normalized variances is $
     T_{m}(\eta_{\bar{\bar{y}}''\bar{\bar{y}}''} - \eta_{\bar{y}''\bar{y}''}) \leq (\eta_{\bar{\bar{x}}''\bar{\bar{x}}''} - \eta_{\bar{x}''\bar{x}''})$.
Similarly, we can show using the same method that $
  (\eta_{\bar{\bar{x}}''\bar{\bar{x}}''} - \eta_{\bar{x}''\bar{x}''}) \leq (\eta_{\bar{\bar{y}}''\bar{\bar{y}}''} - \eta_{\bar{y}''\bar{y}''})$.
Combining these two inequalities, we have 
\begin{align}
    T_{m} \leq \frac{\eta_{\bar{\bar{x}}''\bar{\bar{x}}''} - \eta_{\bar{x}''\bar{x}''}}{\eta_{\bar{\bar{y}}''\bar{\bar{y}}''} - \eta_{\bar{y}''\bar{y}''}} \leq 1 .
\label{EQ: no-feedback 3-step intrinsic noise bound}
\end{align}
With this inequality we can ``close'' the system of equations Eq.~\eqref{EQ: no-feedback 3-step fluctuation-balance in terms of bared variables} to set the limits of $\eta_{\bar{x}''\bar{x}''}$ and $\eta_{\bar{y}''\bar{y}''}$ in 
terms of the measurable (co)variances
\begin{widetext}
\begin{align}
    \begin{split}
    \frac{\eta_{x''y''}(1+T_{m}) - T_{m}\left(\eta_{y''y''} - \eta_{x''x''}\right)}{1+T_{m}} \leq &\eta_{\bar{x}''\bar{x}''} 
    \leq \frac{(1+T_{m})\eta_{x''y''} + \eta_{x''x''} - T_{m}\eta_{y''y''}}{2} \\
    \frac{\eta_{x''y''}(1+T_{m}) - \eta_{x''x''} + T_{m}\eta_{y''y''}}{2T_{m}} \leq &\eta_{\bar{y}''\bar{y}''} 
    \leq \frac{\eta_{x''y''}\left(1+T_{m}\right) - \eta_{x''x''} + \eta_{y''y''}}{1 + T_{m}} .
    \end{split}
    \label{EQ: no-feedback-3step-extrinsic-noise-bounds}
\end{align}
\end{widetext}
We can then substitute these in the open-loop constraint Eq.~\eqref{EQ: no-feedback 3-step intuitive no-feedback bound} to obtain the open-loop constraint in terms of the measurable 
(co)variances. Doing so we obtain the left bound of Fig.~\ref{FIG: 3 step class of systems with maturation (manuscript)}B.\\

\noindent \textbf{Proof of the upper correlation bound:} We use the law of total variance to write 
\begin{align*}
    \eta_{x''x''} = \eta_{int,x''} + \eta_{\bar{x}''\bar{x}''}\quad \text{\&} \quad \eta_{y''y''} = \eta_{int,y''} + \eta_{\bar{y}''\bar{y}''}
\end{align*}
where $\eta_{int,x''} = \frac{1}{\langle x'' \rangle^{2}}\mathrm{E}\big[\Var\left(X''|\mathbf{u}[-\infty,t]\right)\big]$ and $\eta_{int,y''} = \frac{1}{\langle y'' \rangle^{2}}\mathrm{E}\big[\Var\left(Y''|\mathbf{u}[-\infty,t]\right)\big]$. Thus we have $\eta_{x''x''} \geq \eta_{\bar{x}''\bar{x}''}$ and $\eta_{y''y''} \geq \eta_{\bar{y}''\bar{y}''}$. Substituting in Eq.~\eqref{EQ: no-feedback 3-step etaxy covariance equation} gives 
\begin{align*}
    \eta_{x''y''}(1+T_{m}) \leq \eta_{x''x''} + T_{m}\eta_{y''y''},
\end{align*}
which after dividing by $\sqrt{\eta_{x''x''}\eta_{y''y''}}$ results in the upper correlation bound. 

{
\section{Dynamics from static fluorescent protein variability}\label{Appendix section on time-scales for fluorescent reporters}
\label{SEC: Appendix dynamics fluorescent proteins}
When there is no feedback, we can write down the time-evolution equations for the averages of the conditional probability space, and from these we can derive Eq.~\eqref{EQ: no-feedback 3-step double bar variances from Fourier}
\begin{align*}
 &\text{Var}(\bar{\bar{x}}'') = \frac{1}{2\pi}\int_{-\infty}^{\infty}\frac{\mathcal{F}[\mathcal{R}_{F_{x}}]}{(1/\tau''^{2} + \omega^{2})(1 + \omega^{2})}d\omega ,
 \\ &\text{Var}(\bar{\bar{y}}'') = \frac{1}{2\pi}\int_{-\infty}^{\infty}\frac{\mathcal{F}[\mathcal{R}_{F_{y}}]}{(1/\tau''^{2} + \omega^{2})(1 + \omega^{2}T_{m}^{2})}d\omega  .
\end{align*}
We then have
\begin{widetext}
\begin{align*}
    \text{Var}(\bar{\bar{y}}'') &= \frac{1}{2\pi}\int_{-\infty}^{\infty}\frac{\mathcal{F}[\mathcal{R}_{F_{y}}]}{(1/\tau''^{2} + \omega^{2})(1 + \omega^{2}T_{m}^{2})}d\omega   \\
    &= \frac{1}{2\pi}\int_{-\infty}^{\infty}\left[\int_{-\infty}^{\infty}\mathcal{R}_{F_{y}}(t)\text{cos}(\omega t)dt\right]\left(\frac{1}{(1/\tau''^{2} + \omega^{2})(1+(\omega T_{m})^{2})}\right) d\omega   \\
    &= \frac{1}{2\pi}\int_{-\infty}^{\infty}\mathcal{R}_{F_{y}}(t) \left(\int_{-\infty}^{\infty}\frac{\text{cos}(\omega t)}{(1/\tau''^{2} + \omega^{2})}\left(\frac{1}{1+(\omega T_{m})^{2}}\right) d\omega \right)dt   \\
    &= \frac{1}{\pi}\int_{0}^{\infty}\mathcal{R}_{F_{y}}(t) \left(\int_{-\infty}^{\infty}\frac{\text{cos}(\omega t)}{(1/\tau''^{2} + \omega^{2})}\left(\frac{1}{1+(\omega T_{m})^{2}}\right) d\omega \right)dt   \\
    &= \int_{0}^{\infty}\mathcal{R}_{F_{y}}(t) \tau''^{2} \left( \frac{T_{m}e^{-t/T_{m}} - \tau''e^{-t/\tau''} }{T_{m}^{2} - \tau''^2} \right)dt  ,
\end{align*}
\end{widetext}
where the second step comes from the fact that $\mathcal{R}_{F_{y}}(t)$ is    and symmetric so we can omit the $\text{sin}(\omega t)$ part of the Fourier transform, and the 4th 
step comes from the fact that the integrand is symmetric in $t$. We thus have, after normalizing with the averages, 
\begin{align*}
 \eta_{\bar{\bar{x}}''\bar{\bar{x}}''} = \int_{0}^{\infty}\eta_{FF}A_{F_{x}}(t) \left( \frac{\tau_{mat,x}e^{-\frac{t}{\tau_{mat,x}}} - \tau''e^{-\frac{t}{\tau''}} }{\tau_{mat,x}^{2} - \tau''^2} \right)dt 
 , \\
 \eta_{\bar{\bar{y}}''\bar{\bar{y}}''} = \int_{0}^{\infty}\eta_{FF}A_{F_{y}}(t) \left( \frac{\tau_{mat,y}e^{-\frac{t}{\tau_{mat,y}}} - \tau''e^{-\frac{t}{\tau''}} }{\tau_{mat,y}^{2} - \tau''^2} \right)dt 
.
\end{align*}
where  $A_{F_{x}}(t) = A_{F_{y}}(t)$ is the autocorrelation of the translation rate $F(\mathbf{u}(t), \mathbf{u}_{x})$. When $A_{F_{x}} \geq 0$, we find that the integrands of the above integrals are 
always non-negative. Upon comparing them we find that 
\begin{align*}
 T_{m}\eta_{\bar{\bar{y}}''\bar{\bar{y}}''} \leq \eta_{\bar{\bar{x}}''\bar{\bar{x}}''} .
\end{align*}
Using Eq.~\eqref{EQ: 3-step no-feedback double bar extrinsic}, and the fact that $\langle x'' \rangle = \langle y'' \rangle$ from Eq.~\eqref{EQ: 3-step flux-balance}, the 
above inequality becomes $T_{m}\eta_{y''y''} \leq \eta_{x''x''}$, which in terms of the CVs becomes Eq.~\eqref{EQ: two-step no oscillation constraint (manuscript)} of the main text. Note, that the exact same analysis can be used to derive 
\begin{align}
 T_{m}\eta_{\bar{y}''\bar{y}''} \leq \eta_{\bar{x}''\bar{x}''} ,
 \label{EQ: fp strong constraint}
\end{align}
which constrains the reporter (co)variances when the autocorrelation of $\bar{F}(t) = \mathrm{E}\big[F(\mathbf{u}(t), \mathbf{u}_{x})|\mathbf{u}[-\infty,t]\big]$ is stochastic. Note that $\bar{F}(t)$ is the translation rate when we average out all the mRNA intrinsic fluctuations. This latter constraint will be used in the section on stronger constraints on fluorescent reporters using a third reporter.

{
\section{Behaviour of specific example systems}\vspace{-.75em}
\label{SEC: Appendix particular systems}
To gain an intuition for where systems fall in the allowable region we present several example systems in the main text. For example, the arrowed curves in Fig.~\ref{FIG: Class 1 space of solutions -- feedback (manuscript)}B correspond to two toy models subject to an upstream component Z that undergoes Poisson fluctuations
\begin{align*}
\vspace{-.5em}
 z \myrightarrow{\lambda} z + 1\quad  & \quad z \myrightarrow{z/\tau_{z}}  z - 1 .
 \vspace{-.5em}
\end{align*}
The blue curve in Fig.~\ref{FIG: Class 1 space of solutions -- feedback (manuscript)}B corresponds to the system $R(\mathbf{u}(t)) = kz$, 
with $k = 10$, $\tau_{x} = 1$, $\tau_{y} = T$, $\langle z \rangle = \tau_{z}\lambda = 100$. As we increase the speed of the upstream fluctuations by decreasing $\tau_{z}$, the dual reporter correlations move downwards and left towards the bound of Eq.~\eqref{EQ: No oscillation constraint (manuscript)}. This makes intuitive sense because in that regime X and Y have little time to adjust to changing Z levels and de-correlate.}

{In contrast, the grey curve in Fig.~\ref{FIG: Class 1 space of solutions -- feedback (manuscript)}B corresponds to a system with feedback of Y onto its own production, with $R(\mathbf{u}(t)) = kz/(1 + \epsilon y)$, with $k = 10$, $\tau_{x} = 1$, $\tau_{y} = T$, $\tau_{z} = 1$, and $\langle z \rangle = 100$. For $\epsilon = 0$, we have no feedback and the grey line coincides with the previous system (blue curve) with $\langle x \rangle = \langle y \rangle /T = 1000$. As we increase the strength of the feedback $\epsilon$,  the correlations of this example system moves outside the region constrained by Eq.~\eqref{EQ: No feedback constraint (manuscript)} when \mbox{$\epsilon > 0, 0.07, 0.01$ for $T = 1, 0.5, 0.1$ respectively}. Maximizing the discriminatory power of the approach corresponds to minimizing the area in which systems without feedback could lie. Choosing $T = 1$ or $T \ll 1$ would thus be ideal to detect feedback in this system.}

{In Fig.~\ref{FIG: Class 1 space of solutions -- oscillations (manuscript)}C the red curve corresponds to a system stochastically driven through a Poisson variable Z with $\tau_{z} = 1$, $\langle z \rangle = 100$ and $R(\mathbf{u}(t)) = kz$ with $k = 10$ so that $\langle x \rangle = 1000$. In contrast, the blue curve corresponds to a system driven by an oscillation with $R(\mathbf{u}(t)) = K(\sin(\omega t + \phi) + 2)$, where $\phi$ is a random variable that de-synchronizes the ensemble and $K = 500$ so that $\langle x \rangle = 1000$. We set the oscillation period $2\pi/\omega = 2\tau_{x}$ to model an oscillation set by the cell-cycle (for example, the maturation time of mEGFP is roughly half of the \emph{E.~coli} cell-cycle \cite{Balleza2018}). We analyzed both types of systems for fixed $\tau_{x} = 1$ while varying $\tau_{y} = T$ to see which choice of $T$ maximizes the discriminatory power of the constraint. We find that for this particular example, the oscillating system crossed the dashed black curve when $\omega = 1/\sqrt{\tau_{x} \tau_{y}}$.}

\section{The effects of stochastic undercounting on dual reporter correlations}
\label{SEC: Undercounting}

\subsection{Undercounting mRNA}
\vspace{-0.1cm}
First we analyze the effects of undercounting on mRNA-levels of co-regulated genes. 
We would like to know how the derived bounds change when the reporter abundances $X$ and $Y$ are detected with fixed probabilities $p_{x}$ and $p_{y}$ respectively. This is done by introducing two new variables that correspond to the experimental readouts of the reporter abundances: $X_{r}$ and $Y_{r}$. In particular, $X_{r}$ corresponds to the detected number of $X$ molecules when each molecule is detected with probability $p_{x}$ (and similarly for $Y_{r}$ and $p_{y}$). In terms of these variables, open-loop systems are constrained by the following inequalities 
\begin{align}
\begin{split}
    -&\left(\eta_{x_{r}x_{r}} - \frac{p_{y}}{p_{x}}T\eta_{y_{r}y_{r}}\right) \leq \eta_{x_{r}y_{r}}\left(\frac{p_{y}}{p_{x}} - T\right), \\  &\left(\eta_{x_{r}x_{r}} - \frac{p_{y}}{p_{x}}T\eta_{y_{r}y_{r}}\right) \leq \eta_{x_{r}y_{r}}\left(1 - \frac{p_{y}}{p_{x}}T\right) .
    \end{split}
    \label{EQ: Undercounting-1step-nofeedback-bound}
\end{align}
Systems that break this inequality must be connected in some kind of feedback loop. When the detection probabilities are the same for both reporters $p_{x} = p_{y}$, the constraint reduces to the open-loop constraint given by Eq.~\eqref{EQ: No feedback constraint (manuscript)} of the manuscript. However, when $p_{x} \neq p_{y}$, the no-feedback bound differs from Eq.~\eqref{EQ: No feedback constraint (manuscript)} due to the $p_{y}/p_{x}$ term. As we will see in a later section, this ratio can be measured using a third reporter. 

Moreover, open-loop systems that are driven by a stochastic upstream signal must obey the following inequality 
\begin{align}
    \frac{p_{y}}{p_{x}}T\eta_{y_{r}y_{r}} + \frac{(1+T)\left(1 - \frac{p_{y}}{p_{x}}\right)}{2}\eta_{x_{r}y_{r}} \leq \eta_{x_{r}x_{r}} .  
    \label{EQ: Undercounting-1step-no-oscillation-bound}
\end{align}
Open-loop systems that break this bound must be driven by some kind of oscillation. Note that when $p_{x} = p_{y}$, the constraint reduces Eq.~\eqref{EQ: No oscillation constraint (manuscript)}  of the manuscript. However, when $p_{x} \neq p_{y}$, the no-oscillation bound differs from Eq.~\eqref{EQ: No oscillation constraint (manuscript)}. \\

\noindent \textbf{Derivation of Eq.~\eqref{EQ: Undercounting-1step-nofeedback-bound} and Eq.~\eqref{EQ: Undercounting-1step-no-oscillation-bound}:} We consider the following system, which is analogous to the class of systems in Fig.~\ref{FIG: Class 1 space of solutions -- feedback (manuscript)}A in the paper with the addition of a step which will model a binomial read-out of the components $X$ and $Y$

\begin{table}[H]
\begin{tabular}{llll}
Arbitrary mRNA dynamics :
\end{tabular}
\end{table}

\begin{align*}
&x \myrightarrow{R(\mathbf{u}(t))} x + 1  \quad   x \myrightarrow{x/\tau_{x}}  x - 1  \\
&y \myrightarrow{R(\mathbf{u}(t))} y + 1  \quad  
y \myrightarrow{y/\tau_{y}} y - 1  
\end{align*}

\begin{table}[H]
\begin{tabular}{llll}
Read-out mock process :
\end{tabular}
\end{table}

\begin{align*}
&x_{r} \myrightarrow{\lambda_{x} (x-x_{r})} x_{r} + 1 \quad x_{r} \myrightarrow{\beta_{x} x_{r}} x_{r} - 1  \\
&y_{r} \myrightarrow{\lambda_{y} (y - y_{r})} y_{r} + 1 \quad 
y_{r} \myrightarrow{\beta_{y} y_{r}} y_{r} - 1
\end{align*}
Here $X_{r}$ and $Y_{r}$ correspond to the experimental read-outs of $X$ and $Y$ respectively. For large values of $\lambda_{i}$ and $\beta_{i}$, 
the components $X_{r}$ and $Y_{r}$ correspond to binomial read-outs of the mRNA reporters $X$ and $Y$, respectively. $X_{r}$ has a counting success 
rate of $p_{x}=\lambda_{x}/(\lambda_{x} + \beta_{x})$, i.e. each $X$ mRNA molecule has a probability of $p_{x}$ to be detected experimentally (similarly for $Y$ and $p_{y}$). This approach again allows for arbitrary mRNA dynamics and can be solved for relations between (co)variances in exactly the same way as in the previous sections.

In particular, once the first and second moments of this system reach stationarity, general fluctuation balance relations \cite{hilfinger2015a} lead to (co)variance relations, which in the limit where $\beta_x \gg 1/\tau_{x}$ and $\beta_{y} \gg 1/\tau_{y}$, but where $p_{x}$ and $p_{y}$ are kept constant (this is not an approximation but is satisfied by construction of the mathematical mock system to define $X_{r}$ as a binomial cut of $X$), these (co)variance relations are
\begin{align}
\begin{split}
        &\eta_{x_{r}x_{r}} = \frac{1}{\langle x_{r} \rangle} + \eta_{xR}, \qquad 
    \eta_{y_{r}y_{r}} = \frac{1}{\langle y_{r} \rangle} + \eta_{yR}, \\
    &\eta_{x_{r}y_{r}} = \eta_{xy} = \frac{1}{1+T}\eta_{xR} + \frac{T}{1+T}\eta_{yR} .
\end{split}
    \label{EQ: Undercounting-1step-covariance-relations}
\end{align}
These are identical to the (co)variances relations given by Eq.~\eqref{EQ: 1-step fluctuation-balance relations} which are for the variables $X$ and $Y$, with the exception of the averages. Undercounting can only serve to further de-correlate the reporter read-outs, and so we would expect for the upper bound on $\rho_{xy}$ in Fig.~\ref{FIG: Class 1 space of solutions -- feedback (manuscript)}B to hold for the read-outs. Indeed, from the above equations
\begin{align*}
    \eta_{x_{r}y_{r}} =\dots \qquad \qquad& \\\frac{1}{1+T}\eta_{x_{r}x_{r}} + \frac{T}{1+T}\eta_{y_{r}y_{r}} &- \left(\frac{1}{\langle x_{r} \rangle} + \frac{T}{\langle y_{r} \rangle}\right)\left(\frac{1}{1+T}\right) \\
    &\leq \frac{1}{1+T}\eta_{x_{r}x_{r}} + \frac{T}{1+T}\eta_{y_{r}y_{r}} ,
\end{align*}
which corresponds to the bound in Fig.~\ref{FIG: Class 1 space of solutions -- feedback (manuscript)}B of the paper. Note that the bound holds for all detection probabilities.

Next, we will generalize the open-loop constraints to the system with the binomial read-out step. The constraint on open-loop systems derived for the system without the undercounting steps still holds for the $X$ and $Y$ components, as these don't depend in anyway on $X_{r}$ and $Y_{r}$. In terms of the conditional averages $\bar{x}$ and $\bar{y}$, this bound is given by Eq.~\eqref{EQ: 1-step no-feedback conditioned averages}. We would now like to write $\eta_{\bar{x}\bar{x}}$ and $\eta_{\bar{y}\bar{y}}$ in terms of the read-out (co)variances to write this inequality in terms of $\eta_{x_{r}x_{r}}$, $\eta_{y_{r}y_{r}}$, and $\eta_{x_{r}y_{r}}$. Recall from Eq.~\eqref{EQ: no-feedback 1-step barbar equals xR} that $\eta_{\bar{x}\bar{x}} = \eta_{xR}$ and $\eta_{\bar{y}\bar{y}} = \eta_{yR}$, so we need to solve for $\eta_{xR}$ and $\eta_{yR}$ in terms of the read-out (co)variances. 
First note that since $X_{r}$ is a binomial read-out of $X$, we have $\langle x_{r} \rangle = p_{x} \langle x \rangle$, and similarly for $Y$. As a result, the flux-balance relation given be Eq.~\eqref{EQ: 1-step flux-balance relations} becomes
\begin{align}
     \frac{\langle y_{r} \rangle}{\langle x_{r} \rangle} = T\frac{p_{y}}{p_{x}}.
    \label{EQ: Undercounting-1step-intrinsic-ratio}
\end{align}
We can now solve Eq.~\eqref{EQ: Undercounting-1step-covariance-relations} and  Eq.~\eqref{EQ: Undercounting-1step-intrinsic-ratio} for $\eta_{xR}$ and $\eta_{yR}$ as we have 4 equations and 4 unknowns
\begin{align}
    \begin{split}
    &\eta_{xR} = \frac{\left(\eta_{x_{r}y_{r}}(1+T) - T\eta_{y_{r}y_{r}}\right)\frac{p_{y}}{p_{x}} + \eta_{x_{r}x_{r}} }{{\left(1+\frac{p_{y}}{p_{x}}\right)}},\\
   &\eta_{yR} = \frac{\eta_{x_{r}y_{r}}(1+T) - \eta_{x_{r}x_{r}} + \frac{p_{y}}{p_{x}}T\eta_{y_{r}y_{r}}}{{T\left(1+\frac{p_{y}}{p_{x}}\right)}}.  
   \end{split}
   \label{EQ: Undercounting-1step-extrinsic-noise}
\end{align}
Note that when the detection probabilities are the same for both reporters, these expressions become identical to those that were derived without the undercounting step in Eq.~\eqref{EQ: 1-step eta bar in terms of covariances}. Recalling that $\eta_{xR} = \eta_{\bar{x}\bar{x}}$ and $\eta_{yR} = \eta_{\bar{y}\bar{y}}$, we substitute Eq.~\eqref{EQ: Undercounting-1step-extrinsic-noise} into Eq.~\eqref{EQ: 1-step no-feedback conditioned averages} which gives us the open-loop constraint given by Eq.~\eqref{EQ: Undercounting-1step-nofeedback-bound}. 

Next, we will generalize the no-oscillation bound to the system with the binomial read-out step. In terms of the conditional averages of $X$ and $Y$ the no-oscillation bound is given by Eq.~\eqref{EQ: Appendix eta time-scale result}. We then substitute the expressions from Eq.~\eqref{EQ: Undercounting-1step-extrinsic-noise} into Eq.~\eqref{EQ: Appendix eta time-scale result} and this bound becomes Eq.~\eqref{EQ: Undercounting-1step-no-oscillation-bound}.

\subsection{Undercounting fluorescent proteins}
\label{SEC: Undercounting fluorescent proteins}
Next, we analyze the effects of stochastic undercounting on co-regulated fluorescent proteins. This could model, for example, fluorescent proteins with chromophores that sometimes do not undergo maturation properly and thus are not detected.
Here we use the same approach as the previous section by adding two additional variables $X''_{r}$  and $Y''_{r}$ that correspond to the experimental read-outs of the fluorescent proteins. In particular, $X''_{r}$ corresponds to the detected number of $X''$ molecules when each molecule is detected with probability $p_{x}$, and similarly for $Y''_{r}$ and $p_{y}$. In terms of these variables, open-loop systems  are constrained by the following inequalities
following constraint
\begin{widetext}
\begin{align}
    \begin{split}
    T_{m}\left(\text{min}\left(T_{m},\frac{p_{y}}{p_{x}}\right)\eta_{y_{r}''y_{r}''} - \eta_{x_{r}''x_{r}''}\right) &\leq \eta_{x_{r}''y_{r}''}\left(\text{min}\left(T_{m},\frac{p_{y}}{p_{x}}\right) - T_{m}^{2}\right) ,\\   \eta_{x_{r}''x_{r}''} - \text{max}\left(\frac{p_{y}}{p_{x}},1\right)\eta_{y_{r}''y_{r}''} &\leq \left(1 - \text{max}\left(\frac{p_{y}}{p_{x}},1\right) \right)\eta_{x_{r}''y_{r}''} ,\\
    \quad 0 &\leq \eta_{x_{r}''y_{r}''} .
    \end{split}
    \label{EQ: no-feedback 3-step intuitive no-feedback bound-final}
\end{align}
\end{widetext}
Systems that break this inequality must be connected in some kind of feedback loop. When the detection probabilities are the same for both reporters $p_{x} = p_{y}$, the constraint reduces to Eq.~\eqref{EQ: two-step no-feedback constraint (manuscript)} of the main text.

Moreover, open-loop systems that are driven by a stochastic upstream signal must obey the following inequality
\begin{align}
\begin{split}
    \left(T_{m} - \text{min}\left(T_{m},\frac{p_{y}}{p_{x}}\right)\right)\eta_{x_{r}''y_{r}''} \\
    \leq \eta_{x_{r}''x_{r}''} - &\text{min}\left(T_{m}, \frac{p_{y}}{p_{x}}\right)\eta_{y_{r}''y_{r}''}.
    \end{split}
\label{EQ: Undercounting-3step-no-oscillation-bound}
\end{align}
Open-loop systems that break this bound must be driven by some kind of oscillation. When $p_{x} = p_{y}$, the constraint reduces to Eq.~\eqref{EQ: two-step no oscillation constraint (manuscript)} of the main text.\\

\noindent \textbf{Derivation of Eq.~\eqref{EQ: no-feedback 3-step intuitive no-feedback bound-final} and Eq.~\eqref{EQ: Undercounting-3step-no-oscillation-bound}:} We consider the following system, which is analogous to the class of systems in Fig.~\ref{FIG: Class of systems for FPs} with the addition of a step which will model a binomial read-out of the components $X''$ and $Y''$
\begin{table}[H]
\centering
\fontsize{14}{12}
\begin{tabular}{l}
Arbitrary fluorescent protein dynamics:
\end{tabular}
\end{table}
\vspace{-0.7cm}
\begin{align*}
x'\myrightarrow{F(\mathbf{u}(t), x)} x' + 1 \quad (x',x'') \myrightarrow{x'/\tau_{mat,x}} (x' - 1, x'' + 1)   \\
y'\myrightarrow{F(\mathbf{u}(t), y)} y' + 1 \quad (y',y'') \myrightarrow{y'/\tau_{mat,y}} (y' - 1, y'' + 1)
\\  x'' \myrightarrow{x''/\tau''} x'' -1 \quad \quad y'' \myrightarrow{y''/\tau''} y'' -1 \qquad \quad
\end{align*}
\vspace{-0.5cm}

\begin{table}[H]
\centering
\fontsize{14}{12}
\begin{tabular}{l}
Read-out mock process:
\end{tabular}
\end{table}

\vspace{-0.7cm}
\begin{align*}
x''_{r}\myrightarrow{\lambda_{x}(x''-x''_{r})} x''_{r} + 1\quad 
x''_{r} \myrightarrow{\beta_{x}x''_{r}} x''_{r} - 1 \\
y''_{r}\myrightarrow{\lambda_{y}(y''-y''_{r})} y''_{r} + 1\quad 
y''_{r} \myrightarrow{\beta_{y}y''_{r}} y''_{r} - 1
\end{align*}
Just like in the mRNA case, for large values of $\lambda_{i}$ and $\beta_{i}$ the additional read-out steps model a mock process that turn $X''_{r}$ and $Y''_{r}$ into instantaneous binomial read-outs of the fluorescent protein abundances $X''$ and $Y''$. Here $X''_{r}$ corresponds to the abundance of FPs that are fluorescing, where each $X''$ has a probability $p_{x} = \lambda_{x}/(\lambda_{x} + \beta_{x})$ of having matured properly (similarly for $Y''_{r}$). Following the same steps as in the previous section, we get the following (co)variance relations  
\begin{align}
\begin{split}
    &\eta_{x''_{r}y''_{r}}  = \eta_{x''y''}, \\
    \eta_{x''_{r}x''_{r}} = \frac{1}{\langle x''_{r} \rangle} +& \eta_{x''x'},  \quad 
    \eta_{y''_{r}y''_{r}} = \frac{1}{\langle y''_{r} \rangle} + \eta_{y''y'} .
\end{split}
    \label{EQ: Undercounting-3step-covariance-relations-1}
\end{align}
Moreover, since $X''_{r}$ is a binomial read-out of $X''$ we have $\langle x''_{r} \rangle = p_{x}\langle x'' \rangle$, and so from the above and the fluctuation balance equations given by Eq.~\eqref{EQ: 1-step fluctuation-balance relations} we have 
\begin{align}
    \eta_{x''_{r}x''_{r}} = \frac{(1-p_{x})}{\langle x''_{r}\rangle} + \eta_{x''x''} \quad \text{\&} \quad  \eta_{y''_{r}y''_{r}} = \frac{(1-p_{y})}{\langle y''_{r} \rangle} + \eta_{xx} .
    \label{EQ: Undercounting-3step-eta-xr-to-eta-x}
\end{align}
As a result we have
\begin{align*}
    \eta_{x''_{r}y''_{r}} = \eta_{x''y''} 
                        &\leq \frac{1}{1+T_{m}}\eta_{x''x''} + \frac{T_{m}}{1+T_{m}}\eta_{y''y''} \\&\leq \frac{1}{1+T_{m}}\eta_{x''_{r}x''_{r}} + \frac{T_{m}}{1+T_{m}}\eta_{y_{r}''y_{r}''} ,
\end{align*}
where the second step is from the upper bound in Fig.~\ref{FIG: 3 step class of systems with maturation (manuscript)}B given by $
\eta_{x''y''}(1+T_{m}) \leqslant \eta_{x''x''} + T_\mathrm{m}\eta_{y''y''}$. In terms of the read-out reporter correlation and CVs the above equations corresponds to the top bound in Fig.~\ref{FIG: 3 step class of systems with maturation (manuscript)}B of the paper. 

Next, we will generalize the open-loop constraints to the system with the binomial read-out step. When there is no feedback we can use Eq.~\eqref{EQ: no-feedback 3-step fluctuation-balance in terms of bared variables} along with Eq.~\eqref{EQ: Undercounting-3step-covariance-relations-1} and Eq.~\eqref{EQ: Undercounting-3step-eta-xr-to-eta-x} to write 
\begin{align}
\begin{split}
    \eta_{x''_{r}y''_{r}} = \frac{1}{1+T_{m}}\eta_{\bar{x}''\bar{x}''} + \frac{T_{m}}{1+T_{m}}\eta_{\bar{y}''\bar{y}''} , \qquad\\ 
    \eta_{x''_{r}x''_{r}} = \frac{1}{\langle x''_{r} \rangle} + \eta_{\bar{\bar{x}}''\bar{\bar{x}}''},  \qquad 
    \eta_{y''_{r}y''_{r}} = \frac{1}{\langle y''_{r} \rangle} + \eta_{\bar{\bar{y}}''\bar{\bar{y}}''} .
\end{split}
    \label{EQ: Undercounting-3step-covariance-relations}
\end{align}
Recall that in terms of the conditional averages $\bar{x}''$ and $\bar{y}''$ the open-loop constraint is given by Eq.~\eqref{EQ: no-feedback 3-step intuitive no-feedback bound}, which in combination with the first equation on the right in Eq.~\eqref{EQ: Undercounting-3step-covariance-relations} can be written as 
\begin{align}
    T_{m}\eta_{\bar{y}''\bar{y}''} \leq \eta_{x''_{r}y''_{r}} \leq \eta_{\bar{y}''\bar{y}''} .
    \label{EQ: no-feedback 3-step intuitive no-feedback bound-2}
\end{align}
We generally cannot solve for $\eta_{\bar{x}''\bar{x}''}$ and $\eta_{\bar{y}''\bar{y}''}$ in terms of $\eta_{x''_{r}x_{r}''}$, $\eta_{y_{r}''y_{r}''}$, and $\eta_{x_{r}''y_{r}''}$ like we did for the previous case in Eq.~\eqref{EQ: Undercounting-1step-extrinsic-noise} because the above system of equations is underdetermined. However, we can derive bounds on these extrinsic contributions using the inequality Eq.~\eqref{EQ: no-feedback 3-step intrinsic noise bound}. Once the reporter averages have reached stationarity, we have $\langle x_{r}'' \rangle = \frac{p_{y}}{p_{x}}\langle y_{r}'' \rangle$. This balance relation, together Eq.~\eqref{EQ: no-feedback 3-step intrinsic noise bound} and Eqs.~\eqref{EQ: Undercounting-3step-covariance-relations} and \eqref{EQ: no-feedback 3-step intuitive no-feedback bound} lead to the following bounds on the extrinsic contributions
\begin{widetext}
\begin{align}
    \begin{split}
    \frac{\text{max}(\frac{p_{y}}{p_{x}},1)\eta_{x_{r}''y_{r}''}(1+T_{m}) - T_{m}\left(\text{max}(\frac{p_{y}}{p_{x}},1)\eta_{y_{r}''y_{r}''} - \eta_{x_{r}''x_{r}''}\right)}{T_{m} + \text{max}(\frac{p_{y}}{p_{x}},1)} &\leq \eta_{\bar{x}''\bar{x}''} \\
    \eta_{\bar{x}''\bar{x}''} \leq& \; \eta_{x_{r}''y_{r}''} + \frac{T_{m}\left(\eta_{x_{r}''x_{r}''} - \text{min}(T_{m},\frac{p_{y}}{p_{x}})\eta_{y_{r}''y_{r}''}\right)}{\text{min}(T_{m},\frac{p_{y}}{p_{x}}) + T_{m}} \\
    \frac{\eta_{x_{r}''y_{r}''}(1+T_{m}) - \eta_{x_{r}''x_{r}''} + \text{min}\left(T_{m},\frac{p_{y}}{p_{x}}\right)\eta_{y_{r}''y_{r}''}}{\text{min}\left(T_{m}, \frac{p_{y}}{p_{x}}\right) + T_{m}} &\leq \eta_{\bar{y}''\bar{y}''} \\
    \eta_{\bar{y}''\bar{y}''} \leq& \frac{\eta_{x_{r}''y_{r}''}\left(1+T_{m}\right) - \eta_{x_{r}''x_{r}''} + \text{max}\left(\frac{p_{y}}{p_{x}},1\right)\eta_{y_{r}''y_{r}''}}{\text{max}\left(\frac{p_{y}}{p_{x}},1\right) + T_{m}} .
    \end{split}
    \label{EQ: Undercounting-3step-extrinsic-noise}
\end{align}
\end{widetext}
Substituting the last two inequalities into Eq.~\eqref{EQ: no-feedback 3-step intuitive no-feedback bound-2} leads to the constraint given be Eq.~\eqref{EQ: no-feedback 3-step intuitive no-feedback bound-final}. 

Next, we will generalize the "no-oscillation" bound to the system with the binomial read-out step. Recall that, in terms of the conditional averages, this bound is given by Eq.~\eqref{EQ: fp strong constraint}. Substituting the second and third inequalities from Eq.~\eqref{EQ: Undercounting-3step-extrinsic-noise} we obtain the inequality given by Eq.~\eqref{EQ: Undercounting-3step-no-oscillation-bound}.

\section{Genes with proportional transcription rates}
\label{SEC: Proportional rates}
Here we consider the more general class of systems in which two components are produced with arbitrary production rates that are proportional with some proportionality 
constant $\alpha$. 
We find that the system becomes identical to the same system in which $\alpha = 1$ but where there is systematic undercounting with unequal detection probabilities (see Sec.~\ref{SEC: Undercounting}). We will first present the derived bounds for the $\alpha \neq 1$ case followed by the derivations. Note that the results will depend on the proportionality constant $\alpha$, which can be measured using a third reporter as shown Appendix.~\ref{SEC: Additional reporters}. 

\subsection{mRNA with proportional transcription rates}
We consider the following class of systems which is analogous to the class of systems in Fig.~1A in the paper with the exception that now the transcription rates of the two mRNA are not equal but proportional
\begin{align}
\begin{split}
 x \myrightarrow{R(\mathbf{u}(t))} x + 1\quad  & \quad x \myrightarrow{x/\tau_{x}}  x - 1 \\
 y \myrightarrow{\alpha R(\mathbf{u}(t))} y + 1\quad  & \quad y \myrightarrow{y/\tau_{y}} y - 1
\end{split}
\label{EQ: Proportional-rates-1step-reactions}
\end{align}
First we will present the analogue of the constraint on open-loop systems given by Eq.~\eqref{EQ: No feedback constraint (manuscript)} in the paper. In particular, systems from the above class of systems in which the components X and Y do not directly, or indirectly, affect their own transcription rate must satisfy the following inequalities
\begin{align}
\begin{split}
    -\left(\eta_{xx} - \alpha T\eta_{yy}\right) &\leq \eta_{xy}\left(\alpha - T\right), \\ \left(\eta_{xx} - \alpha T\eta_{yy}\right) &\leq \eta_{xy}\left(1 - \alpha T\right) .
\end{split}
    \label{EQ: Proportional-1step-nofeedback-bound}
\end{align}
Systems that break these inequalities must be connected in some kind of feedback loop. Note that this equation is identical to Eq.~\eqref{EQ: Undercounting-1step-nofeedback-bound} with the replacement $\frac{p_{y}}{p_{x}} \rightarrow \alpha$. 

Next we will present the analogue of the constraint on stochastic systems given by Eq.~\eqref{EQ: No oscillation constraint (manuscript)} in the main text. Open-loop systems from the above class in which the transcription rate is stochastic must satisfy the following inequality
\begin{align}
    \alpha T\eta_{yy} + \frac{(1+T)\left(1 - \alpha \right)}{2}\eta_{xy} \leq \eta_{xx} .  
    \label{EQ: Proportional-1step-no-oscillation-bound}
\end{align}
Open-loop systems that break this bound must be driven by some kind of oscillation. Again, note that this equation is identical to Eq.~\eqref{EQ: Undercounting-1step-no-oscillation-bound} with the replacement $\frac{p_{y}}{p_{x}} \rightarrow \alpha$.\\

\noindent \textbf{Derivation of Eqs.~\eqref{EQ: Proportional-1step-nofeedback-bound} and \eqref{EQ: Proportional-1step-no-oscillation-bound}:} Once the averages and the (co)variances of the components X and Y reach stationarity, general fluctuation balance relations \cite{hilfinger2015a} lead to the following (co)variance relations
\begin{align*}
 &\eta_{xx} = \frac{1}{\lb x \rb} + \eta_{xR} ,\quad \eta_{yy} = \frac{1}{\lb y \rb} + \eta_{yR},   \\ &\eta_{xy} = \frac{1}{1+T}\eta_{xR}  + \frac{T}{1+T}\eta_{yR} . 
\end{align*}
These are identical to the fluctuation balance equations for the analogous class of system with $\alpha = 1$, given by Eq.~\eqref{EQ: 1-step fluctuation-balance relations}, which is intuitively explained from the fact that we are considering 
normalized (co)variances which cancel out the proportionality constant. In fact, when the components X and Y do not directly, or indirectly, affect their own transcription rate, we can again condition on the upstream history, in which case $\eta_{\bar{y}\bar{y}}$ is 
equal to the same value it would take when $\alpha = 1$, and so the open-loop constraint of Eq.~\eqref{EQ: 1-step no-feedback conditioned averages} and the no-oscillation bound of Eq.~\eqref{EQ: Appendix eta time-scale result} hold for all $\alpha$. However, the averages will not have the same asymmetry as the $\alpha = 1$ case. In particular, once the averages reach stationarity we obtain the following flux balance relations \cite{hilfinger2015a}
\begin{align}
 \langle R \rangle = \langle x \rangle/\tau_{x} \quad \text{\&} \quad \alpha \langle R \rangle = \langle y \rangle /\tau_{y}  \quad \Rightarrow \quad 
 \frac{\langle y \rangle}{\langle x \rangle} = \alpha T ,
 \label{EQ: Proportional-1step-intrinsic-ratio}
\end{align}
which, with the above fluctuation balance relations, is analogous to Eq.~\eqref{EQ: Undercounting-1step-covariance-relations} and Eq.~\eqref{EQ: Undercounting-1step-intrinsic-ratio} with the exchange $\frac{p_{y}}{p_{x}} \to \alpha$. The following results thus follow 
\begin{align}
    \begin{split}
   &\eta_{\bar{x}\bar{x}} = \frac{\eta_{xy}(1+T)\alpha + \eta_{xx} - \alpha T\eta_{yy}}{1+\alpha }, \\
   &\eta_{\bar{y}\bar{y}} = \frac{\eta_{xy}(1+T) - \eta_{xx} + \alpha T\eta_{yy}}{T\left(1+\alpha \right)} .
   \end{split}
   \label{EQ: Proportional-rates-1step-extrinsic-noise}
\end{align}
Substituting the above expressions into Eq.~\eqref{EQ: 1-step no-feedback conditioned averages} gives us Eq.~\eqref{EQ: Proportional-1step-nofeedback-bound}, and substituting the above expressions into Eq.~\eqref{EQ: Appendix eta time-scale result} gives us Eq.~\eqref{EQ: Proportional-1step-no-oscillation-bound}.

\subsection{Fluorescent proteins with proportional translation rates}
We consider the following class of systems 
\begin{align}
\begin{split}
x'\myrightarrow{F(\mathbf{u}(t), \mathbf{u}_{x})} x' + 1 \quad (x',x'') \myrightarrow{x'/\tau_{mat,x}} (x' - 1, x'' + 1)  \\
y'\myrightarrow{\alpha F(\mathbf{u}(t), \mathbf{u}_{y})} y' + 1 \quad (y',y'') \myrightarrow{y'/\tau_{mat,y}} (y' - 1, y'' + 1)  \\
x'' \myrightarrow{x''/\tau''} x'' -1 \quad y'' \myrightarrow{y''/\tau''} y'' -1 \qquad \quad
\end{split}
\label{EQ: Proportional-rates-3step-reactions}
\end{align}
which is analogous to the class of systems in Fig.~\ref{FIG: Class of systems for FPs} with the exception that now the translation rates of the two fluorescent proteins are not equal but proportional with proportionality constant $\alpha$. First we will present the analogue of the constraint on open-loop systems given by Eq.~(\eqref{EQ: two-step no-feedback constraint (manuscript)}) in the paper. In particular, systems from the above class of systems in which the downstream components $\mathbf{u}_{x}$, $\mathbf{u}_{y}$, X', Y', X'',Y'' do not directly, or indirectly, affect their own transcription and translation rates must satisfy the following inequalities
\begin{align}
    \begin{split}
    \text{min}\left(T_{m},\alpha \right)\eta_{y''y''} - \eta_{x''x''} &\leq \eta_{x''y''}\left(\frac{\text{min}\left(T_{m},\alpha\right) - T_{m}^{2}}{T_{m}}\right) \\   \eta_{x''x''} - \text{max}\left(\alpha,1\right)\eta_{y''y''} &\leq \left(1 - \text{max}\left(\alpha,1\right) \right)\eta_{x''y''} \\
    \quad 0 &\leq \eta_{x''y''} .
    \end{split}
    \label{EQ: Proportional-3step-nofeedback-bound-final}
\end{align}
Systems that break these inequality must be connected in some kind of feedback loop. Note that this equation is identical to Eq.~\eqref{EQ: no-feedback 3-step intuitive no-feedback bound-final} with the exchange $\frac{p_{y}}{p_{x}} \to \alpha$. 

Next we will present the analogue of the constraint on stochastic systems given by Eq.~\eqref{EQ: two-step no oscillation constraint (manuscript)} in the paper. In particular, systems from the above class without feedback in which the translation rate is stochastic must satisfy the following inequality
\begin{align}
    \left(T_{m} - \text{min}\left(T_{m},\alpha \right)\right)\eta_{x''y''} \leq \eta_{x''x''} - \text{min}\left(T_{m}, \alpha \right)\eta_{y''y''}   .  
\label{EQ: Proportional-3step-no-oscillation-bound}
\end{align}
Open-loop systems that break this bound must be driven by some kind of oscillation. Again, note that this equation is identical to Eq.~\eqref{EQ: Undercounting-3step-no-oscillation-bound} with the exchange $\frac{p_{y}}{p_{x}} \to \alpha$.\\

\noindent \textbf{Derivation of Eqs.~\eqref{EQ: Proportional-3step-nofeedback-bound-final} and \eqref{EQ: Proportional-3step-no-oscillation-bound}:} Following the same analysis that was done in Sec.~\ref{Appendix section on open-loop constraint for fluorescent reporters}, we can derive the exact same equations as Eq.~\eqref{EQ: no-feedback 3-step fluctuation-balance in terms of bared variables}, \eqref{EQ: no-feedback 3-step intuitive no-feedback bound}, and \eqref{EQ: no-feedback 3-step intrinsic noise bound}, only now they apply to (co)variances of the class of systems with $\alpha \neq 1$. These equations don't change when $\alpha \neq 1$ because we are considering normalized (co)variances which cancel out the proportionality constant. The flux balance relation given by Eq.~\eqref{EQ: 3-step flux-balance} will be different, however, as now one of the averages will be scaled up or down by the factor $\alpha$. In particular, at stationarity we have the following flux balance relations \cite{Hilfinger2011}
\begin{align*}
 &\langle x'' \rangle / \tau'' = \langle x' \rangle /\tau_{max,x} = \langle F \rangle, \\ &\langle y'' \rangle / \tau'' = \langle y' \rangle /\tau_{max,y} = \alpha \langle F \rangle \quad \Rightarrow \quad \langle y'' \rangle / \langle x'' \rangle = \alpha .
\end{align*}
The system of equations becomes mathematically identical to the system of equations we used to derive the bounds in Sec.~\ref{SEC: Undercounting fluorescent proteins} with the exchanges $\frac{p_{y}}{p_{x}} \to \alpha$.  Eqs.~\eqref{EQ: Proportional-3step-nofeedback-bound-final} and \eqref{EQ: Proportional-3step-no-oscillation-bound} thus follow from the derivations in that section where we exchange $\frac{p_{y}}{p_{x}} \to \alpha$.

\subsection{Systematic undercounting of systems with proportional transcription rates}
\label{SEC: Proportional both}
We found in the previous sections that the equations from which we derive our bounds when $\alpha \neq 1$ are mathematically identical to the analogous equations when $\alpha = 1$ but where each reporter is detected with a fixed probability such that $\alpha = \frac{p_{y}}{p_{x}}$. This is because in both these cases the fluctuation-balance relations between the reporter (co)variances remain unchanged compared to the original class of systems, but the averages will change according to Eq.~\eqref{EQ: Undercounting-1step-intrinsic-ratio} and Eq.~\eqref{EQ: Proportional-1step-intrinsic-ratio}. Therefor, when $\alpha \neq 1$ and there is systematic undercounting such that $p_{y} \neq p_{x}$, the only difference in our equations would be that the averages scale differently. In fact, the equations become analogous to those where there is no systematic undercounting but where the transcription rates are proportional with proportionality factor $\tilde{\alpha} = \alpha \frac{p_{y}}{p_{x}}$. All the bounds presented in Sec.~\ref{SEC: Proportional rates} thus hold but with the exchange $\alpha \to \tilde{\alpha}$. This ``proportionality constant'' will often be unknown, and so in the next section we show how it can be measured 
using a third reporter. 

\section{Additional reporters}
\label{SEC: Additional reporters}

\subsection{Measuring unknown life-time ratios}
\label{SEC: Appendix inferring unknown lifetime ratios}
First, we will show how to measure the mRNA life-time ratio $T$ using a third reporter. Recall, for the class of systems shown in Eq.~\eqref{EQ: Definition of 1-step Dual-Reporter System} of the main text, the following fluctuation balance relations given by  Eq.~\eqref{EQ: 1-step fluctuation-balance relations} must hold. We now allow for an additional reporter $Z$ in the system, where $Z$ has a life-time of $\tau_{z}$, is produced with the same probabilistic birthrate $R(\mathbf{u}(t))$, and is allowed to feedback and affect the cloud of components $\mathbf{u}(t)$. In such a case, any 
pair of components in $X$, $Y$, $Z$ will have a fluctuation balance relations like Eq.~\eqref{EQ: 1-step fluctuation-balance relations}. In particular, we have 
\begin{widetext}
\begin{align*}
 \begin{split}
  \eta_{xx} = \frac{1}{\langle x \rangle} + \eta_{xR} \quad \quad  \eta_{yy} = \frac{1}{\langle y \rangle} + & \eta_{yR} \quad \quad  \eta_{zz} = \frac{1}{\langle z \rangle} + \eta_{zR} \\
  \eta_{xy} = \frac{1}{1 + T_{yx}}\eta_{xR} + \frac{T_{yx}}{1 + T_{yx}}\eta_{yR} \quad \quad \eta_{yz} = \frac{1}{1 + T_{zy}}\eta_{yR} + & \frac{T_{zy}}{1 + T_{zy}}\eta_{zR} \quad \quad \eta_{xz} = \frac{1}{1 + T_{zx}}\eta_{xR} + \frac{T_{zx}}{1 + T_{zx}}\eta_{zR} ,
 \end{split}
\end{align*}
\end{widetext}
where $T_{ij} = \tau_{i}/ \tau_{j}$. Recall that the averages are related here by a flux balance $\langle i \rangle / \langle j \rangle = T_{ij}$, which gives us 
3 more equations. If all we can measure are the reporter (co)variances, we are left with 9 unknowns: the three $\eta_{iR}$, the three averages 
$\langle i \rangle$, and the three life-time ratios $T_{ij}$. This is a system of 9 linear equations, 
so we can solve for all unknowns. In particular, we find 
\begin{align}
 T_{ij} = \frac{\eta_{jj} - \eta_{jk}}{\eta_{ii} - \eta_{ik}} \quad \text{ for } \quad i,j,k\in \{x,y,z\} .
\end{align}
Note that the above expression depends only on the reporter variances and the covariances. Three seperate dual reporter experiments can thus be done to measure all the $\eta_{ij}$ 
and infer all the life-time ratios. If only one life-time ratio is being measured, then only two experiments are needed as the above expression only involve the covariance between two reporters and their normalized variances. 

Next, we would like to derive a similar result for ratios of fluorescent protein maturation times. This can be done by using fluorescent reporters that all share the same (though distinct) promoter. Since such reporters are not involved in their regulation, we can use equation Eq.~\eqref{EQ: no-feedback 3-step etaxy covariance equation} which holds for pairs $X''$, $Y''$ of co-regulated fluorescent reporters without feedback 
\begin{align*}
\eta_{x''y''} = \frac{1}{1+T_{m}}\eta_{\bar{x}''\bar{x}''} + \frac{T_{m}}{1 + T_{m}}\eta_{\bar{y}''\bar{y}''} .
\end{align*}
As we increase the number of reporters, the number of such equations increases faster than the number of unknowns, and eventually we are able to solve for all the unknowns in terms 
of the reporter (co)variances. Unlike the mRNA case we cannot use the other fluctuation balance equations of the form $\eta_{x''x''} = 1/\langle x'' \rangle + \eta_{\bar{\bar{x}}''\bar{\bar{x}''}}$ because 
each of these equations adds an additional unknown $\eta_{\bar{\bar{x}}''\bar{\bar{x}''}}$ to the system of equations. With three reporters we thus do not have enough equations to solve for all the unknowns. 
With 4 reporters, however, we are able to solve for all maturation life-times given that we have one of the life-time ratios, and with 5 reporters we are able to solve for all life-time ratios. \\

\subsection{Measuring birthrate proportionality constants and reporter detection probabilities} 
\label{SEC: Measuring alpha and detection probabilities}

In Sec.~\ref{SEC: Undercounting} we have shown how systematic undercounting affects the derived results, and in Sec.~\ref{SEC: Proportional rates} we have shown how proportional production rates affect the derived results. We found that both cases have the same effect and can be treated together in full generality, see Sec.~\ref{SEC: Proportional both}. That is, when both are treated together we found that the constraints presented in the paper change where now they also depend on the factor $\tilde{\alpha} = \alpha\frac{p_{y}}{p_{x}}$, where $\alpha$ is the proportionality constant between the two production rates and $\frac{p_{y}}{p_{x}}$ is the ratio of detection probabilities. Here we will show how this $\tilde{\alpha}$ can be measured using three dual reporter experiments. The derived bounds from the previous sections can then be used with asymmetrical systematic undercounting and with naturally occurring proportional transcription rates.   

Like the previous subsection, we allow for an additional reporter $Z$ in the system, where the production rate of each reporter is $\alpha_{i}R$ for $i \in \{x,y,z\}$. 
Moreover, we allow for stochastic undercounting where each reporter is detected with fixed probability $p_{i}$ for $i \in \{x,y,z\}$. 
The reporter read-outs will have the following fluctuation balance relations (see Sec.~\ref{SEC: Undercounting} and Sec.~\ref{SEC: Proportional rates})
\begin{widetext}
\begin{align*}
 \begin{split}
  \eta_{x_{r}x_{r}} = \frac{1}{\langle x_{r} \rangle} + \eta_{xR} \quad \quad  \eta_{y_{r}y_{r}} = \frac{1}{\langle y_{r} \rangle} + & \eta_{yR} \quad \quad  \eta_{z_{r}z_{r}} = \frac{1}{\langle z_{r} \rangle} + \eta_{zR} \\
  \eta_{x_{r}y_{r}} = \frac{1}{1 + T_{yx}}\eta_{xR} + \frac{T_{yx}}{1 + T_{yx}}\eta_{yR} \quad \quad \eta_{y_{r}z_{r}} = \frac{1}{1 + T_{zy}}\eta_{yR} + & \frac{T_{zy}}{1 + T_{zy}}\eta_{zR} \quad \quad \eta_{x_{r}z_{r}} = \frac{1}{1 + T_{zx}}\eta_{xR} + \frac{T_{zx}}{1 + T_{zx}}\eta_{zR} ,
 \end{split}
\end{align*}
\end{widetext}
where $T_{ij} = \tau_{i}/ \tau_{j}$. As explained in Sec.~\ref{SEC: Proportional both} these are the exact same fluctuation balance equations as Eq.~\eqref{EQ: 1-step fluctuation-balance relations}, but now the averages will be related by $\langle i_{r} \rangle / \langle j_{r} \rangle = T_{ij}\frac{\tilde{\alpha}_{i}}{\tilde{\alpha}_{j}}$, where $\tilde{\alpha}_{i} = \alpha_{i}p_{i}$ . When 
the life-time ratios $T_{ij}$ are known, this gives us 9 equations and 9 unknowns (3 $\eta_{iR}$, 3 averages $\langle i_{r} \rangle$, and 3 $\tilde{\alpha_{i}}$ ratios) and so we can solve for all the $\tilde{\alpha}_{i}$ ratios. In particular, we find %
\begin{widetext}
\begin{align}
 \frac{\tilde{\alpha}_{i}}{\tilde{\alpha}_{j}} = T_{ji}\left(\frac{T_{ik} + T_{jk}T_{ij}}{T_{jk} + T_{ik}T_{ji}}\right) \left(\frac{(T_{ik} + T_{jk}T_{ij})\eta_{i_{r}i_{r}} - T_{ik}(1+T_{ij})\eta_{ij} - (1+T_{ik})\eta_{ik} + (1+T_{ik})\eta_{jk}}{(T_{jk} + T_{ik}T_{ji})\eta_{j_{r}j_{r}} - T_{jk}(1+T_{ji})\eta_{ji} - (1+T_{jk})\eta_{jk} + (1+T_{jk})\eta_{ik}}\right)
,
\end{align}
\end{widetext}
for $i,j,k \in \{x,y,z\}$. If the life-time ratios $T_{ij}$ are not known, then we have 3 additional unknowns and we have more unknowns than equations. However, including a fourth reporter gives us 
4 new unknowns 4 additional unknowns ($\langle u \rangle$, $\tilde{\alpha}_{u}$, $\tau_{u}$ and $\eta_{uR}$ for a fourth reporter $U$) and 7 additional equations analogous to the ones above, which means we can solve the system of equations for four reporters for all the life-time ratios $T_{ij}$ and the proportionality 
constants $\tilde{\alpha}_{i}$. Note that only (co)variance measurements are needed, and so 4 separate experiments involving two co-regulated reporters can be done to infer the $\tilde{\alpha}_{i}$.

For fluorescent proteins we cannot solve for the $\tilde{\alpha}$ this way because the system of equations will always be underdetermined. However, once the averages reach 
stationarity, flux balance leads to $\langle x'' \rangle /\langle y'' \rangle = \alpha$, where the translation rate of $y'$ is $\alpha F$ and that of $x'$ is $F$. 
If the experiments report absolute numbers of molecules or concentrations, and the averages can be measured, then $\alpha$ can be measured. However, averages are 
often not known because fluorescence measurements involve an unknown scaling factor between the numbers/concentrations and the light intensity. To get by this, one 
could do two single reporter experiments, where the same fluorescent reporter is used for the two genes. The ratio of the two averages would then cancel out this unknown 
scaling factor and would result in the actual ratio of components $\langle x''_{1} \rangle / \langle x''_{2} \rangle = \tilde{\alpha}$, from which $\tilde{\alpha}$ can be inferred. 

\subsection{Stronger constraints on fluorescent reporters using a third reporter}
\label{SEC: Stronger constraints}
The open-loop constraint for fluorescent proteins given by Eq.~\eqref{EQ: two-step no-feedback constraint (manuscript)} in the main text encompasses a larger region in the $\rho_{xy}$ and $CV_{x}/CV_{y}$ plane than the equivalent constraint for mRNA reporters given by Eq.~\eqref{EQ: No feedback constraint (manuscript)}. This is due to the fact that there are unspecified degrees of freedom in the class of fluorescent protein systems that are unknown given the variability of two downstream reporters. Specifically, the fluctuation-balance relations for open-loop mRNA reporters are given by Eq.~\eqref{EQ: 1-step no-feedback fluctuation-balance}
\begin{align}
\begin{split}
\eta_{xx} = \eta_{x,int} + \eta_{\bar{x}\bar{x}}, \qquad \eta_{yy} = \eta_{y,int} + \eta_{\bar{y}\bar{y}}, \\	 
 \eta_{xy} = \frac{1}{1+T}\eta_{\bar{x}\bar{x}}  + \frac{T}{1+T}\eta_{\bar{y}\bar{y}}  , \qquad
 \end{split}
\label{EQ: FP three reporters mrna eqns}
\end{align}
whereas the equivalent relations for fluorescent proteins are given by
\begin{align}
\begin{split}
 \eta_{x''x''} = \eta_{x'',int} +\eta_{\bar{x}''\bar{x}''} , \qquad 
 \eta_{y''y''} = \eta_{y'',int} +\eta_{\bar{y}''\bar{y}''}, \\ 
 \eta_{x''y''}  = \frac{1}{1+T_{m}}\eta_{\bar{x}''\bar{x}''} + \frac{T_{m}}{1+T_{m}}\eta_{\bar{y}''\bar{y}''}. \qquad 
\end{split}
 \label{EQ: FP three reporters FP eqns}
\end{align}
In both systems, the open-loop constraint on the conditional averages is identical, see Eqs~\eqref{EQ: 1-step no-feedback conditioned averages} and \eqref{EQ: no-feedback 3-step intuitive no-feedback bound},
\begin{align}
   T^{2}\eta_{\bar{x}\bar{x}} \leq \eta_{\bar{x}\bar{x}} \leq \eta_{\bar{x}\bar{x}} \quad \text{\&} \quad T_{m}^{2}\eta_{\bar{x}''\bar{x}''} \leq \eta_{\bar{x}''\bar{x}''} \leq \eta_{\bar{x}''\bar{x}''} .
   \label{EQ: FP three reporters no-feedback constraints}
\end{align}
The difference lies in the fact that the intrinsic system for the mRNA reporters is specified as there is only one step downstream from the shared birthrate. Specifically, $\eta_{x,int} = 1/\langle x \rangle$ and $\eta_{y,int} = 1/\langle y \rangle$, and so $\eta_{x,int} = T\eta_{y,int}$. This last equation allows us to solve Eq.~\eqref{EQ: FP three reporters mrna eqns} for $\eta_{\bar{x}\bar{x}}$ and $\eta_{\bar{y}\bar{y}}$ in terms of the measurable (co)variances which allows us to take full advantage of the open-loop constraint given by Eq.~\eqref{EQ: FP three reporters no-feedback constraints}. For fluorescent proteins this is not the case because the mRNA intrinsic systems are not specified (for example, the half-lives of the mRNA are not specified). As a result, we cannot close the system of equations to solve for $\eta_{\bar{x}''\bar{x}''}$ and $\eta_{\bar{y}''\bar{y}''}$ in terms of measurable (co)variances. We can, however, bound the intrinsic noise terms using Eq.~\eqref{EQ: no-feedback 3-step intrinsic noise bound}, which in addition to the open-loop constraint of Eq.~\eqref{EQ: FP three reporters no-feedback constraints} leads to the broader constrained region shown in the paper. We can visualize this difference by comparing the regions bounded by the open-loop constraints in Fig.~\ref{FIG: Class 1 space of solutions -- feedback (manuscript)}B and Fig.~\ref{FIG: 3 step class of systems with maturation (manuscript)}B of the paper. Systems that lie on the $\rho = 0$ line are dominated by intrinsic variability. For the mRNA region in Fig.~\ref{FIG: Class 1 space of solutions -- feedback (manuscript)}B, these systems must lie at the point $CV_{x}/CV_{y} = \sqrt{T}$, whereas for the FP region in Fig.~\ref{FIG: 3 step class of systems with maturation (manuscript)}B they can lie anywhere within $[ \sqrt{T},1 ]$. If the mRNA intrinsic systems were specified, the relation between $\eta_{x'',int}$ and $\eta_{y'',int}$ would be specified and systems on the $\rho = 0$ line would have to be located at some point like in the mRNA class of systems. Here we show how probing the system with a third fluorescent reporter allows us to account for this missing degree of freedom. In particular, we will show how with three reporters we can measure $\eta_{\bar{x}''\bar{x}''}$ and $\eta_{\bar{y}''\bar{y}''}$ so that we can fully take advantage of the open-loop constraint. 

We allow for a third fluorescent protein reporter Z$''$ that shares the same transcription rate and an identical mRNA intrinsic system as $X''$ and $Y''$. We let the maturation time of Z$''$ be $\tau_{mat,z}$, and we let $T_{m,ij} := \tau_{mat,i}/\tau_{mat,j}$ for $i,j \in \{x,y,z\}$. When each reporter does not affect it's own transcription or translation rates, the right-most fluctuation balance relation given by Eq.~\eqref{EQ: FP three reporters FP eqns} holds for each pair of reporters: 
\begin{align}
\begin{split}
   &\eta_{x''y''} =  \frac{\eta_{\bar{x}''\bar{x}''}}{1+T_{m,yx}} + \frac{T_{m,yx}\eta_{\bar{y}''\bar{y}''}}{1+T_{m,yx}}, \\
   &\eta_{y''z''} =  \frac{\eta_{\bar{y}''\bar{y}''}}{1+T_{m,zy}} + \frac{T_{m,zy}\eta_{\bar{z}''\bar{z}''}}{1+T_{m,zy}}, \\
   &\eta_{x''z''} =  \frac{\eta_{\bar{x}''\bar{x}''}}{1+T_{m,zx}} + \frac{T_{m,zx}\eta_{\bar{z}''\bar{z}''}}{1+T_{m,zx}} .
   \end{split}
\end{align}
With three separate dual reporter experiments, the three reporter covariances $\eta_{x''y''}$, $\eta_{x''z''}$ and $\eta_{y''z''}$ can be measured. If the maturation time ratios $T_{m,ij}$ are known, we have three equations and three unknowns, and so we can fully solve for $\eta_{\bar{x}''\bar{x}''}$, $\eta_{\bar{y}''\bar{y}''}$ and $\eta_{\bar{z}''\bar{z}''}$, 
\begin{widetext}
\begin{align}
    \eta_{\bar{i}''\bar{i}''} = \frac{T_{m,ik}(1 + T_{m,ij})\eta_{ij} + (1+T_{m,ik})\eta_{ik} - (1+T_{m,jk})\eta_{jk}}{T_{m,ik} + T_{m,jk}T_{m,ik}} \quad\quad \text{where} \quad i,j,k \in \{x,y,z\} ,
\end{align}
\end{widetext} 
and so we can fully take advantage of the open-loop constraint given by Eq.~\eqref{EQ: FP three reporters no-feedback constraints}.

This is also the case for the constraint on stochastic systems given by Eq.~\eqref{EQ: two-step no oscillation constraint (manuscript)} in the paper. In terms of the conditional averages, the strongest form of this constraint is given by Eq.~\eqref{EQ: fp strong constraint}
\begin{align}
    T_{m}\eta_{\bar{y}''\bar{y}''} \leq \eta_{\bar{x}''\bar{x}''} .
    \label{EQ: FP three reporters no-oscillation constraint}
\end{align}
Using the above approach we can do three separate dual reporter experiments involving 3 pairs of fluorescent reporters to solve for $\eta_{\bar{x}''\bar{x}''}$ and $\eta_{\bar{y}''\bar{y}''}$ and fully take advantage of the constraint given by Eq.~\eqref{EQ: FP three reporters no-oscillation constraint}.

Note that $\eta_{\bar{x}''\bar{x}''}$ is the variability in $X''$ that originates from the shared cloud of components $\mathbf{u}(t)$, so $\eta_{x''x''} - \eta_{\bar{x}''\bar{x}''}$ is the intrinsic noise that originates from the random nature of the transcription, translation, and maturation steps. The variance can also be decomposed as $\eta_{x''x''} = \frac{1}{\langle x'' \rangle } + \eta_{\bar{\bar{x}}''\bar{\bar{x}}}$, where $\frac{1}{\langle x'' \rangle } $ is the intrinsic noise that originates from the translation and maturation step, and $\eta_{\bar{\bar{x}}''\bar{\bar{x}}}$ is the variability originating from the intrinsic mRNA fluctuations as well as from the cloud of components $\mathbf{u}(t)$. If the protein abundances can be measured, along with $\eta_{\bar{x}''\bar{x}''}$ according to the above, then we can determine how much variability is generated through the different steps of gene expression.

\section{Concentrations of growing and dividing cells}
\label{SEC: Appendix concentrations}

\subsection{Derivation of Eq.~\eqref{EQ: No feedback constraint (manuscript)} and Eq.~\eqref{EQ: No oscillation constraint (manuscript)} for mRNA concentrations} 
In growing and dividing cells, the molecular abundances still follow the same production and degradation reactions during the ``cell-cycle'' 
but are now affected by cell-division. 
In particular, considering systems as defined in Eq.~\eqref{EQ: Definition of 1-step Dual-Reporter System}, X and Y share an unspecified birthrate $R(\mathbf{u}(t),V)$ which can now depend on the cell volume $V$ and the abundances of the components $\mathbf{u}(t)$. Furthermore, they are assumed to undergo first order degradation with life-times $\tau_{x}$ and $\tau_{y}$ respectively. 

However, now the cell volume increases and undergoes cell division at times $\{t_{i}\}$, which can vary over the cell ensemble.
At these moments the volume is reduced by a factor $V \to a_{i}V$ at $t_{i}$, as we follow one of the daughter cells.
For perfect symmetric division, for example, we have $a = 1/2$. To allow for stochastic division errors in cell volume, we let the $a_{i}$ vary over the ensemble and the division times through an unspecified distribution --- 
all we specify is that on average we have $\langle a_{i} \rangle = 1/2$. At the division times the cell content splits between the two daughter cells, and we assume that the mRNA numbers are split according to a binomial distribution with probability $a_{i}$. This is summarized as
\begin{align*}
\big(\;V,\; x, \;y \;\big) \xrightarrow{\text{At time } t_{i}} \big(\;a_{i}V,\; B(x,a_{i}),\; &B(y,a_{i}) \;\big) \\
&\text{for } t_{1},\: t_{2}, \: t_{3} \dots 
\end{align*}

We first consider systems in which the cellular components $X$ and $Y$ are affected by, but
do not affect, the otherwise unspecified environmental variables $\mathbf{u}(t)$ and $V$. In this scenario, we can analyze the average stochastic
dual reporter dynamics conditioned on the history of their upstream influences \cite{Hilfinger2012}:
$
\bar{x}(t) = \mathrm{E}\big[X_t|\mathbf{u}[-\infty,t], V[-\infty,t] \big]$ and  $\bar{y}(t) = \mathrm{E}\big[ Y_t|\mathbf{u}[-\infty,t] , V[-\infty,t]\big]$.
Since all these systems have the same volume history, they will all undergo division at the same times $\{t_{i}\}$ with the same splitting factors $\{a_{i}\}$. When the systems are not at one of these division points, 
the time evolution is specified completely by the reactions in Eq.~\eqref{EQ: Definition of 1-step Dual-Reporter System}. In particular, for $t_{i} < t < t_{i+1}$, the time evolution of the conditional averages is given by
\begin{align}
 \frac{d\bar{x}}{dt} = R(t) - \bar{x}/\tau_{x} \quad \text{\&} \quad \frac{d\bar{y}}{dt} =& R(t) - \bar{y}/\tau_{y} \label{EQ: Concentrations-1step-differential-equation-numbers}\\ 
 &\text{when}  \quad  t_{i} < t < t_{i+1}, \nonumber
\end{align}
where $R(t) = R(\mathbf{u}(t))$. At $t = t_{i+1}$ we have $V(t_{i+1}) \to a_{i+1}V(t_{i+1})$, $\bar{x}(t_{i+1}) \to a_{i+1}\bar{x}(t_{i+1})$, 
and $\bar{y}(t_{i+1}) \to a_{i+1}\bar{y}(t_{i+1})$. This gives us the boundary conditions for the above differential equation.

Now we consider the stochastic concentrations, defined as $X_{c} := X/V$ and $Y_{c} := Y/V$. Conditioned on the history of the upstream influences and the volume, we have 
\begin{align*}
 \bar{x}_{c}(t) &=  \mathrm{E}\Big[\frac{X_t}{V_{t}}\Big|\mathbf{u}[-\infty,t], V[-\infty,t] \Big]\\
	        &=  \frac{1}{V(t)}\mathrm{E}\big[X_t|\mathbf{u}[-\infty,t], V[-\infty,t] \big]
	        =  \frac{\bar{x}}{V(t)} ,
\end{align*}
where we can pull out the $V(t)$ from the expectation brackets because it is specified by the conditioning and becomes equivalent to a constant at time $t$.
In $t_{i} < t < t_{i+1}$, we use Eq.~\eqref{EQ: Concentrations-1step-differential-equation-numbers} and the product rule to find 
\begin{align}
 \frac{d\bar{x}_{c}}{dt} = \frac{d}{dt}\left(\frac{\bar{x}}{V(t)}\right) = R_{c}(t) - x_{c}/\tau_{x} - \bar{x}_{c}\frac{V'(t)}{V(t)} ,
 \label{EQ: Concentrations-1step-differential-equation-concentrations-general}
\end{align}
where $R_{c} := R/V$ is interpreted as the production rate of the concentration $X_{c}$. At division time $t_{i+1}$ we have $\bar{x}_{c} \to \frac{a_{i+1}\bar{x}}{a_{i+1}V} = \bar{x}_{c}$. 
Thus $\bar{x}_{c}$ is unchanged at the division times and is continuous, meaning Eq.~\eqref{EQ: Concentrations-1step-differential-equation-concentrations-general} holds for all~$t$. 

Eq.~\eqref{EQ: Concentrations-1step-differential-equation-concentrations-general} holds for any volume dynamics.
Here we assume that cellular volume grows exponentially, i.e., $V'(t) = V(t)\cdot \ln(2)/\tau_{c}$, where $\tau_{c}$ is the average cell cycle time of a cell.  
When this is the case Eq.~\eqref{EQ: Concentrations-1step-differential-equation-concentrations-general} becomes 
\begin{align}
 \frac{d\bar{x}_{c}}{dt} = R_{c}(t) - x_{c}/\tau_{x_{c}} \quad \text{\&} \quad \frac{d\bar{y}_{c}}{dt} = R_{c}(t) - y_{c}/\tau_{y_{c}} ,
 \label{EQ: Concentrations-1step-differential-equation-concentrations-exponential}
\end{align}
where $\tau_{x_{c}} := \left(\frac{1}{\tau_{x}} + \frac{\ln(2)}{\tau_{c}}\right)^{-1}$ is the ``life-time'' of the concentration $X_{c}$, and the right equation follows by symmetry with similarly defined $\tau_{y_{c}}$. 

Eq.~\eqref{EQ: Concentrations-1step-differential-equation-concentrations-exponential} can be understood as a dilution approximation, which is exact for the conditional ensemble average. Because it is mathematically identical to Eq.~\eqref{EQ: no-feedback 1-step differential equations} the same bounds on the average dynamics follow. To relate the results to the stochastic dynamics of $X_{c}$ and $Y_{c}$, we use the law of total variance and Eq.~\eqref{EQ: Concentrations-1step-differential-equation-concentrations-exponential} to derive the (co)variance relations
\begin{align}
    &\eta_{x_{c}x_{c}} =  \frac{\left \langle x_{c}/V \right \rangle}{\langle x_{c} \rangle^{2}} + \eta_{\bar{x}_{c}\bar{x}_{c}},  \; \; \; \; \eta_{y_{c}y_{c}} =  \frac{\left \langle y_{c}/V \right \rangle}{\langle y_{c} \rangle^{2}} + \eta_{\bar{y}_{c}\bar{y}_{c}}, \nonumber \\
    &\eta_{x_{c}y_{c}} = \frac{1}{1+T_{c}}\eta_{\bar{x}_{c}\bar{x}_{c}} + \frac{T_{c}}{1 + T_{c}}\eta_{\bar{y}_{c}\bar{y}_{c}},
\label{EQ: Appendix concentrations (co)variance relations (manuscript)}
\end{align}
and the flux balance relation
\begin{align}
    \langle y_{c} \rangle = T_{c}\langle x_{c} \rangle,
\label{EQ: Appendix concentrations flux balance (manuscript)}
\end{align}
where $T_{c} := \tau_{y_{c}}/\tau_{x_{c}}$ (see below for detailed derivation of Eqs.~\eqref{EQ: Appendix concentrations (co)variance relations (manuscript)} and \eqref{EQ: Appendix concentrations flux balance (manuscript)}). For systems where $R_{c}(t)$ does not vary periodically as a function of the cell-cycle,
$X_{c}$ and $Y_{c}$ are independent of $V(t)$
and Eqs.~\eqref{EQ: Appendix concentrations (co)variance relations (manuscript)} and \eqref{EQ: Appendix concentrations flux balance (manuscript)} become mathematically identical to Eqs.~\eqref{EQ: 1-step fluctuation-balance relations} and \eqref{EQ: 1-step flux-balance relations} from which the constraints derived for $\bar{x}$ and $\bar{y}$ are translated to X and Y. Following the same steps as in Appendix \ref{Appendix section on open-loop constraint} and \ref{Appendix section on time-scales} thus gives the same constraints, which concludes the proof that bounds derived for molecular abundances of stochastically driven systems in  class Eq.~\eqref{EQ: Definition of 1-step Dual-Reporter System} also apply to their cellular concentrations.

When the average dynamics of $X_{c}$ and $Y_{c}$ is not independent of $V$, i.e., when $R_{c}(t)$ is cell-cycle dependent, we cannot close the system of equations given by Eqs.~\eqref{EQ: Appendix concentrations (co)variance relations (manuscript)} and \eqref{EQ: Appendix concentrations flux balance (manuscript)}. However, with a third reporter we are able to close the system of equations to derive bounds similar to the ones presented in the main text for molecular abundances as discussed in Appendix \ref{SEC: Appendix exact constraints on volume dependent genes}.\\

\noindent \textbf{Derivation of Eqs.~\eqref{EQ: Appendix concentrations (co)variance relations (manuscript)} and \eqref{EQ: Appendix concentrations flux balance (manuscript)}}: Taking the ensemble average of Eq.~\eqref{EQ: Concentrations-1step-differential-equation-concentrations-exponential} over all possible histories, we get
\begin{align*}
 \mathrm{E}\left[ \frac{d\bar{x}_{c}}{dt} \right] = \mathrm{E}[ R_{c}(t) ] - \mathrm{E}[ \bar{x}_{c} ]/\tau_{x_{c}} = \langle R_{c} \rangle + \langle x_{c} \rangle/\tau_{x_{c}} .
\end{align*}
Note that $\mathrm{E}\left[ \frac{d\bar{x}_{c}}{dt} \right] = \frac{d}{dt}\mathrm{E}[\bar{x}_{c}] = \frac{d}{dt}\langle x_{c} \rangle$, so once
the averages reach stationarity the left hand side goes to zero and we are left with the following flux-balance relations
\begin{align}
 \langle R_{c} \rangle = \langle x_{c} \rangle / \tau_{x_{c}} \quad \text{\&} \quad \langle R_{c} \rangle = \langle y_{c} \rangle / \tau_{y_{c}} ,
 \label{EQ: Concentrations-1step-flux-balance}
\end{align}
from which Eq.~\eqref{EQ: Appendix concentrations flux balance (manuscript)} follows.
Moreover, Eq.~\eqref{EQ: Concentrations-1step-differential-equation-concentrations-exponential} is identical to analogous equations derived for the molecular numbers Eq.~\eqref{EQ: no-feedback 1-step differential equations}, and so we can follow the same analysis to derive the following expression
\begin{align*}
 \eta_{x_{c}y_{c}} = \frac{1}{1+T_{c}}\eta_{\bar{x}_{c}\bar{x}_{c}} + \frac{T_{c}}{1+T_{c}}\eta_{\bar{y}_{c}\bar{y}_{c}}.
\end{align*}

We now use the law of total variance to decompose $\eta_{x_{c}x_{c}}$ and $\eta_{y_{c}y_{c}}$ into two terms as follows: 
\begin{align*}
    \eta_{x_{c}x_{c}} = &\frac{\mathrm{E}\big[\text{Var}(X_{c}|\mathrm{u}[-\infty,t], V[-\infty,t])\big]}{\langle x_{c} \rangle^{2}} \\ \quad &+ \frac{\text{Var}\left(\mathrm{E}\big[X_{c}|\mathrm{u}[-\infty,t], V[-\infty,t]\big]\right)}{\langle x_{c} \rangle^{2}} 
    \\=& \eta_{x_{c},int} + \eta_{\bar{x}_{c}\bar{x}_{c}} ,
\end{align*}
where we call the first term of the expansion $\eta_{x_{c},int}$. We thus have the following
\begin{align}
 &\eta_{x_{c}x_{c}} = \eta_{x_{c},int} + \eta_{\bar{x}_{c}\bar{x}_{c}}, \quad \eta_{y_{c}y_{c}} = \eta_{y_{c},int} + \eta_{\bar{y}_{c}\bar{y}_{c}}, \nonumber\\   &\eta_{x_{c}y_{c}} = \frac{1}{1+T_{c}}\eta_{\bar{x}_{c}\bar{x}_{c}} + \frac{T_{c}}{1+T_{c}}\eta_{\bar{y}_{c}\bar{y}_{c}}. \quad
\label{EQ: Concentrations-1step-3-equations}
\end{align}

The intrinsic variance is given by $\mathrm{E}\left[ \text{Var}(X_{c}|\mathbf{u}[-\infty,t], V[-\infty,t]) \right]$. 
That is, it's the ensemble average of the conditional variance. Consider the conditional variance
\begin{align*}
\text{Var}(X_{c}|\mathbf{u}[-\infty,t], &V[-\infty,t]) 
			      \\ &= \text{Var}\left(\frac{X}{V}\Big| \mathbf{u}[-\infty,t], V[-\infty,t]\right) 
			    \\&= \frac{1}{V(t)^{2}}\text{Var}\left(X| \mathbf{u}[-\infty,t], V[-\infty,t]\right)  ,
\end{align*}
 We can thus find the conditional variance of $X$, 
divide the result by $V(t)$, and then take the ensemble average to get the intrinsic variance. 
To find the conditional variance of $X$, we consider an ensemble of $N$ cells with the same histories 
$\mathbf{u}[-\infty,t]$ and $V[-\infty,t]$. The variance of one of the cells, 
say $X_{1}$, will give us $\text{Var}(X|\mathbf{u}[-\infty,t], V[-\infty,t])$. To find this variance, 
we start by considering the total number of molecules in the hypothetical ensemble
$X_{T}^{(N)} = X_{1} + X_{2} + \dots + X_{N}$. We now use the law of total variance conditioned on $X_{T}^{(N)}$
\begin{align}
 \text{Var}(X_{1}) = \mathrm{E}[ \text{Var}(X_{1}|X_{T}^{(N)}) ] + \text{Var}(\mathrm{E}[X_{1}|X_{T}^{(N)}]) .
\label{EQ: Consentrations-1step-xtotal-law-of-total-variance}
\end{align}

Note that the expectations and variances here are over the hypothetical ensemble of cells with the same history. 
The trick here is to notice that the death rate of each $X_{i}$ is $x_{i}/\tau_{x}$,
and so the death rate of $X_{T}$ is $x_{T}/\tau_{x}$. This is 
equivalent to saying that each molecule has a probability per unit time of degrading 
given by $1/\tau_{x}$, and this probability is independent of all other molecules 
comprising $X_{T}$ or any of the $X_{i}$. Similarly, each cell in this hypothetical ensemble has the same synchronized birthrate $R(t)$, so the 
probability of some molecule in $X_{T}$ to have been born in any of the cells is 
the same and independent of how many $X$ molecules are in each cell. Lastly, the process by which 
a molecule is ``degraded'' due to cell division is also independent of all other molecules 
in each cell, as we model cell division by a binomial spit where each molecule 
has an equal and independent probability of remaining in the cell. All this to say, each molecule has a probability $1/N$ of being in $X_{1}$, and 
this is independent of how many molecules of $X_{T}$ we know are in $X_{1}$. 
This is a binomial distribution $P(X_{1}|X_{T}^{(N)}) = B(X_{T}^{(N)}, 1/N)$, therefor $\mathrm{E}[X_{1}|X_{T}^{(N)}] = X_{T}^{(N)}/N $ and $\text{Var}(X_{1}|X_{T}^{(N)}) = \mathrm{E}[X_{1}|X_{T}^{(N)}](1 - 1/N)$. Eq.~\eqref{EQ: Consentrations-1step-xtotal-law-of-total-variance} becomes
\begin{align}
 \text{Var}(X_{1})  =& (1 - 1/N)  \langle X_{1} \rangle
 + \text{Var}\left(\frac{X_{T}^{(N)}}{N}\right) .
 \label{EQ: Concentrations-1step-xtotal-law-of-total-variance-2}
\end{align}
The $X_{i}$ are independent and identically distributed random variables. They thus obey the weak law of large numbers, so the second term in Eq.~\eqref{EQ: Concentrations-1step-xtotal-law-of-total-variance-2} goes to zero as $N \to \infty$. Thus $\text{Var}(X_{1}) = \langle X_{1} \rangle = \mathrm{E}[X|\mathbf{u}[-\infty], V[-\infty,t]]$, and so
\begin{align*}
 \text{Var}(X_{c}|\mathbf{u}[-\infty,t], &V[-\infty,t]) \\
 &= \frac{1}{V(t)^{2}}\mathrm{E}[X|\mathbf{u}[-\infty], V[-\infty,t]]  
 \\&= \mathrm{E}\left[\frac{X}{V^{2}}\Big|\mathbf{u}[-\infty], V[-\infty,t]\right] 
 \\&= \mathrm{E}\left[\frac{X_{c}}{V}\Big|\mathbf{u}[-\infty], V[-\infty,t]\right].
\end{align*}
We can now calculate the intrinsic variance of the real ensemble which we write in terms of the normalized variances 
\begin{align}
 \eta_{x_{c},int} = \left \langle \frac{X_{c}}{V} \right \rangle \frac{1}{\langle x_{c} \rangle^{2}} \quad \text{\&} \quad  \eta_{y_{c},int} = \left \langle \frac{Y_{c}}{V} \right \rangle \frac{1}{\langle y_{c} \rangle^{2}} ,
 \label{EQ: Concentrations intrinsic terms general}
\end{align}
which together with Eq.~\eqref{EQ: Concentrations-1step-3-equations} leads to Eq.~\eqref{EQ: Appendix concentrations (co)variance relations (manuscript)}.
When the reporter concentrations are independent of the cell volume, $X_{c}$ will be independent of $V$, and so $\langle X_{c}/V\rangle = \langle x_{c} \rangle \langle \frac{1}{V} \rangle$, and similarly for $y$, which gives us
\begin{align}
 \eta_{x_{c},int} = \left \langle \frac{1}{V} \right \rangle \frac{1}{\langle x_{c} \rangle} \quad \text{\&} \quad  \eta_{y_{c},int} = \left \langle \frac{1}{V} \right \rangle \frac{1}{\langle y_{c} \rangle} .
  \label{EQ: Concentrations intrinsic terms volume independent}
\end{align}
This equation, along with Eq.~\eqref{EQ: Concentrations-1step-differential-equation-concentrations-exponential} and 
Eq.~\eqref{EQ: Concentrations-1step-3-equations}, comprise the exact same equations as Eqs.~\eqref{EQ: 1-step flux-balance relations}, \eqref{EQ: no-feedback 1-step differential equations}, and \eqref{EQ: 1-step no-feedback fluctuation-balance}, from which the bounds for the class of systems in Eq.\eqref{EQ: Definition of 1-step Dual-Reporter System} of the main text were derived. 
The only difference is that now the (co)variances will be for the molecular concentrations, and the life-times $\tau_{x_{c}}$ and $\tau_{y_{c}}$ are for the concentrations as 
they include a contribution due to dilution from the growing volume. Note that if $R_{c}$ is stochastic (non-oscillatory), then it must be independent of the cell volume $V$, and so the bound on stochastic systems given by Eq.~\eqref{EQ: No oscillation constraint (manuscript)} holds in general for concentrations.

\subsection{Derivation of Eq.~\eqref{EQ: two-step no oscillation constraint (manuscript)} and Eq.~\eqref{EQ: two-step no-feedback constraint concentrations (manuscript)} for volume independent fluorescent protein concentrations}

Next, we consider the analogue of the class of systems in Fig.~\ref{FIG: Class of systems for FPs}. The fluorescent protein numbers are still described by the reactions in Fig.~\ref{FIG: Class of systems for FPs}, with the addition that the mRNA and proteins are also each ``degraded'' by cell division
\begin{align*}
(V, z) \xrightarrow{\text{At time } t_{i}} &(a_{i}V,  B(z,a_{i})), \\
&\text{for } t_{1}, t_{2}, \dots, \text{ with } z \in \{x', y', x'', y''\}.
\end{align*}
When the fluorescent protein components $X'$, $X''$, $Y'$, $Y''$ do not affect the environmental variables $\mathbf{u}(t)$ and $V$, we 
can analyze the average stochastic dual reporter dynamics conditioned on the history of their upstream influences 
\begin{align*}
\bar{x}'(t) &= \mathrm{E}\big[X'_t|\mathbf{u}[-\infty,t], V[-\infty,t] \big], \\ \bar{y}'(t) &= \mathrm{E}\big[ Y'_t|\mathbf{u}[-\infty,t] , V[-\infty,t]\big], \\
\bar{\bar{x}}'(t) &= \mathrm{E}\big[X'_t|\mathbf{u}[-\infty,t], V[-\infty,t], \mathbf{u}_{x}[-\infty,t] \big], \\ \bar{\bar{y}}'(t) &= \mathrm{E}\big[ Y'_t|\mathbf{u}[-\infty,t] , V[-\infty,t], \mathbf{u}_{y}[-\infty,t]\big] .
\end{align*}
We follow the same analysis as was done in the previous section to derive the following differential equations 
\begin{align}
\begin{split}
 \frac{d\bar{x}'_{c}}{dt} = F_{c}(t) - \bar{x}'_{c}/\tau_{x'_{c}} \quad & \quad \frac{d\bar{y}'_{c}}{dt} = F_{c}(t) - \bar{y}'_{c}/\tau_{y'_{c}} \\
 \frac{d\bar{x}''_{c}}{dt} = \bar{x}'_{c}/\tau_{mat,x} - \bar{x}''_{c}/\tau''_{c} \quad & \quad \frac{d\bar{y}''_{c}}{dt} = \bar{y}'_{c}/\tau_{mat,y} - \bar{y}''_{c}/\tau''_{c} 
\end{split} 
\label{EQ: Concentrations-3step-differential-equations-1}
\end{align}
\begin{align}
\begin{split} 
 \frac{d\bar{\bar{x}}'_{c}}{dt} = F_{x,c}(t) - \bar{\bar{x}}'_{c}/\tau_{x'_{c}} \quad & \quad \frac{d\bar{\bar{y}}'_{c}}{dt} = F_{y,c}(t) - \bar{\bar{y}}'_{c}/\tau_{y'_{c}} \\
 \frac{d\bar{\bar{x}}''_{c}}{dt} = \bar{\bar{x}}'_{c}/\tau_{mat,x} - \bar{x}''_{c}/\tau''_{c} \quad & \quad \frac{d\bar{\bar{y}}''_{c}}{dt} = \bar{\bar{y}}'_{c}/\tau_{mat,y} - \bar{y}''_{c}/\tau''_{c} 
\label{EQ: Concentrations-3step-differential-equations-2}
\end{split}
\end{align}
where $\tau_{x'_{c}} = (\frac{1}{\tau_{mat,x}} + \frac{\ln(2)}{\tau_{c}})^{-1}$, $\tau''_{c} = \left(\frac{1}{\tau''} + \frac{\ln(2)}{\tau_{c}}\right)^{-1}$, 
$F_{c}(t) = \frac{1}{V(t)}\mathrm{E}[F(\mathbf{u}(t), \mathbf{u}_{x})|\mathbf{u}[-\infty,t], V[-\infty,t]]$ and $F_{x,c} = \frac{1}{V(t)}F(\mathbf{u}(t), \mathbf{u}_{x}(t))$.  
We take the ensemble average of Eq.~\eqref{EQ: Concentrations-3step-differential-equations-1} over the different histories, which once the first moments reach stationarity, leads to the following flux balance relations 
\begin{align}
 \langle x''_{c} \rangle /\tau''_{c} = \frac{\tau_{x'_{c}}}{\tau_{mat,x}}\langle F_{c} \rangle \quad \text{\&} \quad \langle y''_{c} \rangle /\tau''_{c} = \frac{\tau_{y'_{c}}}{\tau_{mat,y}}\langle F_{c} \rangle .
 \label{EQ: FP concentration flux balance}
\end{align}
Eqs.~\eqref{EQ: Concentrations-3step-differential-equations-1} and \eqref{EQ: Concentrations-3step-differential-equations-2} are identical to the analogous differential equations derived in the molecular number system, with the exception that the degradation time of $\bar{x}'_{c}$ is given by 
$\tau_{x'_{c}}$ instead of $\tau_{mat,x}$ due to the added degradation that comes from dilution from the growing volume. Nevertheless, we can follow the same analysis to derive the following expression
\begin{align}
 \eta_{x''_{c}y''_{c}} = \eta_{\bar{x}''_{c}\bar{y}''_{c}} = \frac{1}{1 + T_{m}^{c}}\eta_{\bar{x}''_{c}\bar{x}''_{c}} + \frac{T_{m}^{c}}{1 + T_{m}^{c}}\eta_{\bar{y}''_{c}\bar{y}''_{c}} ,
\label{EQ: Concentrations-3step-fluctuation-balance}
\end{align}
where $T_{m}^{c} := \tau_{y'_{c}}/\tau_{x'_{c}}$. The last four equations are mathematically identical to the analogous equations derived for co-regulated fluorescent proteins without feedback, from which, along with Eq.~\eqref{EQ: 3-step no-feedback double bar extrinsic}, we derived the bounds in the fluorescent protein system. We thus need to show that an equation identical to Eq.~\eqref{EQ: 3-step no-feedback double bar extrinsic} holds for the reporter concentrations. From there the constraints follow from our previous proofs. To do this, we can use the law of total variance to decompose the variance of $X_{c}$ by conditioning on the history of $\mathbf{u}$, $V$, and $\mathbf{u}_{x}$, which gives us
\begin{align*}
 &\text{Var}(X''_{c}) \\
 &= \mathrm{E}\left[ \text{Var}(X''_{c}|\mathbf{u}_{x}[-\infty,t], \mathbf{u}[-\infty,t], V[-\infty,t]) \right ] 
 +  \text{Var}(\bar{\bar{x}}''_{c}) .
\end{align*}
We would now like to apply the same analysis as was done for the mRNA concentration system in order to derive an expression 
for the first term on the right. We thus look at an ensemble of $N$ cells with the same histories $\mathbf{u}[-\infty,t]$, $V[-\infty,t]$, and $\mathbf{u}_{x}[-\infty,t]$, 
which corresponds to each cell having the same synchronized translation rate $F(t)$. In such a case, we use the same trick as was done for the mRNA system in the previous section to show
\begin{align*}
\mathrm{E}\left[ \text{Var}(X''_{c}|\mathbf{u}_{x}[-\infty,t], \mathbf{u}[-\infty,t], V[-\infty,t]) \right ] = \left \langle \frac{X''_{c}}{V}\right \rangle .
\end{align*}
When the reporter concentrations are independent of the cell volume, we have $\left \langle \frac{X_{c}}{V}\right \rangle = \left \langle \frac{1}{V}\right \rangle \langle x''_{c} \rangle$, and so in 
terms of the normalized variances we have 
\begin{align}
\begin{split}
 &\eta_{x''_{c}x''_{c}} = \left \langle \frac{1}{V}\right \rangle \frac{1}{\langle x''_{c} \rangle} +  \eta_{\bar{\bar{x}}_{c}''\bar{\bar{x}}_{c}'' } \\ &\eta_{y''_{c}y''_{c}} = \left \langle \frac{1}{V}\right \rangle \frac{1}{\langle y''_{c} \rangle} +  \eta_{\bar{\bar{y}}_{c}''\bar{\bar{y}}_{c}'' } . 
 \end{split}
 \label{EQ: FP concentration eta}
\end{align}
This is the equation that we were seeking as it is mathematically analogous to Eq.~\eqref{EQ: 3-step no-feedback double bar extrinsic}. The factor on front of the average terms does not change the analysis as it is only the ratio of these terms that came into our derivations. This ratio is different than in the molecular numbers system due to the added degradation that comes from dilution from the growing volume. In particular
\begin{align}
 \frac{\eta_{x''_{c}x''_{c}} - \eta_{\bar{\bar{x}}_{c}''\bar{\bar{x}}_{c}''}}{\eta_{y''_{c}y''_{c}} - \eta_{\bar{\bar{y}}_{c}''\bar{\bar{y}}_{c}''}} = \frac{T_{m}^{c}}{T_{m}} .
\label{EQ: Concentrations-3step-intrinsic-relation}
\end{align}
We can follow the exact same steps as was done for the fluorescent protein number system to derive the analogous constraints in terms of the reporter concentrations. The only difference would be in the right-most open loop bound which shifts as a result of this added "dilution degradation" of the immature protein concentrations. 

\subsection{Volume dependent genes --- exact constraints using a third reporter}
\label{SEC: Appendix exact constraints on volume dependent genes}
We were unable to formally prove the open-loop constraint for volume dependent genes. Mathematically, the problem lies in the fact that when the reporter concentrations are volume dependent, the intrinsic terms $\eta_{x_{c},int}$ and $\eta_{y_{c},int}$ are given by Eq.~\eqref{EQ: Concentrations intrinsic terms general} and not by Eq.~\eqref{EQ: Concentrations intrinsic terms volume independent}. These introduce two new unknowns to the system of equations, namely $\left \langle \frac{X_{c}}{V} \right \rangle \frac{1}{\langle x_{c} \rangle^{2}}$,  and $\left \langle \frac{Y_{c}}{V} \right \rangle \frac{1}{\langle y_{c} \rangle^{2}}$, which makes the system of equations under-determined.  When the concentrations are volume independent, Eq.~\eqref{EQ: Concentrations-1step-flux-balance} and Eq.~\eqref{EQ: Concentrations intrinsic terms volume independent} imply that the ratio of intrinsic noise terms are given by  $\eta_{x_{c},int}/\eta_{y_{c},int} = T_{c}$, which allows us to close the system of equations and solve for $\eta_{\bar{x}_{c}\bar{x}_{c}}$ and $\eta_{\bar{y}_{c}\bar{y}_{c}}$ in terms of the measurable (co)variances which are bounded by the sought after constraints.  Note that $\sqrt{\eta_{x_{c},int}/\eta_{y_{c},int}}$ sets the point along the $\rho_{x_{c}y_{c}}$ line in Fig.~\ref{FIG: Concentrations (manuscript)}B (left panel) where the orange lines converge. For volume dependent concentrations, numerical simulations show that $\sqrt{\eta_{x_{c},int}/\eta_{y_{c},int}} \neq \sqrt{T_{c}}$, but it holds to good approximation with a divergence of up to around 6\% for the specific models that we simulated. 

Though we cannot prove the open-loop constraint using two reporters, we can infer the missing degree of freedom by introducing an additional reporter into the system and derive an open-loop constraint on volume dependent genes. In particular, Eq.~\eqref{EQ: 1-step no-feedback conditioned averages} and Eq.~\eqref{EQ: Appendix eta time-scale result} still hold in terms of concentrations of growing and dividing cells (similarly for the fluorescent reporters), even when the reporter concentrations are cell volume dependent. It's only because the $\eta_{x_{c},int}$ and $\eta_{y_{c},int}$ terms become more complicated for volume dependent systems that we cannot close the system of equations and solve for $\eta_{\bar{x}_{c}\bar{x}_{c}}$ and $\eta_{\bar{y}_{c}\bar{y}_{c}}$ in terms of the measurable (co)variances. Thus, if we can measure $\eta_{\bar{x}_{c}\bar{x}_{c}}$ and $\eta_{\bar{y}_{c}\bar{y}_{c}}$ then we can take full advantage of the constraints. With a third reporter we are able to solve for $\eta_{\bar{x}_{c}\bar{x}_{c}}$ and $\eta_{\bar{y}_{c}\bar{y}_{c}}$ in terms of the measurable (co)variances as explained in Sec.~\ref{SEC: Stronger constraints}. In that section we showed how to measure $\eta_{\bar{x}''\bar{x}''}$ and $\eta_{\bar{y}''\bar{y''}}$ using three separate dual reporter experiments involving three dual reporter pairs, but the exact same reasoning can be used to measure $\eta_{\bar{x}_{c}\bar{x}_{c}}$ and $\eta_{\bar{y}_{c}\bar{y}_{c}}$ (or the fluorescent protein equivalents), as the same covariance equations used in that section hold exactly for the reporter concentrations. Thus even if we are unable to formally prove the open-loop constraint for volume dependent systems with two reporters, with a third reporter we are able to measure the missing degree of freedom to prove the constraint.

\section{Experimental data analysis}
\label{SEC: Appendix Experimental data analysis}

\begin{figure*}
\centering
\includegraphics[width=0.9\columnwidth]{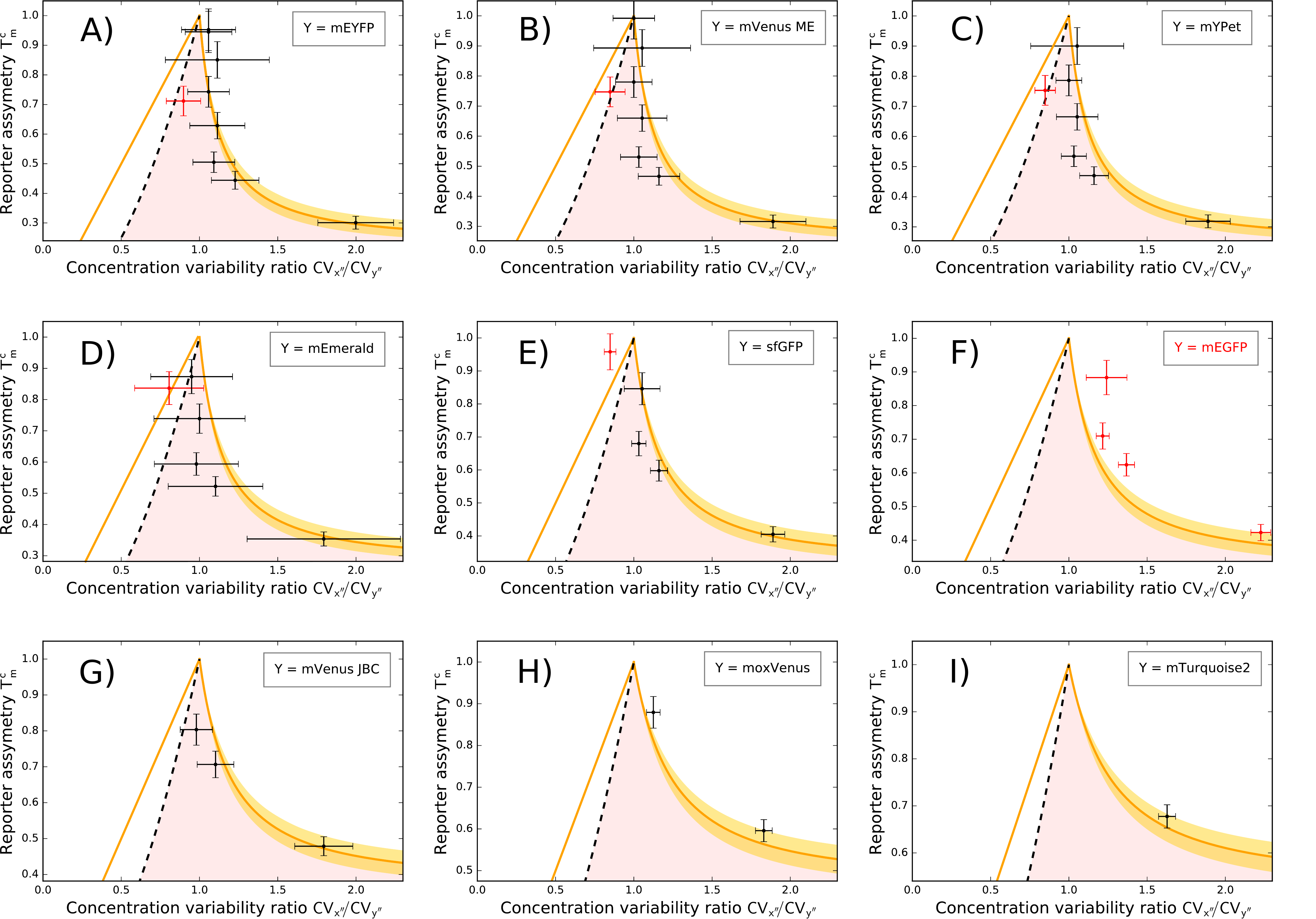}
\caption{ {\textbf{Utilizing constraints on flow-cytometry data from constitutively expressed fluorescent proteins.} Here we plot all of the possible pairs from Table 1 in the Supplemental Material \cite{SI}. We plot 9 separate plots instead of combining the data into a single plot because for a given $T_{m}^{c}$ the right bound given by Eq.~\eqref{EQ: two-step sequential no-feedback constraint (manuscript)} is only determined if we specify one of the maturation times. Thus for fixed Y reporter, the right bound is well defined for any given X reporter, with the yellow corridor indicating the estimated uncertainty in $\tau_c$ and $\tau_{mat,y}$.   Because all fluorescent proteins were constitutively expressed and were not fused to cellular proteins we expect the experimental data to be consistent with the class of gene expression models that do not exhibit feedback and are not periodically driven (pink region). All variability ratios with respect to all reference fluorescence proteins confirm the above picture with the exception for mEGFP (with ratios indicated in red) for which the data violate the expected behaviour. With the exception of the indicated mEGFP outlier all data fall along the right hand boundary.}}
\label{FIG: All data}
\end{figure*} 

Balleza et al.~quantified the maturation dynamics of 50 different fluorescent proteins (FPs) in  \emph{E.~coli} using time-lapse microscopy \cite{Balleza2018}. Their data show that the maturation step of roughly a third of the tested FPs are well described by first-order kinetics as assumed in our class of gene expression models. Additionally, the authors quantified heterogeneity in fluorescence levels when the respective FPs were expressed under the constitutive promoter \emph{proC}. These data were obtained from flow-cytometry measurements of clonal \emph{E.~coli} MG1655 (CGSC 6300) populations in M9-rich media at 37$^{\circ}$C.  From their publicly available raw cell-to-cell heterogeneity data \cite{online-data}, we selected the ten FPs that were best modelled by first-order maturation kinetics (see Supplemental material \cite{SI}).

FP abundance CVs were obtained from the reported flow-cytometry fluorescence distributions. Under the assumption that reporter concentrations are cell-volume independent, i.e., transcription rates are cell-cycle independent, then the concentration CVs are related to the CVs in abundances and the CV in cell volume $CV_{V}$ \cite{galbusera2020using} 
\begin{equation}
    CV_{\text{concentration}}^{2} = \frac{CV_{\text{abundance}}^{2}}{1 + CV_{V}^{2}} - \frac{CV_{V}^{2}}{1 + CV_{V}^{2}}.
    \label{EQ: Appendix concentration CVs from abundance CVs (manuscript)}
\end{equation}
We estimated $CV_{V}$ using separate time-lapse microscopy data that quantifies \emph{E.~coli} growth dynamics \cite{wang2010robust}.
From the publicly available time-traces of \emph{E.~coli} length we computed 
an average volume variability of \emph{E.~coli} MG1655 (CGSC 6300) grown in LB media of \mbox{$CV_{V} = 0.261\pm0.005$}. The resulting concentration variability data are presented in Fig.~\ref{FIG: Concentrations (manuscript)}C and a complete list of concentration CVs obtained this way along with the chosen FPs and their maturation times are presented in Table 1 in the Supplemental Material \cite{SI}.

{
As shown in Fig.~\ref{FIG: All data}, all fluorescent protein pairs fall within the region expected for constitutively expressed genes, with the exception of mEGFP. The data fall along the right orange bound given by the right inequality in Eq.~\eqref{EQ: two-step sequential no-feedback constraint (manuscript)}. According to Eq.~\eqref{EQ: FP concentration flux balance}, systems that lie along this boundary satisfy $CV_{x''_{c}}/CV_{y''_{c}} = \sqrt{\langle y''_{c} \rangle / \langle x''_{c} \rangle}$
where $x_{c}''$ and $y_{c}''$ denote the concentrations of the fluorescent proteins. Systems that lie along this bound thus have CVs that scale inversely with the square root of the averages. According to Eq.~\eqref{EQ: FP concentration eta}, the concentration CVs are given by
\begin{align}
\begin{split}
    &CV_{x_{c}''}^{2} = \left \langle \frac{1}{V}\right \rangle \frac{1}{\langle x''_{c}\rangle}\\
    &+   \int_{0}^{\infty}\eta_{FF}A_{F_{x_{c}}}(t) \left( \frac{\tau_{mat,x}e^{-t/\tau_{mat,x}} - \tau_{c}''e^{-t/\tau_{c}''} }{\tau_{mat,x}^{2} - \tau_{c}''^2} \right)dt,
\end{split}
    \label{EQ: CV concentrations general}
\end{align}
where $F_{x_{c}}$ is the production rate of the immature protein concentrations. The first term on the right, which scales inversely to the average, corresponds to noise that originates at the translation step, the maturation step, fluctuations in protein degradation, and binomial splitting of proteins at cell division. The term on the right corresponds to noise that originates from transcription as well as mRNA and translation rate fluctuations. Therefor, data that lie along the right orange boundary in Fig.~\ref{FIG: All data} corresponds to variability with negligible transcription noise contributions. 
}

{
In order to confirm this observation and also analyze the mEGFP discrepancy, we plot in Fig.~\ref{FIG: All data} the fluorescent protein concentration CVs as a function of their maturation times. In the regime where the the second term on the right of Eq.~\eqref{EQ: CV concentrations general} is negligible, and where we assume that the fluorescent protein concentrations are degraded solely by dilution from the growing and dividing cells, we can use Eq.~\eqref{EQ: FP concentration flux balance} to write
\begin{align}
    CV_{x_{c}''} = A\sqrt{\frac{\ln(2)}{\tau_{c}}\left( 1 + \text{ln}(2)\frac{\tau_{mat,x}}{\tau_{c}}\right)},
    \label{EQ: Negligible transcription noise model}
\end{align}
where $A = \sqrt{\left \langle \frac{1}{V}\right \rangle \frac{1}{\langle F\rangle}}$ is an unknown parameter. From this equation we see that as the maturation time gets larger, so does the CV. This is because as the maturation time gets larger, the fraction of matured proteins is reduced, which results in larger intrinsic fluctuations that originate from the maturation step and from cell partitioning of matured proteins. On the other hand, we can look at the regime in which there is negligible translation noise (where the second term on the right of Eq.~\eqref{EQ: CV concentrations general} dominates). We model the autocorrelation of the protein concentration translation rate as a decaying exponential $A_{F_{x_{c}}}(t) = e^{-t/\tau_{F}}$, and we assume that the protein concentrations are degraded solely from dilution, in which case we have
\begin{align}
    CV_{x_{c}''} =  B \sqrt{\frac{\tau_{F}(\tau_{c}''\tau_{x_{c}'}+\tau_{F}(\tau_{c}'' + \tau_{x_{c}'}))}{(\tau_{F} + \tau_{c}'')(\tau_{F} + \tau_{x_{c}'})(\tau_{c}'' + \tau_{x_{c}'})}},
    \label{EQ: Negligible translation noise model} 
\end{align}
where $\tau_{c}'' = \tau_{c}/\ln(2)$, $\tau_{x_{c}'} = \left(1/\tau_{mat,x} + \ln(2)/\tau_{c}\right)^{-1}$, and $B = \sqrt{\eta_{FF}}$ is an unknown parameter. In Fig.~\ref{FIG: CV plot} we plot Eq.~\eqref{EQ: Negligible transcription noise model} and Eq.~\eqref{EQ: Negligible translation noise model} with $A = 1.3$, $B = 1.2$, $\tau_{F} = \tau_{c}/10$, and $\tau_{c} = 28.5 \pm 2$ min (as reported in \cite{Balleza2018}). We find that the computed CVs from the flow cytometry data sets are well described by Eq.~\eqref{EQ: Negligible transcription noise model}, except for the mEGFP point which displays a CV smaller than the others. The only parameter left in this model to vary is the division time of the cells $\tau_{c}$. When this parameter is changed to $35.6$ min, the model captures the mEGFP data point as shown by the dashed curve in Fig.~\ref{FIG: CV plot}. The discrepancy between mEGFP and the other 9 fluorescent proteins could thus be explained by a slightly slower growth rate in the cell cultures used for the mEGFP data sets. The blue curve, on the other hand, corresponds to the negligible translation noise model given by Eq.~\eqref{EQ: Negligible translation noise model}. We find that as the maturation time gets larger, the CV gets smaller, which is expected as fluorescent proteins with large maturation times have less time to adjust to varying upstream fluctuations and inherit less variability. }

{
The previous analysis indicates that the variability is dominated by noise originating from the translation step, maturation step, fluctuations in protein degradation, and noise originating from binomial splitting of protein numbers at cell divisions, which are all described by the first term on the right in Eq.~\eqref{EQ: CV concentrations general}. However, it is possible that measurement noise which scales inversely with the average is contributing to the measured variability. For example, in \cite{galbusera2020using} it is shown that flow cytometry measurements contain a significant amount of measurement noise when used with bacteria due to their small size. This measurement noise is shown to scale inversely with the average of the fluorescence signal, meaning it could mask itself as biological variability contributing to the first term on the right of Eq.~\eqref{EQ: CV concentrations general}. 
In \cite{galbusera2020using} a rigorous method is developed to measure the true CVs using flow cytometry on bacteria, separating measurement noise and autofluorescence contributions using commonly used calibration beads.}

\begin{figure}[htp]
\includegraphics[width=0.99\columnwidth]{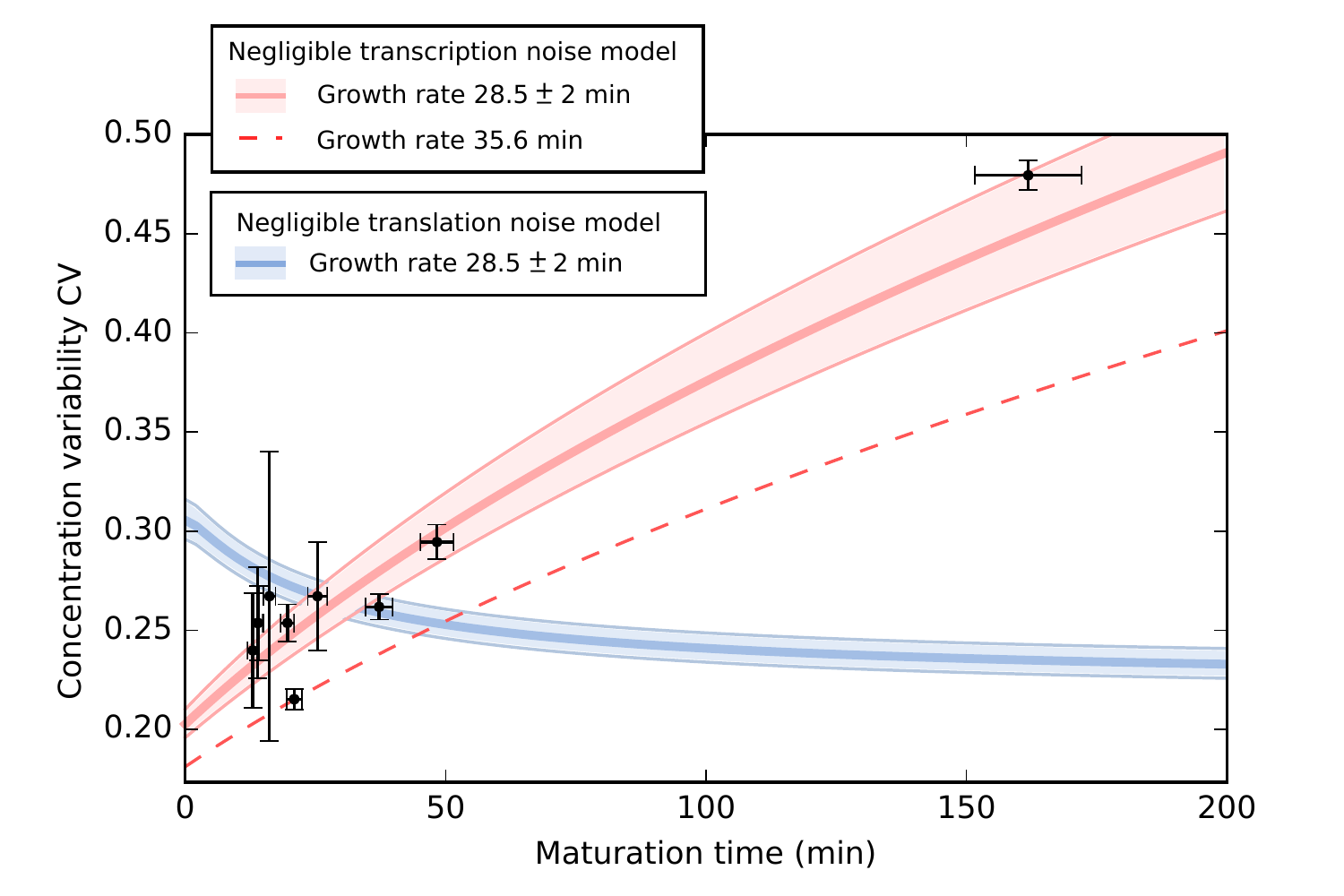}
\caption{ {\textbf{The data is best described by a model with negligible transcription noise.} Plotted are the concentration CVs from Table 1 of the Supplemental Material \cite{SI} as a function of their respective fluorescent protein maturation times. The pink curve corresponds to the model given by Eq.~\eqref{EQ: Negligible transcription noise model} with $A = 1.3$ and $\tau_{c}$ as reported in \cite{Balleza2018}: $28.5\pm2$ min. The pink corridor indicates the result of the uncertainty in $\tau_{c}$. The blue curve corresponds to the model given by Eq.~\eqref{EQ: Negligible translation noise model} in which translation noise is negligible, with $B = 1.2$, $\tau_{F} = 28.5/10$ min, and $\tau_{c} = 28.5\pm2$ min. We find that the data follows the pink curve and is thus best described by the negligible transcription noise model. The data point with the smallest CV corresponds to mEGFP, and the dashed red curve corresponds to the same model as the pink curve but with $\tau_{c}$ increased by a factor of 1.25. A possible explanation for this relatively low CV observed for mEGFP is that the cell cultures expressing mEGFP had slightly slower growth rates.}}
\label{FIG: CV plot}
\end{figure}

\end{document}


\preprint{APS/123-QED}

\title{\large{Inferring gene regulation dynamics from static snapshots of gene expression variability} \\ $\quad$ \\
\large{\emph{Supplementary Material}}}%

\author{Euan Joly-Smith}
\affiliation{Department of Physics, University of Toronto, 60 St.~George St., Ontario M5S 1A7, Canada}
\author{Zitong Jerry Wang}
\affiliation{California Institute of Technology, Division of Biology and Biological Engineering, Pasadena CA 91125, USA}
\author{Andreas Hilfinger}
\affiliation{Department of Physics, University of Toronto, 60 St.~George St., Ontario M5S 1A7, Canada}
\affiliation{Department of Mathematics, University of Toronto, 40 St.~George St., Toronto, Ontario M5S 2E4}
\affiliation{Department of Cell \& Systems Biology, University of Toronto, 25 Harbord St, Toronto, Ontario M5S 3G5}

\maketitle
\numberwithin{equation}{section}

\vspace{-0.8cm}
\tableofcontents{}

\makeatletter\@input{xx.tex}\makeatother

\setcounter{page}{1}

\section{Inferring the timescale of upstream variability --- numerical bound}

Using Eq.~\eqref{EQ: no-oscillation 1-step eta bar autocorrelation integral}, we have 
\begin{align*}
 \eta_{\bar{x}\bar{x}} - a \eta_{\bar{y}\bar{y}} = 
 \int_{0}^{\infty} \eta_{RR}A_{R}(t) \left( \frac{e^{-t/\tau_{x}}}{\tau_{x}} - a \frac{e^{-t/\tau_{y}}}{\tau_{y}}\right)  dt \quad \text{where} \quad T \leq a \leq 1 \quad .
\end{align*}
 
\begin{figure*}[htb!]
\vspace{-0.4cm}
\centering
  \includegraphics[width=0.45\columnwidth]{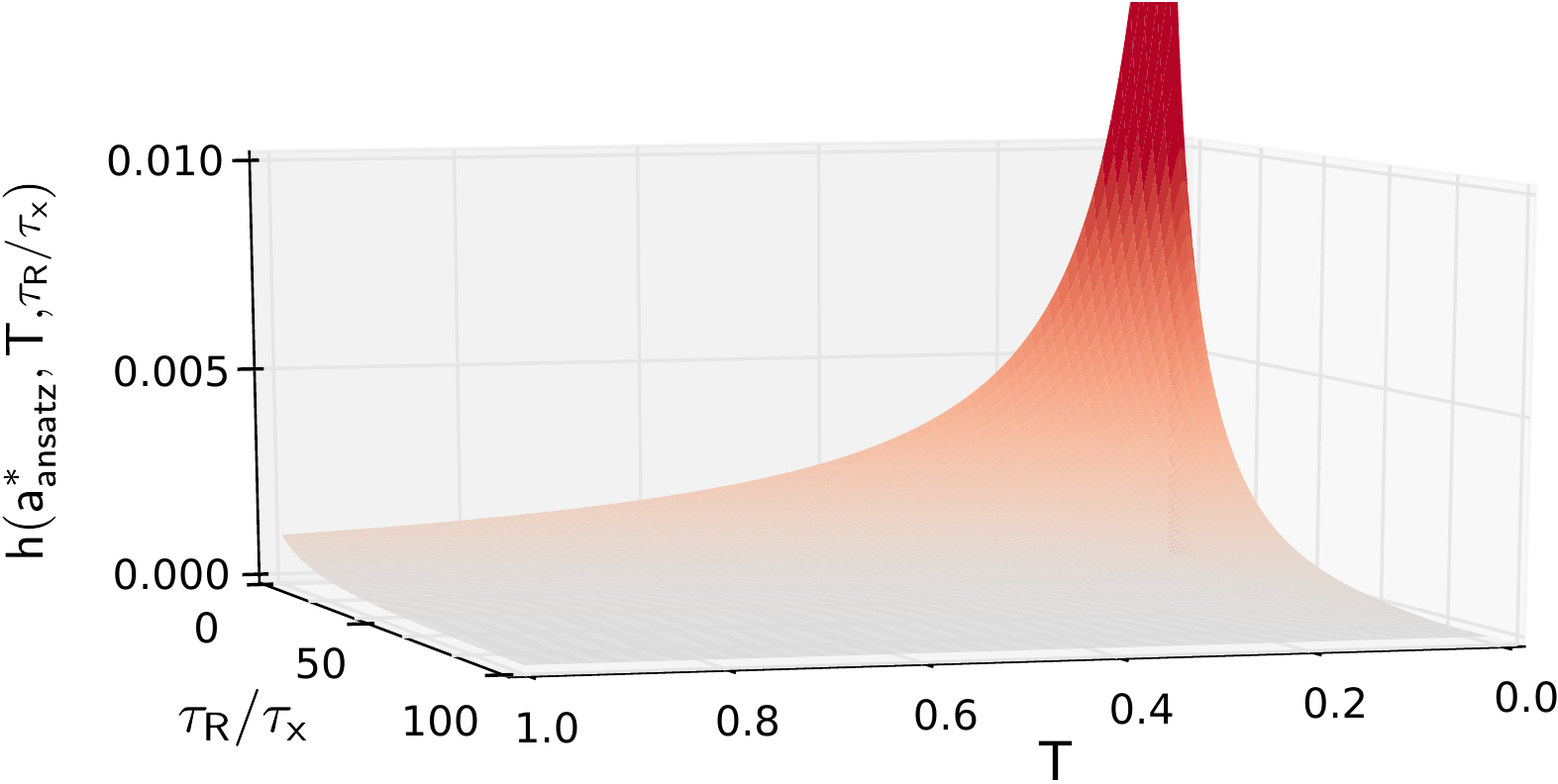}
   \caption{Plot of $h(a^{*}_{ansatz},T,\tau_{R}/\tau_{x})$ over different $T$ and $\tau_{R}/\tau_{x}$ values, where $a^{*}_{ansatz} = (0.9+ T\frac{\tau_{x}}{\tau_{R}})/(0.9 + \frac{\tau_{x}}{\tau_{R}})$. Note that $h$ here is always positive and for most values of the domain it is approximately zero (the peak is $h_{max} \approx 0.02$).}
    \label{FIG: Timescales ansatz}
\end{figure*}

The expression in parentheses is negative for $t \leq t^{*}$ and positive for $t \geq t^{*}$, where $t^{*} = \ln(a/T)\tau_{x}\tau_{y}/(\tau_{x} - \tau_{y})$. We thus have 
\begin{align}
 \eta_{\bar{x}\bar{x}} - a \eta_{\bar{y}\bar{y}} &= 
 \int_{0}^{t^{*}} \eta_{RR}A_{R}(t) \left( \frac{e^{-t/\tau_{x}}}{\tau_{x}} - a \frac{e^{-t/\tau_{y}}}{\tau_{y}}\right)  dt  + \int_{t^{*}}^{\infty} \eta_{RR}A_{R}(t) \left( \frac{e^{-t/\tau_{x}}}{\tau_{x}} - a \frac{e^{-t/\tau_{y}}}{\tau_{y}}\right)  dt \nonumber\\
 &\geq \int_{0}^{t^{*}} \eta_{RR} \left( \frac{e^{-t/\tau_{x}}}{\tau_{x}} - a \frac{e^{-t/\tau_{y}}}{\tau_{y}}\right)  dt  +  \int_{t^{*}}^{\infty} \eta_{RR}A_{R}(t) \left( \frac{e^{-t/\tau_{x}}}{\tau_{x}} - a \frac{e^{-t/\tau_{y}}}{\tau_{y}}\right)  dt \nonumber\\
 &\geq \int_{0}^{t^{*}} \eta_{RR} \left( \frac{e^{-t/\tau_{x}}}{\tau_{x}} - a \frac{e^{-t/\tau_{y}}}{\tau_{y}}\right)  dt  +  \int_{t^{*}}^{\infty} \eta_{RR}e^{-t/\tau_{R}} \left( \frac{e^{-t/\tau_{x}}}{\tau_{x}} - a \frac{e^{-t/\tau_{y}}}{\tau_{y}}\right)  dt \quad , \label{EQ: integral}
\end{align}
where in the second step we use the fact that the expression in parantheses is negative and $A_{R}(t) \leq 1$, and the third step comes from the fact that $A_{R}(t) \geq e^{-t/\tau_{R}}$. 
Without loss of generality we work in units where $\tau_{x} = 1$ and $\tau_{y} = T$, in which case we solve the above integral to obtain 
\begin{align}
  &\eta_{\bar{x}\bar{x}} - a \eta_{\bar{y}\bar{y}} \geq \eta_{RR} h(a,T,\tau_{R}) \\
  \text{where} \quad h(a,T,\tau_{R}) &= \left(1 - a  +   a\left(\frac{a}{T}\right)^{-\frac{1}{1 - T}} - \left(\frac{a}{T}\right)^{-\frac{T}{1 - T}} +  \frac{\tau_{R}\left(\frac{a}{T}\right)^{-\frac{(1+\tau_{R})T}{\tau_{R}(1-T)}}}{1+\tau_{R}} - \frac{a\tau_{R}\left(\frac{a}{T}\right)^{-\frac{(r+T)}{\tau_{R}(1-T)}}}{T+\tau_{R}} \right) \quad .\nonumber
\label{EQ: no-oscillation 1-step timescale a equation}
\end{align}
Now, we can infer that $h(a, T, \tau_{R})$ is zero for some $a^{*} \in [T,1]$. In particular, for $a = T$ we have
\begin{align*}
    h(T,T,\tau_{R}) = \frac{\tau_{R}}{1+\tau_{R}} - \frac{T\tau_{R}}{T + \tau_{R}} = \frac{\tau_{R}^{2}(1  - T)}{(1+\tau_{R})(T + \tau_{R})} > 0 \quad .
\end{align*}
Moreover, we note that the $\tau_{R}$ dependence of $h$ comes from the integral on the right in Eq.~\eqref{EQ: integral}, and since the expression in parenthesis in this integral is always positive, the integral will increase as $\tau_{R}$ increases. As a result, $h(a,T,\tau_{R})$ is a monotonically increasing function of $\tau_{R}$ and so $h(1,T,\tau_{R}) < h(1,T, \tau_{R} \to \infty) = 0$. Thus, we've shown that $h(a,T,\tau_{R})$ is positive for $a = T$ and negative for $a = 1$, and so by the mean value theorem there must exist an $a^{*}\in [T,1]$ s.t. $h(a^{*},T,\tau_{R}) = 0$. This $a^{*}$ can be found numerically for a given $\tau_{R}$ and $T$, and plugging this $a^{*}$ into Eq.~\eqref{EQ: no-oscillation 1-step timescale a equation} gives us the strictness bound on systems that satisfy $e^{-t/\tau_{R}} \leq A_{R}(t)$
\begin{align}
    \eta_{\bar{x}\bar{x}} - a^{*}\eta_{\bar{y}\bar{y}} \geq 0 \quad \text{where} \quad h(a^{*},T,\tau_{R}/\tau_{x}) = 0\quad ,
\end{align}
where the $\tau_{x}$ factor is added for systems with arbitrary units. We were unable to solve for an analytical solution to $a^{*}$ given $T$ and $\tau_{R}$. However, through trial and error, and by looking at the functional form of the solution of systems with exponential auto-correlations, we proposed the ansatz that $h(a^{*}_{ansatz},T,\tau_{R}/\tau_{x}) \geq 0$, where $a^{*}_{ansatz} =  (0.9 + T\tau_{x}/\tau_{R})/(0.9 + \tau_{x}/\tau_{R})$. By numerically plotting $h(a^{*}_{ansatz},T,\tau_{R}/\tau_{x})$ over different $T \in [0,1]$ and $\tau_{R} \in [0,100]$ (see Fig.~\ref{FIG: Timescales ansatz}), we find that $h(a^{*}_{ansatz},T,\tau_{R}/\tau_{x}) \geq 0$ in this regime of parameters, and so through Eq.~\eqref{EQ: no-oscillation 1-step timescale a equation} we have the slightly weaker bound
\begin{align}
    \eta_{\bar{x}\bar{x}} - a^{*}_{ansatz}\eta_{\bar{y}\bar{y}} \geq 0 \quad \text{where} \quad a^{*}_{ansatz} = \left( \frac{0.9 + \frac{T\tau_{x}}{\tau_{R}}}{0.9 + \frac{\tau_{x}}{\tau_{R}}}\right) \quad .
\end{align}

We recall from Eq.~\eqref{EQ: no-feedback 1-step barbar equals xR} that $\eta_{\bar{x}\bar{x}} = \eta_{xR}$ and $\eta_{\bar{y}\bar{y}} = \eta_{yR}$, and we can solve 
for $\eta_{xR}$ and $\eta_{yR}$ in terms of measurable (co)variances through Eq.~\eqref{EQ: 1-step eta bar in terms of covariances}. Substituting these into the above inequality 
gives us
\begin{align}
 (a - T)\rho_{xy} \leq \left( \frac{a + T}{1+T} \right) \left( \frac{CV_{x}}{CV_{y}} - T\frac{CV_{y}}{CV_{x}}\right) \quad \text{where} \quad  a = \left(\frac{0.9 + \frac{T\tau_{x}}{\tau_{R}}}{0.9 + \frac{\tau_{x}}{\tau_{R}}}\right) \quad .
 \label{EQ: stricter bound}
\end{align}
Systems that break the above inequality cannot satisfy $e^{-t/\tau_{R}} \leq A_{R}(t)$, and so the upstream variability must 
fluctuate with timescale smaller than $\tau_{R}$, see Fig.~\ref{FIG: Timescales}.

For fluorescent protein reporters of the class of systems in Fig.~\ref{FIG: 3 step class of systems with maturation (manuscript)}A of the paper, we can follow the same analysis as above. In particular, 
we consider systems in which the following holds
\begin{align*}
 e^{-t/\tau_{R}} \leq A_{\bar{F}}(t) \quad \text{where} \quad \bar{F}(t) = \mathrm{E}\big[ F(\mathbf{u}(t), \mathbf{u}_{x}) | \mathbf{u}[-\infty,t] \big] \quad .
\end{align*}
That is, $\bar{F}(t)$ is the translation rate when we average out all the intrinsic fluctuations originating from the mRNA intrinsic system, and so 
$A_{\bar{F}}(t)$ characterizes the timescale of the upstream fluctuations originating from the cloud of components $\mathbf{u}(t)$. 
Similarly to the above, we have 
\begin{align*}
  \eta_{\bar{x}''\bar{x}''} - a \eta_{\bar{y}''\bar{y}''} 
  \geq \int_{g(t) < 0} \eta_{RR} g(t)  dt 
  +  \int_{g(t) \geq 0} \eta_{RR} e^{-t/\tau_{u}}g(t) dt \quad \\
  \text{where} \quad g(t) = \left( \frac{\tau_{mat,x}e^{-t/\tau_{mat,x}}-\tau''e^{-t/\tau''}}{\tau_{mat,x}^{2} - \tau''^{2}} - a \frac{\tau_{mat,y}e^{-t/\tau_{mat,y}}-\tau''e^{-t/\tau''}}{\tau_{mat,y}^{2} - \tau''^{2}}\right) \quad ,
\end{align*}
In order to find the strongest bound on these systems, the largest $T_{m} \leq a^* \leq 1$ value that would set the right hand side of the above inequality to zero would need to be found. This can be done 
numerically, where this $a^*$ value would depend on $\tau_{R}$, $T$, and $\tau''$. If $\tau'$ is not known, then the largest $a$ that would set the right hand side of the above inequality to be positive for all $\tau'$ can be used. In both of these cases we would end up with 
\begin{align*}
 a \eta_{\bar{y}''\bar{y}''} \leq \eta_{\bar{x}''\bar{x}''}  \quad \text{where} \quad a = a^{*} \text{ (see above paragraph)}\quad . 
\end{align*}
When there is no feedback, recall that Eq.~\eqref{EQ: no-feedback 3-step fluctuation-balance in terms of bared variables} and Eq.~\eqref{EQ: no-feedback 3-step intrinsic noise bound} hold. 
Using these equations with the above inequality we find 
\begin{align}
 (a - T_{m})\rho_{x''y''} \leq \left( \frac{a + T_{m}}{1+T_{m}} \right) \left( \frac{CV_{x''}}{CV_{y''}} - T_{m}\frac{CV_{y''}}{CV_{x''}}\right) \quad  .
\end{align}
Systems that break the above inequality must break the inequality given by $e^{-t/\tau_{R}} \leq A_{F}(t)$, and so the upstream variability must 
fluctuate with timescales smaller than $\tau_{R}$.

\begin{figure*}[h]
\centering
  \includegraphics[width=0.89\columnwidth]{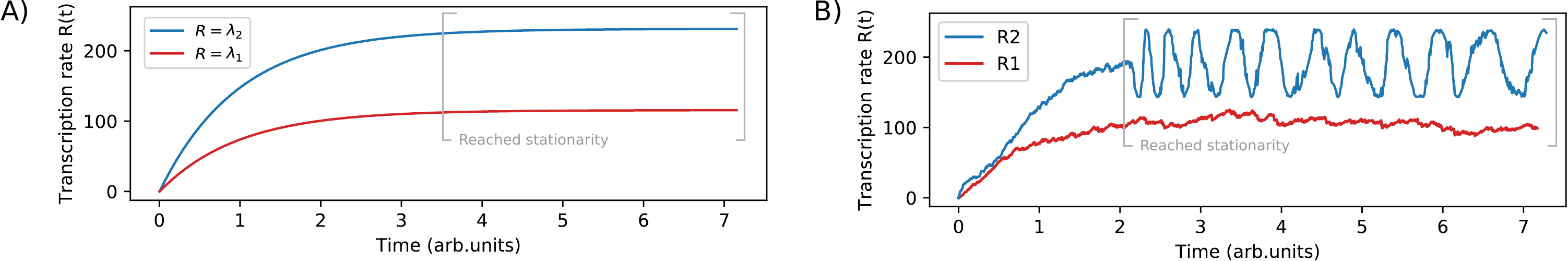}
   \vspace{-.5em}
   \caption{
    {\textbf{Irreversible cell states are non-ergodic processes.} A) Here are possible realizations of the transcription rate $R(\mathbf{u}(t))$ in a particular system within the class of Fig.~\ref{FIG: Class 1 space of solutions -- feedback (manuscript)}A in which the rate R either remains at the constant state of $R=\lambda_{1}$ or $R=\lambda_{2}$ after the initial transient dynamics have decayed. In each realization, the rate has a probability of $1/2$ of growing into the $R = \lambda_{1}$ state or into the $R = \lambda_{2}$ state. This could, for example, model a population of cells in which each cell must irreversibly switch to one of two states. This system is not ergodic, as following an individual cell over time can only provide information on the sub-population that shares that cell's state, and not the whole population. B) A similar system as A, with one of the states having an oscillating transcription rate and the other a stochastic transcription rate. }}
    \label{FIG: Ergodicity}
\end{figure*}

{\section{Examples of non-ergodic systems}}
{
Non-ergodic systems are systems where time averages do not correspond to population averages. Though the cloud of components $\mathbf{u}(t)$ in Fig.~\ref{FIG: Class 1 space of solutions -- feedback (manuscript)}A is dynamic and can vary as a function of time, it is left unspecified and we do not need to assume ergodicity in $R(\mathbf{u}(t))$. As an example, consider the case where $\mathbf{u}(t)$ is such that the rate $R(\mathbf{u}(t))$ evolves from some initial value and eventually reaches one of two values, $\lambda_{1}$ or $\lambda_{2}$, from which it stays constant, see Fig.~\ref{FIG: Ergodicity}A. Here, let's say that the choice of states $\lambda_{1}$ or $\lambda_{2}$ is random with a probability of 1/2. If we follow any single cell over time from some fixed initial state, the transcription rate will evolve to either $R = \lambda_{1}$ or $R = \lambda_{2}$ with equal probability, and then stay constant. If we wait for stationarity, in which each system has left the transient state, then we have two sub-populations. Such a system, though not ergodic, will satisfy the constraints presented in the main text. In particular, this exact system will lie along the right boundary of the open-loop constraint given by Eq.~\eqref{EQ: No feedback constraint (manuscript)} of the main text. In previous sections we've conditioned on the history of the upstream variables $\mathbf{u}(t)$ for systems without feedback. For the system in Fig.~\ref{FIG: Ergodicity}A this corresponds to conditioning on $R = \lambda_{1}$ or $R = \lambda_{2}$. When we take the average over all histories we effectively average over the two $R$ states. 
}

{
We can consider a more general non-ergodic system in which the cells move into one of two states with equal probability, defined by either having a transcription rate $R_{1}(t)$ or $R_{2}(t)$. For example, $R_{1}$ could be a stochastic signal and $R_{2}$ could be an oscillation, see Fig.~\ref{FIG: Ergodicity}B. Here when we condition on the history we are also conditioning on the state of the cell by specifying whether the history is an oscillation or not. When we take the average over all histories we in effect take the average of all histories in each state and then average over each state. }

{
It is worth noting that for such non-ergodic systems with different transcription states, if there is any feedback between the downstream variables $X$ and $Y$ and the cloud of components $\mathbf{u}(t)$ in any of the different cell states, then that constitutes feedback in our framework and could break the open-loop constraint given by Eq.~\eqref{EQ: No feedback constraint (manuscript)}.}

{
Moreover, the bound on stochastic systems given by Eq.~\eqref{EQ: No oscillation constraint (manuscript)} bounds systems with non-negative autocorrelation. The autocorrelation here is the autocorrelation obtained by ensemble averages, defined by 
\begin{equation*}
A(s) : = \frac{\lb R(t+s) R(t) \rb - \lb R(t+s)\rb \lb R(t) \rb}{\Var[R(t)]} ,
\end{equation*}
where outer brackets are averages over the population, and the variance at the bottom is the variance over the population. For a non-ergodic system of the form in Fig.~\ref{FIG: Ergodicity}B, this autocorrelation would be an average between the autocorrelations of the two possible transcription rates. It's thus possible that a system will break the constraint given by Eq.~\eqref{EQ: No oscillation constraint (manuscript)} if only one of the sub-populations has an oscillating transcription rate. }

{\section{Flow cytometry data analysis}}

\begin{figure}[hbt!]
\includegraphics[width=0.8\columnwidth]{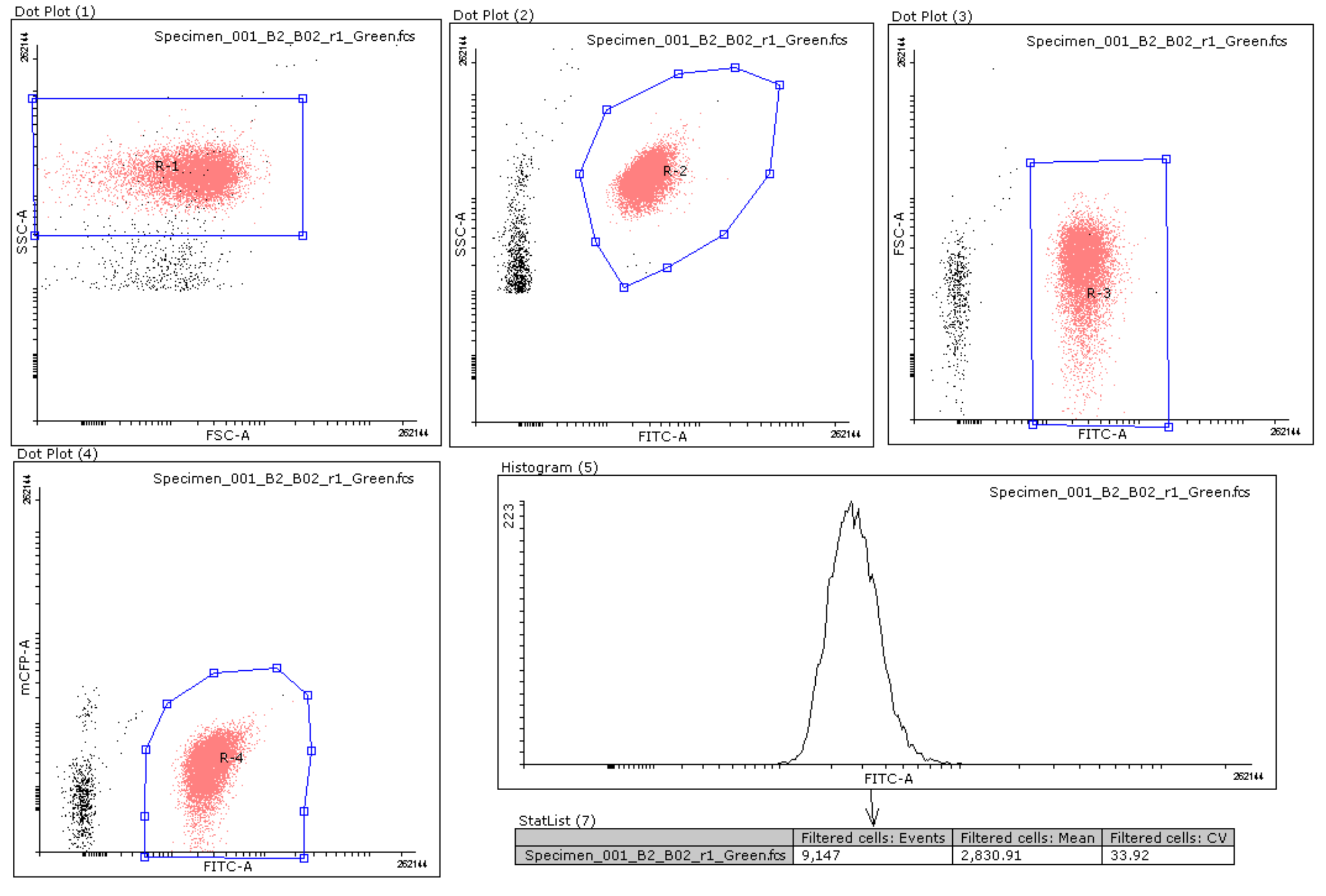}
\caption{ 
{
\textbf{Filtering cells from debris based on scattering and fluorescence profiles.} Here we plot a figure illustrating the filtering strategy that was used to filter out debris from the raw flow cytometry datasets. The side (SSC) and forward (FSC) scattering signals were used to separate the cluster of cell-like objects from debris measurements that lie outside the cluster. Additionally, the green (FITC) and cyan (mCFP) signals were used with the scattering signals to separate the cluster of fluorescent objects. On the bottom right is the resulting green fluorescence distribution for mEGFP, with statistics showing the abundance of post-sorted events (with pre-sorted at 10,000), the mean, and the coefficient of variation.}}
\label{FIG: flow software}
\end{figure} 

\begin{figure}[]
\includegraphics[width=0.68\columnwidth]{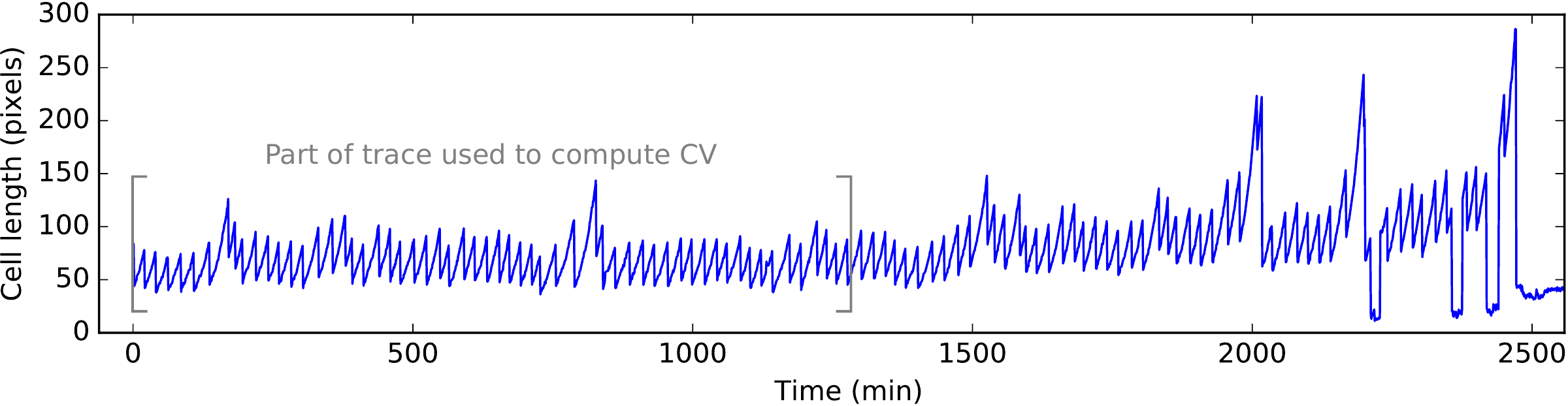}
\caption{{ \textbf{Example time-trace of growing and dividing \emph{E.~coli} cells.} Data taken from \cite{wang2010robust} where individual \emph{E.~coli} cells were followed using time-lapse microscopy and their lengths measured over time. One pixel corresponds to $0.0645\mu m$. Many of these time-traces exhibited large fluctuations near the end, suggesting a possible systematic error. We therefor computed the CVs of the first half of each time-trace.}}
\label{FIG: cell size time traces}
\end{figure}

{
For each fluorescent protein, coefficients of variation (CVs) were obtained for three independent cell cultures according to Fig.~\ref{FIG: flow software}, and the average was taken with an uncertainty given by the standard error. The CVs that were obtained this way correspond to abundance CVs. In order to obtain the concentration CVs, we used Eq.~\eqref{EQ: Appendix concentration CVs from abundance CVs (manuscript)} as described in Appendix \ref{SEC: Appendix Experimental data analysis}. Note that $CV_{V}$ can be estimated from first principles for a given cellular growth dynamics. For example, if we assume that the volume is growing exponentially between divisions with symmetric division times, then $CV_{V} \approx 0.2$ (similarly for a volume growing linearly). This is independent of the cell division time and can be used as a lower bound for $CV_{V}$.  In general, $CV_{V}$ will be bigger than 0.2 due to stochastic effects like asymmetric divisions and division times. We thus estimated $CV_{V}$ using separate data that explicitly quantifies \emph{E.~coli} growth dynamics through time-lapse microscopy \cite{wang2010robust}. In particular, in \cite{wang2010robust} individual \emph{E.~coli} cells are followed and their lengths are measured as a function of time for about 100 divisions, see Fig.~\ref{FIG: cell size time traces}. From the publicly available data we analyzed the length time-traces of the mother cells from one experiment with \emph{E.~coli} MG1655 (CGSC 6300) grown in LB media at 37$^\circ$C. This includes over 100 length time-traces like the one in Fig.~\ref{FIG: cell size time traces}, thus constituting approximately $10^4$ divisions. We noticed that many time-traces exhibited large fluctuations near the end (Fig.~\ref{FIG: cell size time traces}), which suggests a systematic error. Therefor, we computed the length CV from the first half of each time-trace, and then took the average, resulting in $CV_{V} = 0.261\pm0.005$. }

\begin{table}[b]
\includegraphics[width=0.75\columnwidth]{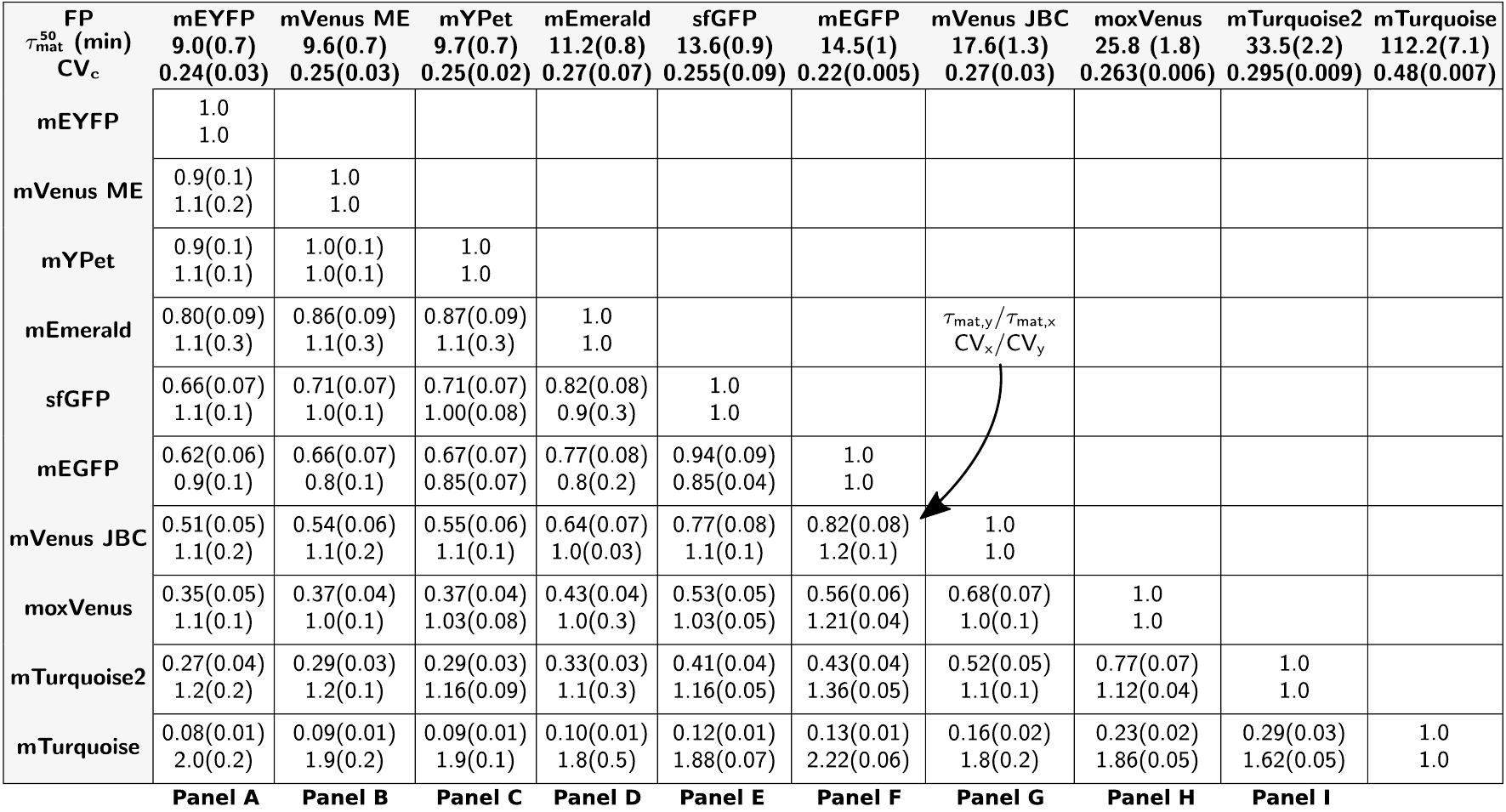}
\caption{{\textbf{Maturation times and concentration CVs for the selected fluorescent proteins.} In the top row we present the 10 selected fluorescent proteins from the publicly available data that are most closely modelled by first-order maturation kinetics, along with their maturation half-life (related to the maturation time through $\tau_{mat} = \tau_{mat}^{50}/\ln(2)$), and the concentration CVs computed from the flow cytometry datasets. Numbers in parentheses are the uncertainties, taken from \cite{Balleza2018} for the maturation times and taken as the standard error of three replicate flow cytometry data sets along with error propagation for the CVs. In the white cells we report the maturation time ratios and the ratio of CVs. Our constraints are unique for a given pair of CVs and maturation times, thus we show only one ratio for a given fluorescent protein pair. Each panel corresponds to a set of reporter pairs where the $Y$ reporter is held fixed and the $X$ reporter is varied (equivalent to varying $\tau_{mat,x}$ while keeping $\tau_{may,y}$ fixed). Each column corresponds to a panel in Fig.~\ref{FIG: All data}.  } }
\label{TAB: Data}
\end{table}

{
Of the 50 fluorescent proteins that were studied in \cite{Balleza2018}, varying maturation kinetics were observed, including first-order exponential kinetics as assumed in Fig.~\ref{FIG: 3 step class of systems with maturation (manuscript)}, second-order kinetics, and other more complex maturation kinetics. As a result, two effective maturation times are reported: $\tau_{mat}^{50}$ and $\tau_{mat}^{90}$, corresponding to the time it takes for 50\% or 90\% of an ensemble of fluorescent proteins to become mature, respectively. The former, $\tau_{mat}^{50}$, is the maturation half-life, which for first-order maturation kinetics as assumed in Fig.~\ref{FIG: 3 step class of systems with maturation (manuscript)} is related to the average maturation time through $\tau_{mat}^{50}/\tau_{mat} = \ln(2)$, and to $\tau_{mat}^{90}$ through $\tau_{mat}^{50}/\tau_{mat}^{90} = \ln(2)/\ln(10)$. In order to pick a subset of fluorescent proteins that are the most closely modelled by a first-order maturation step, we picked the 10 fluorescent proteins for which the experimental ratio $\tau_{mat}^{50}/\tau_{mat}^{90}$ lies closest to $\ln(2)/\ln(10)$. However, we found that this algorithm would capture a few fluorescent proteins whose maturation curves in \cite{Balleza2018} were noisy and could possibly also be described by a second-order process. Therefore, we picked the 10 fluorescent proteins for which the ratio $\tau_{mat}^{50}/\tau_{mat}^{90}$ lies closest to $\ln(2)/\ln(10)$ and that also display a clear first-order trend in the maturation kinetic time-traces reported in \cite{Balleza2018}. The chosen fluorescent proteins with their $\tau_{mat}^{50}$ maturation times and their concentration CVs are presented in Table \ref{TAB: Data}. }

\section{Simulations}
Though the constraints presented in the main text are analytically proven, we performed gillespie simulations to demonstrate {that the entire bounded regions are achievable}. Here we summarize what systems were simulated.

To demonstrate the open-loop constraint we performed gillespie simulations of certain systems that are part of the classes of Fig.~\ref{FIG: Class 1 space of solutions -- feedback (manuscript)}A and Fig.~\ref{FIG: 3 step class of systems with maturation (manuscript)}A of the paper. The systems that were simulated include those driven by a sinusoidal, those driven by an oscillating step function, those driven by a poisson process $z$, and those driven by the square of a poisson process $z^2$. In all these cases many simulations were performed with varying reporter lifetimes and reporter birthrate parameters (like the sinusoidal amplitude and frequency, or the half-life of the upstream poisson process, etc.). The time (co)variances for these particular system realizations was then computed, and these correspond to the blue dots in the paper figures. For closed-loop systems, we performed gillespie simulations of systems driven by rates with feedback. In particular, to fill most of the region in the figures, rates of the form $R(x) \propto (K + x^{n})^{-1}$, $R(y) \propto (K + y^{n})^{-1}$, and $R(y) \propto (K + y^{n} + x^{m})^{-1}$ were done over different $n$, $m$, and $K$. The region outside of the orange lines in Fig.~\ref{FIG: Class 1 space of solutions -- feedback (manuscript)}B and Fig.~\ref{FIG: 3 step class of systems with maturation (manuscript)}B near the $\rho = 1$ line was only accessible by rates of the form $R(x,y) \propto U(x - ky)$, where $U$ is a step function and $k$ is a parameter that was varied. 

To demonstrate the constraint on stochastic systems we performed gillespie simulations of open-loop systems that are known to be stochastic and periodic. These include those driven by a poisson process and those driven by a sinusoidal. We performed many simulations where the parameters of these specified birthrates were varied.

To simulate concentrations of growing and dividing cells, we performed gillespie simulations of molecular numbers that undergo a binomial split at particular times separated by the same period $\ln(2)\tau_{c}$. The volume was then modelled as an exponentially growing volume $V \propto e^{t/\tau_{c}}$ that is reduced by a factor of 1/2 at the same moments that the molecular numbers undergo a binomial split. The reporter numbers and the cell volume were kept track of, and from these the concentration (co)variances were computed from the time averages of $x(t)/V(t)$ and $y(t)/V(t)$. The concentration production rates $R_{c} := R/V$ that we simulated include poisson processes, sinusoidals, rates proportional and inversely proportional to the cell volume, and oscillating step functions.